\documentclass[oneside,a4paper,11pt,explicit]{book}
\def\ARXIVBUILD{1}
\ifdefined\ARXIVBUILD
\else
  \csname input\endcsname{arxiv_volume_III.tex}
\fi

\usepackage[T1]{fontenc}
\usepackage[utf8]{inputenc}
\usepackage{textcomp}
\usepackage{vol3_kultem}

\usepackage{amsmath,amssymb}
\usepackage{booktabs}
\usepackage{tabularx}
\usepackage{array}
\usepackage{longtable}
\usepackage{lscape}
\usepackage{graphicx}
\graphicspath{{./}}
\usepackage[dvipsnames,svgnames,table]{xcolor}
\usepackage[most]{tcolorbox}
\usepackage{float}
\usepackage{hyperref}

\makeatletter
\renewcommand*\l@section{\@dottedtocline{1}{1.5em}{2.8em}}
\renewcommand*\l@subsection{\@dottedtocline{2}{3.8em}{3.7em}}
\renewcommand*\l@subsubsection{\@dottedtocline{3}{7.0em}{4.6em}}
\makeatother

\usetikzlibrary{arrows.meta,calc,positioning,shapes.geometric,
  decorations.pathmorphing,backgrounds,fit}
\numberwithin{equation}{chapter}

\definecolor{ink}{HTML}{151B23}
\definecolor{titlebg}{HTML}{100880}
\definecolor{softink}{HTML}{263441}
\definecolor{muted}{HTML}{5C6875}
\definecolor{paper}{HTML}{F6F7F2}
\definecolor{panel}{HTML}{FFFFFF}
\definecolor{line}{HTML}{D8DEE5}
\definecolor{teal}{HTML}{007C77}
\definecolor{blue}{HTML}{245AA6}
\definecolor{gold}{HTML}{B86B00}
\definecolor{rose}{HTML}{A7354D}
\definecolor{green}{HTML}{23724A}
\definecolor{violet}{HTML}{5A4CA0}

\newcommand{\facility}{3.5-meter Segmented-Mirror Robotic Space Telescope}

\newcolumntype{Y}{>{\raggedright\arraybackslash}X}

\newtcolorbox{leadbox}[2][]{
  enhanced, colback=paper, colframe=#2, boxrule=0.9pt, arc=2mm,
  left=2.2mm, right=2.2mm, top=1.8mm, bottom=1.8mm,
  fonttitle=\sffamily\bfseries, coltitle=white,
  attach boxed title to top left={xshift=2mm,yshift=-2mm},
  boxed title style={colback=#2,arc=1.2mm,boxrule=0pt}, #1 }
\newtcolorbox{metricbox}[1]{
  enhanced, colback=white, colframe=#1, boxrule=0.65pt, arc=1.3mm,
  left=1.8mm, right=1.8mm, top=1.6mm, bottom=1.6mm }

\makeatletter
\renewenvironment{thebibliography}[1]
  {\section*{References}\@mkboth{}{}%
   \list{\@biblabel{\@arabic\c@enumiv}}%
        {\settowidth\labelwidth{\@biblabel{#1}}%
         \leftmargin\labelwidth \advance\leftmargin\labelsep
         \usecounter{enumiv}\let\p@enumiv\@empty
         \renewcommand\theenumiv{\@arabic\c@enumiv}}%
   \small\sloppy\clubpenalty4000\widowpenalty4000\sfcode`\.\@m}
  {\def\@noitemerr{\@latex@warning{Empty `thebibliography' environment}}\endlist}
\makeatother

\newcommand{\wpvolumelabel}{III}
\newcommand{\wpvolumetitle}{Key Scientific Mission: Exoplanet Science with a Coronagraph}
\title{3.5-meter Segmented-Mirror Robotic Space Telescope}
\subtitle{Mission White Paper: \wpvolumelabel. \wpvolumetitle}
\newcommand{\wpauthors}{Juhan Kim$^{1}$, Sang Hyun Lee$^{2,3}$, Yong-Woo Kang$^{2}$, Jeong-Yeol Han$^{2,4}$, Sungwook E. Hong$^{2,4}$, Bongkon Moon$^{2,4}$, Donguk Song$^{2}$, Juhyung Kang$^{2}$, Myeong-Gu Park$^{5}$, Sang Chul Kim$^{2,4}$, Chung-Uk Lee$^{2}$, Sangmo Tony Sohn$^{6}$, Arman Shafieloo$^{2,4}$, David Parkinson$^{2,4}$, Hong Soo Park$^{2,4}$, Dohyeong Kim$^{7}$, Chan Park$^{2}$, Jungjoo Sohn$^{8}$, Young-Beom Jeon$^{2}$, Jong-Hak Woo$^{9}$, Hyung Mok Lee$^{9}$, Hong Bae Ann$^{7}$, Myungkook James Jee$^{10}$, Mansoo Choi$^{2}$, Changbom Park$^{1}$}
\date{2026}
\newcommand{\wpabstract}{%
This volume defines the exoplanet science program enabled by the dedicated high-contrast coronagraph in the baseline science payload of the \facility. The observatory architecture incorporates the optical interfaces, wavefront sensing and control, pointing stability, and operations software required for coronagraphic observations from the outset. The observing strategy gives priority to the nearest stellar systems because they provide the most accessible laboratories for planetary exploration and the most likely destinations of future interstellar missions. The diffraction limit sets a reflected-light horizon of roughly 10--15 pc for planets at 1 AU and roughly 50--80 pc for Jupiter analogs. Within those horizons, the telescope can image nearby giant planets, obtain reflected-light spectra of their atmospheres, survey young systems and circumstellar disks, and support the habitability and biosignature programs that larger future missions will pursue. Published results from Kepler, TESS, and other transit surveys provide comparison samples for the directly imaged population. Stellar activity is monitored only to calibrate host-star variability in the atmospheric analysis. A systematic census of the nearest stellar neighbors provides a lasting reference for exoplanet science and future space exploration.
}
\newcommand{\wpkeywords}{\textbf{Keywords:} exoplanets, coronagraphy, direct imaging, reflected-light spectroscopy, nearby stars, circumstellar disks}
\newcommand{\wpchapteroffset}{2}
\newcommand{\MakeVolumeBody}{%
  \definecolor{Blue1}{HTML}{86A67C}%
  \definecolor{Blue2}{HTML}{5E8158}%
  \definecolor{Blue3}{HTML}{45613F}}

\newcommand{\MakeFrontCover}{%
  \begin{titlepage}
  \thispagestyle{empty}\sffamily\centering
  \noindent\colorbox{Blue3}{\parbox[t]{\dimexpr\textwidth-2\fboxsep\relax}{%
    \vspace{4mm}\centering
    {\color{white}\bfseries\fontsize{24}{29}\selectfont 3.5-meter Segmented-Mirror\\[1.5mm]
      Robotic Space Telescope}\\[3.5mm]
    {\color{white}\Large Mission White Paper}\\[1.5mm]
    {\color{white!88}\large \wpvolumelabel. \wpvolumetitle}%
    \vspace{4mm}}}
  \vfill
  \includegraphics[width=\textwidth]{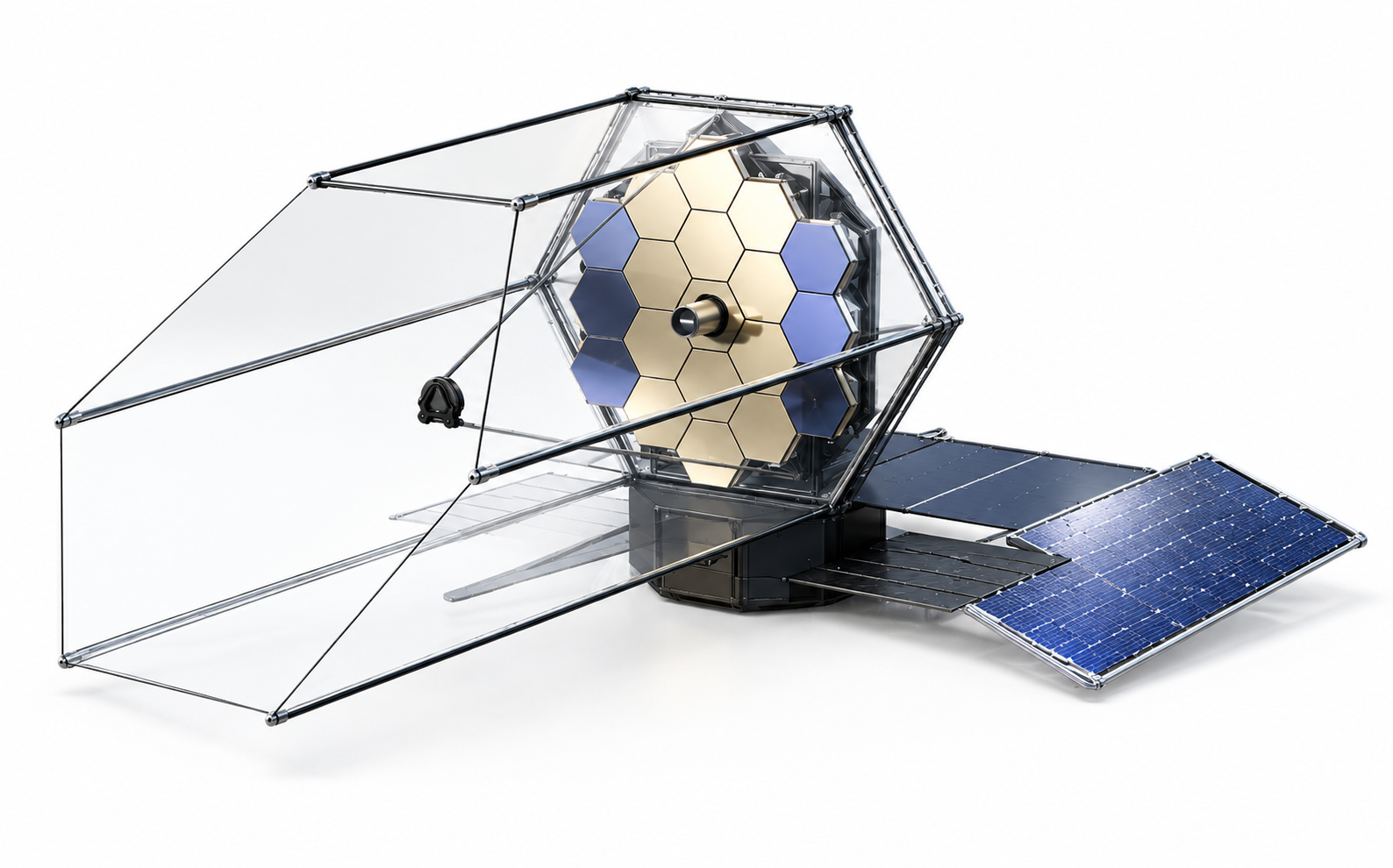}
  \vfill
  {\color{ink}\bfseries\normalsize \wpauthors\par}
  \vspace{3mm}
  {\color{muted}\small
    $^{1}$ Korea Institute for Advanced Study \textperiodcentered\
    $^{2}$ Korea Astronomy and Space Science Institute \\
    $^{3}$ University of Ulsan \textperiodcentered\
    $^{4}$ University of Science and Technology\\
    $^{5}$ Kyungpook National University \textperiodcentered\
    $^{6}$ Space Telescope Science Institute \textperiodcentered\
    $^{7}$ Pusan National University\\
    $^{8}$ Korea National University of Education \textperiodcentered\
    $^{9}$ Seoul National University \textperiodcentered\
    $^{10}$ Yonsei University\par}
  \vspace{5mm}
  {\color{Blue3}\bfseries\Large 2026}
  \vspace{4mm}
  \end{titlepage}}

\newcommand{\MakeColophon}{%
  \clearpage
  \thispagestyle{empty}\sffamily
  \null\vfill
  \noindent{\color{muted}\footnotesize
    {\color{ink}\bfseries 3.5-meter Segmented-Mirror Robotic Space Telescope}\\
    Mission White Paper: \wpvolumelabel. \wpvolumetitle\\[5mm]
    {\color{rose}\bfseries Release date September 17, 2026\\[5mm]}
    \textcopyright\ 2026 Korea Astronomy and Space Science Institute and the
    authors. All rights reserved.\\[2mm]
    Prepared by the mission study team at the Korea Astronomy and Space Science
    Institute, University of Ulsan, University of Science and Technology, the
    Space Telescope Science Institute, the Korea Institute for Advanced Study,
    Pusan National University, Korea National University of Education, Seoul
    National University, Kyungpook National University, and Yonsei University.\\[5mm]
    {\color{ink}\bfseries Suggested citation:} Kim, J., Lee, S.-H., Kang, Y.-W.,
    et al.\ (2026), \textit{3.5-meter Segmented-Mirror Robotic Space Telescope:
    Mission White Paper, \wpvolumelabel. \wpvolumetitle}.\\[2mm]
    {\color{ink}\bfseries Corresponding author:} Sang Hyun Lee
    \textperiodcentered\ \texttt{shlee@kasi.re.kr}.\par}
  \vspace{8mm}
  \clearpage}

\newcommand{\MakeBackCover}{%
  \clearpage
  \ifodd\value{page} \hbox{}\thispagestyle{empty}\newpage \fi
  \thispagestyle{empty}\null
  \clearpage
  \thispagestyle{empty}\sffamily\centering
  \null\vspace*{\stretch{1}}
  \includegraphics[width=\textwidth]{vol3_3.5ST_GPT.png}
  \vspace*{\stretch{2.2}}
  \noindent\colorbox{Blue3}{\parbox[t]{\dimexpr\textwidth-2\fboxsep\relax}{%
    \vspace{3mm}\centering
    {\color{white}\bfseries\Large 3.5-meter Segmented-Mirror Robotic Space Telescope}\\[1.5mm]
    {\color{white!88}\normalsize Mission White Paper: \wpvolumelabel. \wpvolumetitle}%
    \vspace{3mm}}}
  \vspace{3mm}
  \clearpage}

\hypersetup{
  colorlinks=true,
  linkcolor=blue,
  citecolor=teal,
  urlcolor=gold,
  pdftitle={3.5-meter Segmented-Mirror Robotic Space Telescope -- \wpvolumelabel. \wpvolumetitle},
  pdfauthor={Juhan Kim et al.}}

\begin{document}
\frontmatter
\MakeFrontCover
\MakeColophon
\thispagestyle{fancy}
\vspace*{0.5em}
\noindent{\Large\bfseries Abstract}\par\smallskip
\noindent\wpabstract\par\medskip
\noindent\wpkeywords
\clearpage
\tableofcontents

\mainmatter
\renewcommand{\chaptername}{Volume}
\renewcommand{\thechapter}{\Roman{chapter}}
\setcounter{chapter}{\wpchapteroffset}
\MakeVolumeBody
% Explicit inclusion is required for arXiv's dependency scanner.
\chapter{Exoplanet Science with a Coronagraph}\label{app:exoplanets}
\vspace{8mm}
%%%%%%%%%%%%%%%%

%%%%%%%%%%%%%
\section{Exploring the Nearest Stellar Systems}
\label{sec:Nearest_system}

Recent space-observatory concepts have established the value of combining
high-contrast imaging, time-domain monitoring, and optical or near-infrared
spectroscopy \cite{roy2026III,wevers2026III}. This volume defines a more specific
3.5-metre program. It ranks targets by distance, applies one uniform
selection rule to the nearest stellar components, and uses repeated
coronagraphic observations to measure separation, orbital motion, and
spectral properties. The nearest 20 components in 13 systems therefore form
a controlled reference sample rather than a list assembled from previously
known planets or from expected discovery probability \cite{Henry2018,Gaia2022}.

The distance ranking follows a direct geometric argument. At a fixed orbital
radius, a nearby planet has a larger angular separation and a higher photon
rate. The same targets also permit repeated measurements of common proper
motion, orbital curvature, stellar variability, and the stability of the
coronagraphic subtraction \cite{Traub2010,Macintosh2015}. The program thus
tests planetary occurrence and system architecture while preserving the
observational completeness needed to interpret non-detections. This emphasis
on a defined nearby-star sample and on mission-specific observing cadence
defines the observing strategy through a nearby-star sample, repeated
coronagraphic visits, and reflected-light spectroscopy.

\subsection{Scientific Rationale for Exploring the Nearest Stars}
\label{sec:Rationale_Nearest}

The nearest stellar systems provide the most favorable environment for the direct 
imaging of exoplanets. At a given orbital radius, planets around nearby stars exhibit larger angular separations, allowing observations closer to their host stars within the same inner working angle of the coronagraph \cite{Traub2010}. 
Their proximity also results in higher photon fluxes, enabling improved signal-to-noise ratios and more efficient spectroscopic characterization \cite{LUVOIR2019}. 
Furthermore, repeated observations over the mission lifetime allow orbital motion, common proper motion, and temporal variability to be measured with significantly higher precision than for more distant targets \cite{Macintosh2015}. Consequently, nearby stars maximize the scientific return of a
high-contrast imaging mission by providing the best opportunity not only to detect
exoplanets but also to characterize their physical and dynamical properties
\cite{LUVOIR2019}.

Most direct-imaging surveys inevitably introduce target-selection biases by prioritizing stars with known planets, favorable stellar properties, or expected high detection probabilities \cite{Nielsen2019}. 
While such strategies maximize the likelihood of discovery, they do not necessarily provide a representative view of the planetary systems in the Solar neighborhood. In contrast, a distance-ranked sample minimizes prior selection effects by adopting proximity as the sole primary criterion \cite{Henry2018}. 
This approach naturally includes stars of different spectral types, ages, multiplicities, and planetary architectures, regardless of whether planets have already been detected \cite{Henry2018}. 
Such a distance-ranked nearest-neighbor strategy ensures that both detections and non-detections contribute equally to constraining the diversity and occurrence of nearby planetary systems \cite{Bowler2016}.

A complete survey of every nearby star is beyond the scope of a single mission. Direct imaging requires repeated observations to confirm planetary candidates, reject background contaminants, measure orbital motion, and obtain spectroscopic follow-up \cite{Currie2022}. 
These requirements become even more demanding for multiple stellar systems \cite{HabEx2020}. 
The primary exploration sample is therefore defined as the 20 nearest stellar components belonging to 13 stellar systems. This sample size represents a practical
balance between exploration and scientific depth: it is sufficiently small to permit repeated and comprehensive observations throughout the mission, while sufficiently diverse to include single stars, binaries, hierarchical multiple systems, and a broad range of stellar types. Rather than defining a physical boundary, the nearest 20 stars establish a mission-scale benchmark for systematic exploration of the Solar neighborhood.

The nearest stellar sample provides an observational benchmark that extends well beyond the discovery of individual exoplanets. Its proximity enables sensitive searches for long-period giant planets, circumstellar debris structures, and faint companions that remain inaccessible around more distant stars \cite{Bowler2016}. 
Published radial-velocity, astrometric, and transit measurements complement
direct imaging by constraining the orbital architecture of nearby planetary
systems \cite{LUVOIR2019}.
Equally important, the absence of detectable planets around individual stars provides quantitative constraints on planetary occurrence rates and formation scenarios 
\cite{Bowler2016}. 
The resulting dataset will therefore serve as the definitive observational reference for the nearest planetary systems and as a calibration sample for larger statistical surveys extending to greater distances \cite{LUVOIR2019}.

The primary exploration program is designed to balance the goals of systematic exploration and detailed scientific investigation \cite{LUVOIR2019,HabEx2020}. 
Rather than allocating observing time exclusively to either a few well-known planetary systems or a large statistical sample, the mission adopts a tiered strategy.
The nearest 20 stellar components constitute the highest-priority exploration sample, receiving repeated observations throughout the mission to establish the most complete observational legacy of the Solar neighborhood. This foundation is complemented by an extended survey of additional nearby stars and specialized observing programs targeting known planetary systems, young stars, and debris disks
\cite{LUVOIR2019,HabEx2020}. 
Such a strategy combines the completeness expected of an exploration mission with the statistical and astrophysical investigations required for modern exoplanet science.

\subsection{The Primary Sample of the 20 Nearest Stellar Components}

\begin{table*}[ht]
\small
\begin{tabularx}{\textwidth}{l l c c X c}
\hline
\textbf{Stellar System} &
\textbf{Star} &
\textbf{Distance} &
\textbf{Spectral} &
\textbf{Known Planets} &
\textbf{Separation} \\
&
&
\textbf{(pc)} &
\textbf{Type} &
&
\textbf{($^{\prime\prime}$)}\\
\hline
Alpha Centauri & Proxima Cen & 1.301 & M5.5 Ve & b, d (c candidate) & $>$10000 \\
Alpha Centauri & $\alpha$ Cen A & 1.338 & G2 V & None confirmed & 17.5 \\
Alpha Centauri & $\alpha$ Cen B & 1.338 & K1 V & None confirmed & 17.5 \\
Barnard's Star & Barnard & 1.826 & M4 V & b, c, d, e & --- \\
Wolf 359 & Wolf 359 & 2.386 & M6 V & None confirmed & --- \\
Lalande 21185 & GJ 411 & 2.547 & M2 V & b, c & --- \\
Sirius & Sirius A & 2.637 & A1 V & None confirmed & 7.5 \\
Sirius & Sirius B & 2.637 & DA2 & None confirmed & 7.5 \\
GJ 65 & BL Ceti & 2.687 & M5.5 Ve & Candidate & 2.1 \\
GJ 65 & UV Ceti & 2.687 & M6 Ve & Candidate & 2.1 \\
Ross 154 & Ross 154 & 2.973 & M3.5 Ve & None confirmed & --- \\
Ross 248 & Ross 248 & 3.156 & M5.5 V & None confirmed & --- \\
$\epsilon$ Eridani & $\epsilon$ Eri & 3.216 & K2 V & b & --- \\
Lacaille 9352 & GJ 887 & 3.289 & M1 V & b, c & --- \\
Ross 128 & Ross 128 & 3.374 & M4 V & b & --- \\
EZ Aquarii & Aa & 3.401 & M5 V & None confirmed & 0.009 \\
EZ Aquarii & Ab & 3.401 & M5 V & None confirmed & 0.009 \\
EZ Aquarii & B & 3.401 & M5 V & None confirmed & 0.36$^{a}$ \\
61 Cygni & A & 3.497 & K5 V & None confirmed & 24 \\
61 Cygni & B & 3.497 & K7 V & None confirmed & 24 \\
\hline
\end{tabularx}
\vspace{2mm}
\caption{
The 20 nearest stars to the Sun selected as the primary exploration sample of the mission.
The sample contains 20 stellar components belonging to 13 stellar systems.
Distances and spectral classifications were compiled from the volume-complete nearby-star census \cite{reyle21}
and Gaia DR3 \cite{gaiadr3}.
Planetary information follows the NASA Exoplanet Archive.
For multiple systems, the listed separation gives the characteristic angular separation to the nearest stellar companion.
}
\label{tab:nearest20}
\footnotesize
$^{a}$ For hierarchical multiple systems, the listed separation refers to the angular separation from the nearest stellar companion.
\end{table*}

The primary target sample consists of the 20 nearest stellar components to the Sun,
which belong to 13 stellar systems \cite{Henry2018,Gaia2022}. 
The targets were selected strictly in order of distance, without excluding stars on the basis of spectral type, stellar activity, multiplicity, or the prior detection of planets. Consequently, the outer boundary of the sample lies near 3.5 pc, at the distance of the 61 Cygni system, but this distance is a result of the 20-star selection rather than an independently imposed volume limit. Stellar distances, spectral classifications, and system membership were compiled from modern nearby-star catalogues and Gaia astrometry \cite{Henry2018,Gaia2022,Golovin2023}, 
while the status of confirmed and candidate planets was assessed using the NASA Exoplanet Archive and the relevant discovery literature \cite{Akeson2013}. 
For multiple systems, angular separations were adopted from published orbital solutions or representative projected separations \cite{Hartkopf2001}.
Because these separations vary with orbital phase, the values listed in Table~\ref{tab:nearest20} should be interpreted as characteristic angular scales for planning coronagraphic observations rather than fixed instantaneous separations.

\begin{figure}[tp]
\centering
\includegraphics[width=0.82\textwidth]{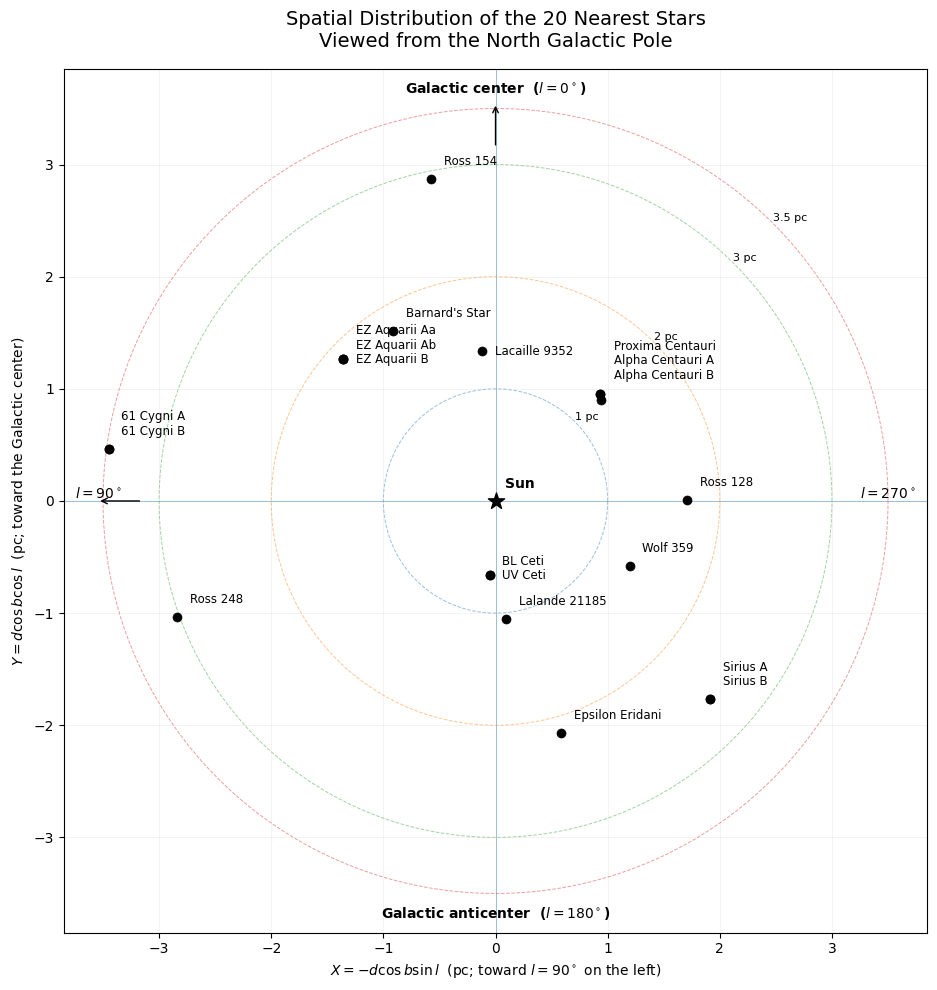}
\caption{Spatial distribution of the 20 nearest stellar components to the Sun projected onto the Galactic plane. The positions of the 20 nearest stellar components are shown in Galactic Cartesian coordinates, assuming the Sun at the origin. The map is viewed from the North Galactic Pole, with the Galactic center ($\ell=0^\circ$) at the top, Galactic longitude $\ell=90^\circ$ to the left, $\ell=180^\circ$ at the bottom, and $\ell=270^\circ$ to the right. Concentric circles indicate heliocentric distances of 1, 2, 3, and 3.5 pc. The coordinates were calculated from the heliocentric distance and Galactic longitude and latitude of each star. Components of multiple stellar systems are plotted individually, although their separations are too small to be resolved on this parsec-scale map.}
\label{fig:20nearist_projected}
\end{figure}

A substantial fraction of the sample can be approached using conventional single-star coronagraphic observing strategies \cite{Guyon2006}. 
Isolated systems such as Barnard's Star, Wolf 359, Lalande 21185, Ross 154, Ross 248, $\epsilon$ Eridani, Lacaille 9352, and Ross 128 are not known to contain bright stellar companions close enough to generate a second dominant diffraction pattern within the high-contrast search region. Proxima Centauri is gravitationally associated with Alpha Centauri AB, but its very wide separation allows it to be treated as an effectively independent coronagraphic target \cite{Beichman2020}. 
These systems are therefore comparatively favorable with respect to stellar multiplicity, although their overall detectability still depends on stellar brightness, activity, planetary angular separation, dust emission, and the contrast between the star and its planets. In particular, the predominance of M dwarfs in the
nearest-star sample offers favorable planet-to-star flux ratios in some observing bands, but their compact habitable zones and frequent magnetic activity introduce separate limitations \cite{Shields2016}. 
Thus, “favorable” in this context refers primarily to the absence of contaminating light from a nearby stellar companion, rather than to uniformly high sensitivity to all classes of planets.

Several high-priority targets are spatially resolved multiple systems whose stellar components can, in principle, be observed separately. Alpha Centauri A and B, Sirius A and B, and 61 Cygni A and B have angular separations larger than the nominal inner working angle of the coronagraph. Nevertheless, a resolved companion does not cease to be an important source of contamination \cite{Guyon2006}. 
Light from the off-axis star can generate diffraction features, scattered light, detector dynamic-range requirements, and time-variable speckle structures across the planetary discovery region \cite{Sirbu2017}. 
The severity of these effects depends on the instantaneous projected separation, position angle, brightness ratio, observing wavelength, and optical field of view. The relatively wide 61 Cygni system is expected to be the least demanding of these resolved binaries, whereas Alpha Centauri and Sirius require more careful control of companion light because of their high apparent brightness and, in the case of Sirius, the extreme contrast between the stellar components \cite{Beichman2020}. 
Observations of these systems should therefore be planned for favorable orbital configurations and telescope orientations, with the usable discovery region and achieved contrast evaluated separately for each component \cite{LUVOIR2019}.

The most challenging targets in the primary sample are the close binary GJ 65 and the hierarchical triple EZ Aquarii. In GJ 65, the two M-dwarf components are separated by only a few arcseconds, so the diffraction and speckle fields of both stars may affect the same high-contrast search region \cite{Thomas2015}. 
EZ Aquarii presents a still more complex configuration: its inner Aa--Ab pair is unresolved at conventional coronagraphic angular resolution, while the third component lies sufficiently close to influence observations of the inner subsystem. The 20 stellar components therefore do not correspond to 20 equivalent and independently suppressible coronagraphic targets. For an unresolved subsystem such as EZ Aquarii Aa--Ab, the appropriate objective is not necessarily an independent circumstellar search around each component, but a search for planets or other companions in dynamically stable circumbinary regions \cite{Holman1999}. 
Close multiple systems may require multi-star wavefront control, restricted discovery zones, specialized reference-star subtraction, observations at selected orbital phases, or a combination of these methods \cite{Sirbu2017}. 
Their sensitivity and completeness must consequently be calculated separately from those of conventional single-star observations \cite{Currie2022}.

The scientific inventory will retain all 20 stellar components, but observing time and operational planning will be assigned primarily at the level of the 13 stellar systems. The number of pointings, visits, telescope orientations, and specialized observing modes will therefore differ among targets according to system architecture and coronagraphic complexity \cite{LUVOIR2019}.
Conventional single-star systems may be observed through a common multi-epoch sequence, whereas resolved or close multiple systems will receive customized strategies and, where necessary, additional calibration time \cite{LUVOIR2019}. 
No target will be removed from the primary sample solely because its multiplicity reduces the expected sensitivity. Excluding difficult systems would compromise the completeness of a distance-ranked census and reintroduce a selection bias in favor of simple stellar environments \cite{Henry2018}. 
Instead, the mission will report target-specific discovery regions, contrast limits, and completeness functions, clearly distinguishing regions that have been observationally excluded from regions that remain technically unconstrained \cite{Bowler2016}. 
This system-level policy preserves the exploration principle of surveying the 20 nearest stars while maintaining a physically realistic allocation of mission resources
\cite{LUVOIR2019}.

Although the present survey is intentionally limited to the 20 nearest stellar components, it is designed as the first stage of a progressively expanding nearby-star imaging program \cite{LUVOIR2019}. 
The initial sample provides the highest angular resolution, the greatest sensitivity to planets at small physical separations, and the most comprehensive characterization of the stellar environments closest to the Solar System \cite{Guyon2006}. 
These observations will establish a homogeneous reference dataset against which future observations of more distant stars can be directly compared. As coronagraph performance, wavefront-control techniques, and post-processing algorithms continue to improve, the survey can naturally expand to increasingly larger distance-limited samples while preserving the same objective target-selection criteria \cite{Currie2022}. 
Consequently, the scientific value of the nearest-star survey extends well beyond its immediate discoveries. It establishes a long-term legacy archive for comparative exoplanet science, provides benchmark targets for future space and ground-based high-contrast instruments, and forms the observational foundation for a complete census of planetary systems in the Solar neighborhood \cite{LUVOIR2019}.

\section{Search for Habitable Worlds around the Nearest FGK Stars}
\label{sec:Search-Habitable}

Following the distance-prioritized exploration of the nearest stellar systems in
Section~\ref{sec:Nearest_system}, this mission places a second major priority on the search for potentially
habitable worlds around the nearest F-, G-, and K-type stars. The program selects 
approximately 20 nearby FGK stars whose habitable zones are fully or substantially 
accessible outside the coronagraphic inner working angle, providing the most favorable 
targets for the direct detection and characterization of Earth-like planets
\cite{kopparapu13,Stark2014}. Repeated high-contrast observations will be used to 
search for planets, constrain their orbits, and identify candidates located within the 
habitable zone. If a potentially habitable planet is discovered, it will become a 
highest-priority target for reflected-light spectroscopy to characterize its atmosphere 
and search for molecular signatures associated with habitability and possible biological 
activity \cite{Schwieterman2018}. In this way, the program extends the exploration of 
our nearest stellar neighbors toward the fundamental question of whether potentially 
life-bearing worlds exist in the immediate neighborhood of the Sun.

\subsection{Scientific Rationale: The Nearest Habitable Worlds}
\label{sec:habitable-rationale}

The search for habitable worlds is most meaningful when it begins with planetary
systems that can be compared directly with the only known example of a life-bearing
planet: the Earth--Sun system. F-, G-, and K-type main-sequence stars provide a natural
starting point because they include solar-type stars and offer relatively stable stellar
environments in which terrestrial planets may receive stellar irradiation comparable to
that received by Earth \cite{kopparapu13}. Their habitable zones also lie at
substantially larger orbital separations than those of lower-mass stars, making potentially
habitable planets more favorable targets for spatially resolved direct imaging
\cite{kopparapu13,Tuchow2024}. This does not imply that planets around M dwarfs
are uninhabitable; rather, FGK systems provide Solar-System-like environments while
avoiding some of the major uncertainties associated with the very close-in habitable
zones of low-mass stars, including strong stellar activity and its effects on planetary
atmospheres \cite{Shields2016}.

Proximity is equally fundamental to this program. For a planet at a given orbital
separation, the apparent star--planet separation increases inversely with distance, making
the habitable zones of the nearest FGK stars more likely to extend beyond the
coronagraphic inner working angle. At the same time, the received planetary flux
increases strongly with proximity, providing the photons required not only for detection
but also for repeated imaging and reflected-light spectroscopy
\cite{Tuchow2024,Stark2019}. The nearest FGK stars therefore offer a uniquely favorable
combination of Solar-System-like habitable environments, spatially resolvable habitable
zones, and sufficient planetary photon flux for atmospheric characterization. They are
consequently the most compelling targets for this mission to progress from the detection
of nearby planets to the characterization of potentially habitable worlds and, ultimately,
the search for atmospheric signatures that may indicate biological activity
\cite{Schwieterman2018}.

\subsection{Target Selection and Observable Habitable Zones}
\label{sec:hz-target-selection}

The selection of targets for direct imaging of potentially habitable planets is
governed by both the physical extent of the habitable zone (HZ) and the angular
separation accessible to the coronagraph. The location of the HZ depends primarily
on stellar luminosity and spectral energy distribution and therefore varies
systematically across the FGK main-sequence spectral sequence
\cite{kopparapu13,kopparapu14}. Representative effective temperatures and
luminosities for the individual spectral subtypes are adopted from empirical
main-sequence stellar calibrations \cite{pecaut13}.

The HZ boundaries are calculated using the effective stellar-flux formulation
developed for main-sequence stars \cite{kopparapu13,kopparapu14}. Defining

\begin{equation}
T' = T_{\rm eff} - 5780~{\rm K},
\label{eq:hz-temperature}
\end{equation}

\noindent
the effective stellar flux corresponding to a given HZ boundary is expressed as

\begin{equation}
S_{\rm eff}
\approx
S_{\rm eff,\odot}
+aT'
+bT'^2
+cT'^3
+dT'^4 ,
\label{eq:seff}
\end{equation}

\noindent
where $S_{\rm eff,\odot}$ is the effective flux for the corresponding HZ boundary
around a solar-temperature star. The coefficients $a$, $b$, $c$, and $d$ are
polynomial fitting coefficients derived from grids of one-dimensional
radiative--convective climate-model calculations. Separate coefficient sets are
provided for the Recent Venus, runaway-greenhouse, maximum-greenhouse, and
Early Mars boundaries \cite{kopparapu13,kopparapu14}. These coefficients
describe the temperature dependence of the calculated climate limits and should
not be interpreted as separate physical processes.

For the present target-selection analysis, a fiducial $1\,M_{\oplus}$ terrestrial
planet is adopted. The mass-dependent runaway-greenhouse inner boundary is
therefore evaluated using the $1\,M_{\oplus}$ coefficients given by
Kopparapu et al.~\cite{kopparapu14}. The Recent Venus, maximum-greenhouse,
and Early Mars limits use the corresponding coefficient sets given in the same
work. This provides a uniform Earth-analog reference for constructing the target
sample before the physical properties of any individual planet are known. Once
a planet is detected and its mass, radius, or other relevant properties become
constrained, its location relative to the HZ can be reassessed using
planet-specific parameters.

\begin{figure}[tp]
\centering
\includegraphics[width=1.0\textwidth]{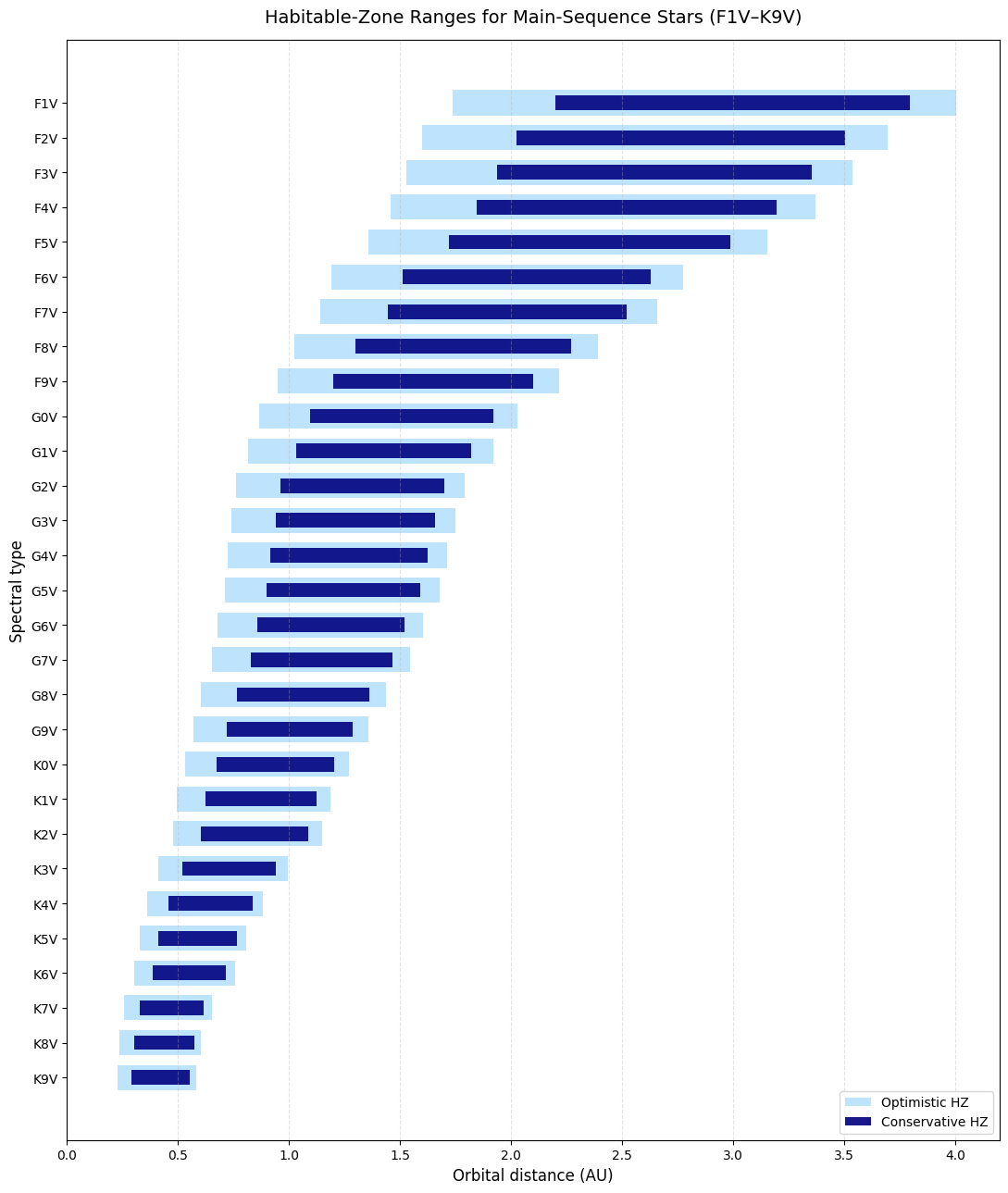}
\caption{Habitable-zone ranges for FGK main-sequence stars.
Optimistic (light blue) and conservative (dark blue) habitable-zone (HZ) boundaries are shown as a function of spectral type from F1V to K9V. The conservative HZ is bounded by the runaway-greenhouse inner limit and the maximum-greenhouse outer limit, whereas the optimistic HZ extends from the Recent Venus inner limit to the Early Mars outer limit. As stellar luminosity decreases toward later spectral types, both HZs move progressively inward. These ranges provide the physical basis for evaluating the angular accessibility of potentially habitable planets with the coronagraph.}
\label{fig:HZ_range}
\end{figure}

For a star of luminosity $L_\star$, the orbital distance corresponding to each
HZ boundary is

\begin{equation}
d_{\rm HZ}
\approx
\left(
\frac{L_\star/L_\odot}{S_{\rm eff}}
\right)^{1/2}
{\rm AU},
\label{eq:hz-distance}
\end{equation}

\noindent
where $L_\star/L_\odot$ is the stellar luminosity relative to the Sun. The
conservative HZ is defined from the runaway-greenhouse inner limit to the
maximum-greenhouse outer limit, whereas the optimistic HZ extends from the
empirical Recent Venus inner limit to the Early Mars outer limit
\cite{kopparapu13,kopparapu14}. As illustrated in
Figure~\ref{fig:HZ_range}, the HZ occurs at larger orbital radii around more
luminous F-type stars and progressively moves inward toward later G- and K-type
stars. The optimistic HZ is retained as a broader discovery region, while the
conservative HZ provides the primary reference for assigning observing priority.

For direct imaging, however, the physical location of the HZ alone is insufficient.
The corresponding angular separation on the sky must also exceed the coronagraphic
inner working angle (IWA). For a HZ boundary at orbital distance $d_{\rm HZ}$
around a star at distance $d_\star$, the corresponding angular scale is

\begin{equation}
\theta_{\rm HZ}({\rm arcsec})
=
\frac{d_{\rm HZ}({\rm AU})}{d_\star({\rm pc})},
\label{eq:hz-angular}
\end{equation}

\noindent
because 1 AU subtends 1 arcsec at a distance of 1 pc. This quantity represents
the angular scale associated with the physical orbital separation; the
instantaneous projected separation of an actual planet may be smaller depending
on orbital inclination and phase.

For the baseline coronagraph, the IWA is adopted as

\begin{equation}
\theta_{\rm IWA}
=
3\frac{\lambda}{D},
\label{eq:IWA}
\end{equation}

\noindent
which corresponds to approximately $0.097$ arcsec for $D=3.5$ m and
$\lambda=550$ nm. The IWA is one of the principal parameters controlling the
accessible discovery space and expected yield of directly imaged terrestrial
planets \cite{Stark2014}.

Combining Equations~\ref{eq:hz-distance}--\ref{eq:IWA} defines a
spectral-type-dependent distance horizon for HZ imaging. For an HZ with inner
and outer boundaries $d_{\rm in}$ and $d_{\rm out}$, respectively, the entire HZ
is geometrically accessible when

\begin{equation}
\frac{d_{\rm in}}{d_\star}
>
\theta_{\rm IWA},
\label{eq:entire-hz}
\end{equation}

\noindent
whereas only part of the HZ is accessible when

\begin{equation}
\frac{d_{\rm in}}{d_\star}
\leq
\theta_{\rm IWA}
<
\frac{d_{\rm out}}{d_\star}.
\label{eq:partial-hz}
\end{equation}

\begin{figure}[tp]
\centering
\includegraphics[width=1.0\textwidth]{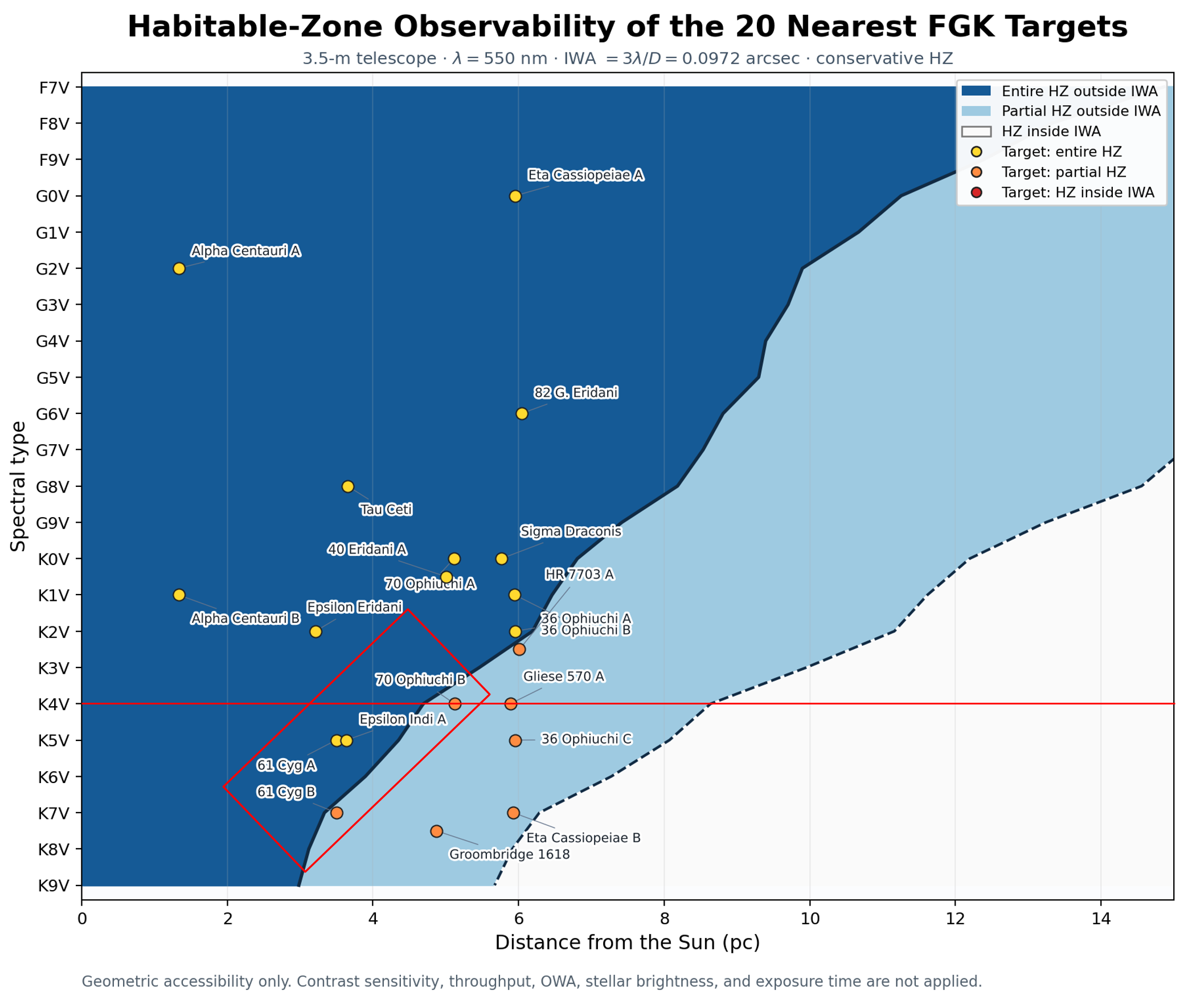}
\caption{Observability of the conservative habitable zones around FGK stars with the 3.5-m telescope.
The accessible distance range is shown as a function of spectral type from F1V to K9V, adopting a coronagraphic inner working angle of $3\lambda/D=0.097$ arcsec at 550 nm. The observability classification is based on the conservative habitable zone (HZ), bounded by the runaway-greenhouse inner limit and the maximum-greenhouse outer limit. Dark blue indicates systems for which the entire conservative HZ lies outside the IWA (Entire HZ Observable), light blue indicates systems for which only part of the conservative HZ is accessible (Partial HZ Observable), and white indicates systems for which the conservative HZ remains inaccessible within the IWA. The solid and dashed curves mark the maximum distances for entire and partial HZ accessibility, respectively. These boundaries define the primary geometric criterion for selecting the nearby FGK target sample.The red horizontal line marks the approximate K4 boundary beyond which Earth-like reflected-light contrasts approach the $10^{-9}$ regime, while the red box highlights the nearby late-K targets discussed further in Section~\ref{sec:earthlike_multi_epoch}.}
\label{fig:fgk20-observability}
\end{figure}

\noindent
If

\begin{equation}
\theta_{\rm IWA}
\geq
\frac{d_{\rm out}}{d_\star},
\label{eq:unobservable-hz}
\end{equation}

\noindent
the HZ is geometrically inaccessible at the adopted wavelength. These three
regimes are illustrated in Figure~\ref{fig:fgk20-observability}. Because the HZs
of more luminous F-type stars occur at larger orbital radii, they can remain
spatially accessible at greater stellar distances, whereas progressively closer
systems are required toward later G- and K-type stars.

The target-selection strategy is deliberately inclusive while maintaining a
conservative observing priority. Systems with either entire or partial HZ
accessibility are retained in the broader discovery sample, and the optimistic
HZ is also considered where geometrically accessible. The highest observing
priority, however, is assigned to the nearest FGK main-sequence stars for which
the entire conservative HZ lies outside the IWA, followed by systems for which
only part of the conservative HZ is accessible. This approach avoids prematurely
excluding nearby systems with potentially habitable discovery space while
concentrating the primary observing effort on the most robust and geometrically
complete HZ targets.

The Entire/Partial classification used here represents a geometric accessibility
criterion only. It does not by itself account for orbital inclination and phase,
planet--star contrast, exozodiacal background, coronagraph throughput, integration
time, or detection completeness. These effects will be incorporated in subsequent
end-to-end performance and observing-schedule analyses.

\subsection{The Primary Sample of 20 Nearby FGK Stars}
\label{sec:20NB_FGK}

%% tabel 
\begin{table}[htbp]
\centering
\caption{The 20 nearest F, G, K type main-sequence stellar 
components and the angular extent of their conservative habitable 
zones relative to the inner working angle (IWA) of the 3.5-m 
space telescope.}
\label{tab:fgk20}
\small
\begin{tabular}{l c r c}
\toprule
Star & Spectral Type & Distance (pc) &
Conservative HZ (IWA units) \\
\midrule
Alpha Centauri A    & G2V    & 1.332 & 7.43--13.10 \\
Alpha Centauri B    & K1V    & 1.332 & 4.83--8.68 \\
Epsilon Eridani     & K2V    & 3.212 & 1.93--3.48 \\
61 Cygni A          & K5V    & 3.496 & 1.21--2.26 \\
61 Cygni B          & K7V    & 3.496 & 0.96--1.81 \\
Epsilon Indi A      & K5V    & 3.638 & 1.17--2.17 \\
Tau Ceti            & G8V    & 3.652 & 2.15--3.83 \\
Groombridge 1618    & K7.5Ve & 4.871 & 0.66--1.26 \\
40 Eridani A        & K0.5V  & 5.010 & 1.33--2.39 \\
70 Ophiuchi A       & K0V    & 5.114 & 1.35--2.42 \\
70 Ophiuchi B       & K4V    & 5.122 & 0.91--1.68 \\
Sigma Draconis      & K0V    & 5.764 & 1.20--2.15 \\
Gliese 570 A        & K4V    & 5.886 & 0.80--1.47 \\
Eta Cassiopeiae B   & K7V    & 5.927 & 0.57--1.07 \\
36 Ophiuchi B       & K1V    & 5.948 & 1.08--1.94 \\
36 Ophiuchi A       & K2V    & 5.951 & 1.04--1.88 \\
36 Ophiuchi C       & K5V    & 5.954 & 0.71--1.32 \\
Eta Cassiopeiae A   & G0V    & 5.954 & 1.89--3.32 \\
HR 7703 A           & K2.5V  & 6.012 & 0.96--1.74 \\
82 G. Eridani       & G6V    & 6.041 & 1.46--2.59 \\
\bottomrule
\end{tabular}

\vspace{0.5em}
\begin{minipage}{0.96\linewidth}
\footnotesize
\textit{Note.} The sample consists of the 20 nearest F-, G-, and K-type
main-sequence stellar components. Distances are based on nearby-star
astrometry, including Gaia and Hipparcos measurements \cite{Gaia2022}.
The conservative habitable-zone boundaries are calculated as described
in Section~\ref{sec:hz-target-selection}. HZ angular separations are
expressed in units of the coronagraph inner working angle,
$\mathrm{IWA}=3\lambda/D$, adopting $D=3.5$ m and $\lambda=550$ nm,
corresponding to $\mathrm{IWA}=0.0972''$. An inner HZ boundary greater
than 1 IWA indicates that the entire conservative HZ is geometrically
accessible, whereas a range spanning 1 IWA indicates partial HZ
accessibility. These values represent geometric accessibility only and
do not include contrast, orbital phase, exozodiacal background,
throughput, or integration-time requirements.
\end{minipage}
\end{table}

Applying the selection criteria described in Section~\ref{sec:hz-target-selection}, we define a primary
sample of the 20 nearest F-, G-, and K-type main-sequence stellar components.
The restriction to main-sequence stars is intentional. Evolved stars, including
subgiants, are excluded even when they are nearby and have F-, G-, or K-type
spectral classifications, because their luminosities and habitable-zone locations
evolve substantially with time \cite{DanchiLopez2013} and do not provide the
long-term, Solar-System-like environments that motivate this survey.

Restricting the sample to main-sequence stars therefore provides a more physically
homogeneous population for evaluating habitable-zone accessibility and for searching
for terrestrial planets under relatively stable stellar irradiation. Within this
main-sequence population, proximity is used as the primary ordering criterion,
since nearby systems provide both larger angular separations between the star and
its habitable zone and higher planetary photon fluxes for subsequent characterization
\cite{Tuchow2024}.

Table~\ref{tab:fgk20} lists the resulting 20 stellar components and gives the angular extent of
each conservative habitable zone in units of the coronagraphic inner working angle
(IWA). The conservative HZ boundaries follow the climate calculations of
Kopparapu et al. \cite{kopparapu13,kopparapu14}, while representative
main-sequence stellar parameters are based on the dwarf sequence of
Pecaut \& Mamajek \cite{pecaut13}. Figure~\ref{fig:fgk20-observability} places the same targets
on the HZ observability map, showing directly how accessibility depends jointly
on spectral type and distance.

The sample is intentionally not restricted to systems for which the entire
conservative HZ is accessible. Instead, it preserves the 20 nearest eligible FGK
main-sequence targets as a broad discovery sample while assigning observing priority
according to HZ accessibility. Systems for which the entire conservative HZ lies
outside the IWA receive the highest priority, followed by systems for which only
a portion of the conservative HZ is accessible. The discovery space can further
extend into the optimistic HZ, bounded by the Recent Venus and Early Mars limits,
where geometrically accessible \cite{kopparapu13}.

This tiered approach recognizes that HZ accessibility in direct imaging is not
simply a binary property. In partially accessible systems, a planet may move outside
the IWA during favorable orbital phases and become detectable through repeated
observations \cite{GuimondCowan2019}. The survey therefore begins with the nearest
systems offering complete access to the conservative HZ and progressively extends
to partial conservative-HZ and optimistic-HZ discovery space. In this way, the
program is inclusive in defining where potentially habitable planets may be found,
while remaining conservative in determining which systems are observed first.

\subsection{From Planet Detection to Atmospheric Characterization}
\label{sec:Detect_Char}

The discovery of a point source near a target star is only the first step toward
identifying a potentially habitable world. The primary FGK targets will therefore
be observed repeatedly rather than through single-epoch imaging. Initial
high-contrast observations will search for planetary candidates outside the
coronagraphic inner working angle, while subsequent epochs will test whether each
candidate shares the proper motion of the host star and exhibits orbital motion.
Multi-epoch astrometry will then be used to constrain the planet's orbit, including
its semimajor axis and orbital geometry, and to distinguish bound companions from
unrelated background sources \cite{GuimondCowan2019}.

Determining the orbit is essential for assessing habitability. A planet observed
at a projected separation corresponding to the habitable zone is not necessarily
located within the HZ in three-dimensional space, while a planet that appears close
to the star at one epoch may move into an observable portion of the HZ at another
orbital phase \cite{GuimondCowan2019}. The derived orbit will therefore be combined
with the stellar luminosity and the conservative and optimistic HZ boundaries
defined in Section~\ref{sec:hz-target-selection}. Particular attention will be given to planets whose orbits lie wholly or substantially within the conservative HZ, while planets occupying the optimistic HZ will remain part of the broader habitability assessment. Repeated observations are especially important for systems with partially accessible HZs, where orbital motion can bring a planet outside the IWA and provide additional opportunities for detection and characterization.

Once a planet is confirmed as a potentially habitable candidate, the observing
strategy will shift from discovery to characterization. Such objects will receive
high-priority follow-up observations, including reflected-light spectroscopy when
sufficient planet--star separation and signal-to-noise ratio are available.
Broadband photometry and spectroscopy can constrain atmospheric composition,
cloud properties, surface albedo, and other planetary properties through atmospheric
retrieval analyses \cite{Feng2018,KawashimaRugheimer2019}. In particular,
reflected-light spectra of terrestrial planets can contain absorption signatures
from molecules such as H$_2$O, O$_2$, O$_3$, CH$_4$, and CO$_2$, although their
detectability depends strongly on spectral coverage, signal-to-noise ratio, clouds,
and atmospheric abundance \cite{KawashimaRugheimer2019}.

Observations at multiple orbital phases may further help disentangle atmospheric,
cloud, and surface contributions to the reflected-light signal and improve the
interpretation of the planetary spectrum. The program thus establishes a continuous
observational pathway from planet detection, through orbital confirmation and HZ
assessment, to atmospheric characterization, with the most promising candidates
advancing to the biosignature investigations described in Section~\ref{sec:Habitab_Biosig}.

\subsection{Search for Habitability and Biosignatures}
\label{sec:Habitab_Biosig}

For planets identified as potentially habitable through the orbital and HZ
assessment described above, the next scientific objective is to determine whether
their atmospheres and surface environments are consistent with habitability and,
ultimately, whether they show evidence suggestive of biological activity
\cite{Schwieterman2018,Catling2018}. Reflected-light spectra will first be used
to establish the broader planetary context, including atmospheric composition,
clouds and aerosols, and wavelength-dependent reflectivity
\cite{Schwieterman2018}.

Particular attention will be given to evidence for H$_2$O, because the presence
of atmospheric water vapor, when interpreted together with the planet's orbit
and incident stellar radiation, provides an important constraint on potentially
habitable conditions. The location of a planet within the HZ alone will therefore
not be treated as evidence of habitability; rather, orbital, atmospheric, and
stellar information will be considered together
\cite{Schwieterman2018,Catling2018}.

The search for biosignatures will likewise rely on multiple lines of evidence
rather than on any single spectral feature. Candidate atmospheric biosignatures
include O$_2$ and O$_3$, while reduced species such as CH$_4$, when detectable,
can provide additional information about atmospheric chemical disequilibrium
\cite{Schwieterman2018,Meadows2018,KrissansenTotton2018}. Of particular interest
are combinations of oxidized and reduced species that are difficult to maintain
simultaneously in substantial abundances without continuous replenishment
\cite{KrissansenTotton2018}.

However, none of these molecules is uniquely biological. Abiotic processes can
produce O$_2$ or O$_3$, and the abundance and persistence of CH$_4$ can also be
controlled by geological and photochemical processes
\cite{Meadows2018,Schwieterman2018}. Biosignature interpretation must therefore
consider the planetary environment, stellar spectral energy distribution and
activity, atmospheric photochemistry, and plausible abiotic false-positive
scenarios \cite{Catling2018,Meadows2018}.

The proximity of the primary FGK sample is especially valuable in this context.
Nearby systems provide the most favorable angular separation and photon flux for
repeated direct imaging and increasingly deep spectroscopic observations, allowing
promising planets to be revisited as their orbits become better constrained and
as favorable observing geometries occur \cite{Tuchow2024}. The goal of this
program is therefore not simply to detect an isolated ``biosignature molecule,''
but to build a physically consistent assessment of each potentially habitable
world \cite{Catling2018}.

A planet showing an accessible HZ orbit, evidence for an atmosphere and
water-bearing conditions, and a combination of atmospheric species difficult
to explain through known abiotic processes would become a compelling target for
intensive follow-up observations. Such nearby worlds would also constitute
natural precursor targets for future larger-aperture missions capable of
substantially deeper atmospheric characterization.

\vspace*{1.0cm}

%%%%%%%%%%%%%%%%%%%%%%%%%%%%%%%%%%%%%%%%%
\section{High-Contrast Observing Capability of the Mission}
\label{sec:context}

Direct imaging of exoplanets is one of the principal techniques for
detecting and characterizing planets around other stars. The method
suppresses the overwhelming light of the host star and isolates the much
fainter light of its planets. A planet shines in reflected starlight or
in its own thermal emission. In both channels it is typically millions
to billions of times fainter than its star. Unlike the radial-velocity
and transit techniques, direct imaging measures the position, the
brightness, and the spectrum of the planet itself
\cite{galicher23,zurlo24}.

The core technology is the coronagraph. A coronagraph blocks or
attenuates the stellar image inside the optical train, suppresses the
diffraction pattern, and darkens the field immediately around the star.
Faint companions and circumstellar disks then become detectable
\cite{galicher23}. Two parameters summarize coronagraph performance.
The \emph{contrast} is the faintest planet-to-star flux ratio that the
system can detect. Modern space-coronagraph designs target contrasts of
roughly $10^{-8}$ to $10^{-9}$. The Roman Space Telescope hosts a
technology-demonstration coronagraph with a formal detection requirement
near $10^{-8}$. Current performance models place its best-effort limit
between $10^{-8}$ and a few times $10^{-9}$
\cite{bailey23,llopsayson25}. An Earth twin in reflected
visible light sits near $10^{-10}$, and a Jupiter twin near $10^{-9}$
\cite{traub10}. The \emph{inner working angle} (IWA) is the smallest
angular separation from the star at which that contrast is delivered.
Typical designs reach an IWA of two to three times the diffraction
scale $\lambda/D$ \cite{galicher23}.

The baseline science payload includes a dedicated coronagraph as a core instrument for high-contrast exoplanet observations. The observatory architecture therefore incorporates the wavefront stability, thermal stability, pointing control, optical interfaces, and operations capabilities required for coronagraphy from the initial mission configuration. The design also preserves flexibility for future improvements in wavefront control, detector technology, and coronagraph performance. The optical path reserves
the required volume. The wavefront-sensing and control budget can be
extended to the levels that coronagraphy demands. The observing and
operations software is structured to accommodate a high-contrast mode.
The mission therefore functions as a technological and operational
precursor for flagship direct-imaging observatories, and its exoplanet
program can grow as the module matures. Recent theoretical work on the
quantum limits of high-contrast detection indicates that substantial
gains beyond classical coronagraphy remain available to future
instruments \cite{deshler24}.

\subsection{Point-spread function, contrast, and wavefront stability}
\label{sec:perf}

\begin{figure}[tp]
\centering
\includegraphics[width=\textwidth]{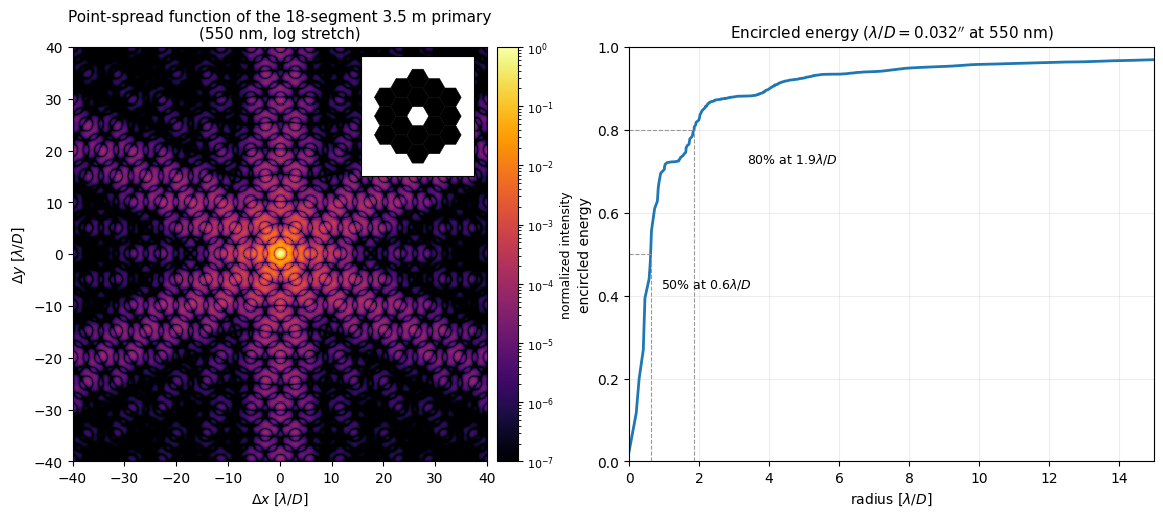}
\caption{Point-spread function of the 18-segment 3.5\,m primary at 550\,nm
(left, logarithmic stretch; the aperture is inset) and its encircled
energy (right). The computation is the squared modulus of the Fourier
transform of the aperture. The segment gaps produce the six-fold
diffraction structure that the coronagraph suppresses.}
\label{fig:psf}
\end{figure}

The 3.5\,m primary is a mosaic of 18 hexagonal segments, and its
diffraction pattern shows the imprint of that geometry.
Figure~\ref{fig:psf} shows the physical-optics point-spread function, the
squared modulus of the Fourier transform of the aperture. The segment
gaps redistribute light into a six-fold pattern of secondary maxima. Half
of the encircled energy falls within $0.6\lambda/D$ and eight-tenths
within $1.9\lambda/D$, where $\lambda/D=0.032''$ at 550\,nm. A coronagraph
should suppress this structure to reach the planetary signal.

\begin{figure}[tp]
\centering
\includegraphics[width=\textwidth]{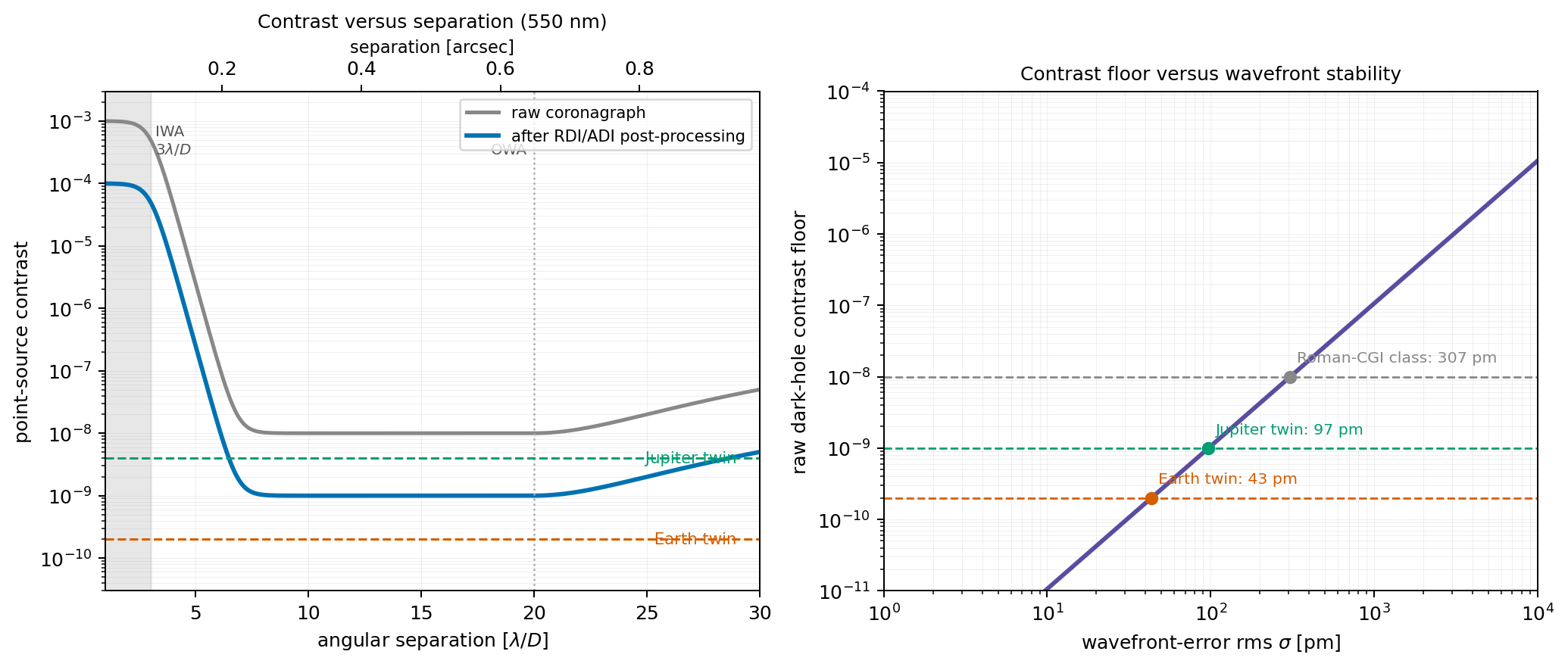}
\caption{Contrast versus angular separation (left) with the inner and
outer working angles and the reflected-light contrasts of Earth and
Jupiter analogs \cite{traub10} (the $4\times10^{-9}$ Jupiter value is the
near-full-phase brightness; at quadrature the ratio is near $10^{-9}$),
and a heuristic contrast scaling versus wavefront-error stability (right) from
the $(2\pi\sigma/\lambda)^2$ scatter relation divided across the
$\sim$$10^{3}$ controllable modes of the dark hole. The scaling provides
physical context and does not replace a coronagraph-specific error budget.
Roman provides the $10^{-8}$ flight-demonstration reference. The
$10^{-9}$ level is a development goal for mature-giant imaging
\cite{bailey23,llopsayson25}.}
\label{fig:contrast}
\end{figure}

Two numbers set the achievable science. Figure~\ref{fig:contrast} (left)
shows an illustrative point-source contrast against angular separation.
The $10^{-8}$ raw contrast represents the flight-demonstration regime of
the Roman Coronagraph. A post-processed contrast near $10^{-9}$ is a
development goal for the present concept rather than a demonstrated
performance level. A Jupiter analog at $4\times10^{-9}$ can lie above the
goal near full phase. An Earth analog at $2\times10^{-10}$ remains below
the goal. Wavefront stability contributes to the contrast floor. A
wavefront error of root-mean-square $\sigma$
scatters a fraction $(2\pi\sigma/\lambda)^2$ of the starlight out of the
core, and spreading that light across the roughly $10^{3}$ controllable
modes of the dark hole leaves a mean contrast near
$(2\pi\sigma/\lambda)^2/N$. Figure~\ref{fig:contrast} (right) uses the
relation only as an order-of-magnitude scaling. The total RMS error does
not define a coronagraph requirement because segment piston, low-order
drift, pointing jitter, amplitude errors, and their temporal spectra
couple differently into the dark hole. An end-to-end propagation model
of the selected coronagraph must establish the actual stability budget.
The tens-of-picometre regime remains a useful scale for the long-term
Earth-analog goal \cite{bailey23}.

%%%%%%%%%%%%%%%%%%%%%%%% B.4

\section{From the $10^{-9}$ Contrast Limit to Nearby Earth-like Worlds}
\label{sec:earthlike_detectability}

\subsection{The Challenge of Earth-like Planet Detection}
\label{sec:earthlike_challenge}

The high-contrast performance discussed in Section~\ref{sec:context} poses a fundamental
challenge for terrestrial-planet science with the 3.5-meter Segmented-Mirror
Robotic Space Telescope. An Earth--Sun analog observed in reflected light has
a characteristic planet-to-star contrast of order $10^{-10}$\cite{trauger07,lawson13}, 
approximately one order of magnitude deeper than the $\sim10^{-9}$ post-processing
development goal considered for the 3.5mST coronagraph. At first sight, this
appears to place Earth--Sun analogs beyond the practical reach of the mission.

The nominal inner working angle (IWA) should not be interpreted as
a sharp boundary beyond which the full limiting contrast is immediately
achieved; coronagraphic sensitivity depends jointly on residual stellar
leakage, planet throughput, and angular separation \cite{nemati23}.
Consequently, a habitable zone lying nominally outside the IWA is
geometrically accessible, but this condition alone does not guarantee that an
Earth-sized planet within that zone can be detected.

These limitations motivate a more specific question: does the solar
neighborhood contain stellar systems in which an Earth-like planet has a more
favorable reflected-light contrast while its habitable zone remains
sufficiently separated from the host star for coronagraphic detection?
Addressing this question requires the contrast limit and the IWA constraint to
be considered simultaneously rather than independently.

\subsection{Earth-like Planet Contrast around Cooler Stars}
\label{sec:earthlike_contrast}

The contrast limit discussed in Section~\ref{sec:earthlike_multi_epoch} does not apply equally to Earth-like planets around all main-sequence stars. In reflected light, the planet-to-star flux ratio can be written as \cite{traub10}

\begin{equation}
\frac{F_p}{F_\star}
=
A_g \Phi(\alpha)
\left(\frac{R_p}{a}\right)^2 ,
\label{eq:reflected_contrast}
\end{equation}

where $A_g$ is the geometric albedo, $\Phi(\alpha)$ is the phase function,
$R_p$ is the planetary radius, and $a$ is the orbital separation. For a
planet receiving the same bolometric irradiation as the Earth, the
Earth-equivalent insolation distance (EEID) scales approximately as \cite{kopparapu13,kopparapu14}

\begin{equation}
a_{\rm EEID}
=
\sqrt{\frac{L_\star}{L_\odot}}\ {\rm AU}.
\label{eq:eeid}
\end{equation}

\begin{figure}[tp]
\centering
\includegraphics[width=1.0\textwidth]{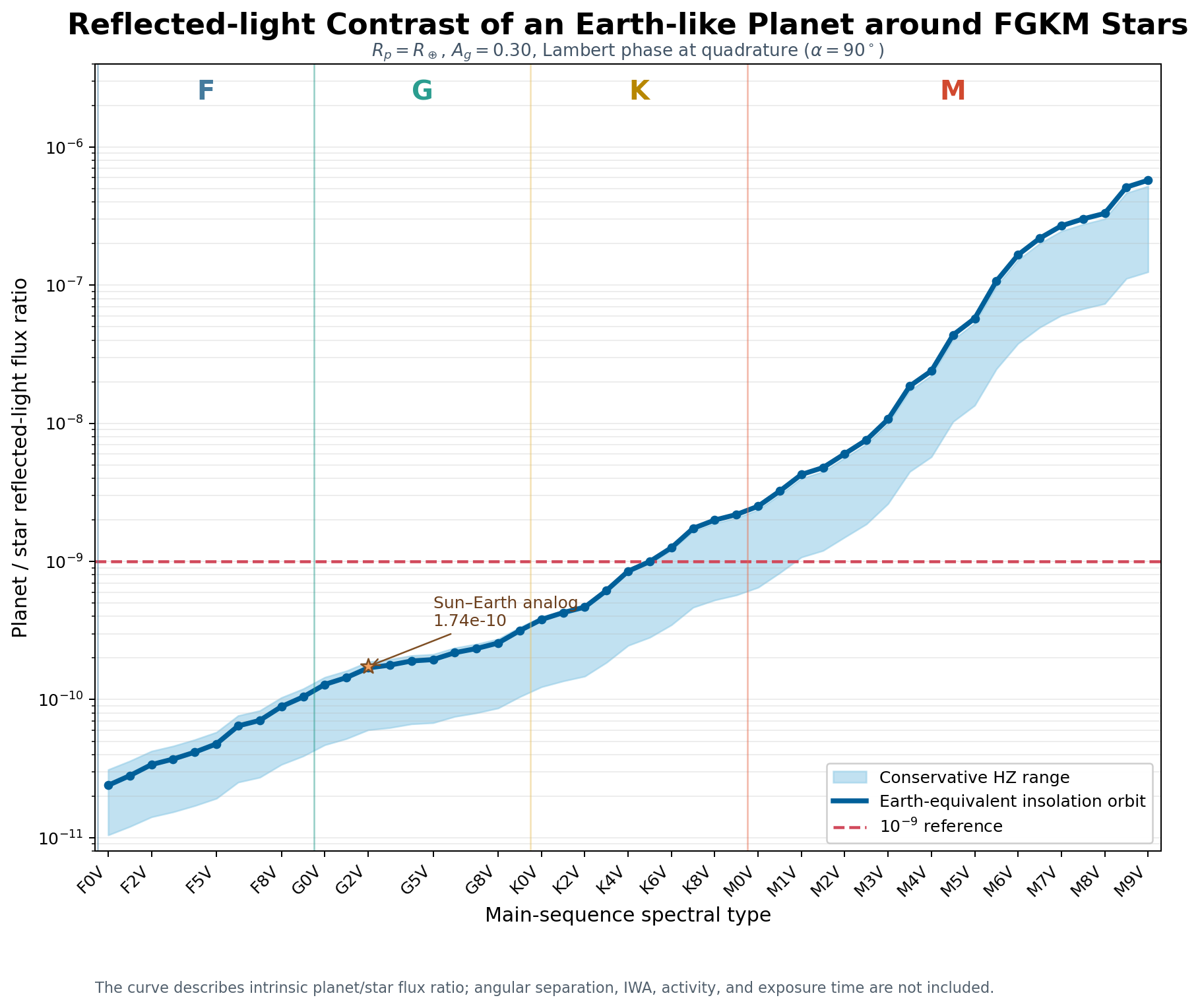}
\caption{Left: separation-dependent coronagraphic contrast, demonstrating that the nominal IWA is not a hard contrast boundary. Right: reflected-light contrast of an Earth-like planet versus host-star spectral type, showing the approach to the $10^{-9}$ regime toward late-K stars.}
\label{fig:earthlike-contrast-trade}
\end{figure}

As the stellar luminosity decreases toward later spectral types, the EEID
moves closer to the host star \cite{kopparapu13}. For a planet with the same radius, 
albedo, and illumination phase, the smaller orbital separation increases the
planet-to-star reflected-light contrast. Thus, while an Earth--Sun analog has
a characteristic contrast of order $10^{-10}$, an Earth-like planet receiving
comparable irradiation around a sufficiently cool K dwarf can approach the
$\sim10^{-9}$ regime. As illustrated in Figure~\ref{fig:earthlike-contrast-trade}, 
this transition occurs approximately in the late-K regime, near K4 and later
under the assumptions adopted here.

This apparent advantage, however, introduces a competing limitation. The same
decrease in stellar luminosity that improves the planet-to-star contrast also
moves the habitable zone inward, reducing its angular separation from the
host star. Moreover, as shown in the left panel of Figure~\ref{fig:contrast}, 
the nominal inner working angle is not a sharp boundary at which the full limiting
contrast is suddenly achieved. Residual stellar light remains significant
immediately outside the IWA, and the achievable contrast improves only
gradually with increasing separation. A planet located only slightly outside
the nominal IWA may therefore remain more difficult to detect than a simple
IWA threshold would imply.

These two effects define a narrow observational trade-off. Earlier-type stars
provide larger habitable-zone angular separations but less favorable
Earth-like planet contrasts, whereas cooler stars provide increasingly
favorable contrasts at progressively smaller angular separations. The most
promising targets for the 3.5mST should therefore occur where these two
requirements overlap: the host star must be cool enough for an Earth-like
planet to approach the $10^{-9}$ contrast regime, but sufficiently nearby
that its habitable zone remains accessible outside the most challenging
near-IWA region.

%------------------------------------------------------------
%------------------------------------------------------------

\subsection{Prime Nearby Targets for Earth-like Planet Searches}
\label{sec:prime_earthlike_targets}

The contrast--separation trade-off described above can now be combined with
the habitable-zone accessibility analysis of Section~\ref{sec:Search-Habitable}. 
Figure~\ref{fig:fgk20-observability} shows the nearby FGK sample in the plane of 
spectral type and distance, together with the boundaries at which the habitable 
zone becomes fully or partially accessible outside the nominal $3\lambda/D$ IWA. 
The late-K regime identified in Section~\ref{sec:earthlike_contrast} 
adds a second requirement to this 
geometric selection: an Earth-like planet must not only reach a sufficiently 
large angular separation, but must also provide a reflected-light contrast 
approaching the $\sim10^{-9}$ performance goal of the coronagraph.

Combining these two criteria substantially narrows the terrestrial-planet
search space. As indicated in Figure~\ref{fig:fgk20-observability}, only a 
small number of nearby late-K stars occupy the region in which both conditions 
become favorable. Among them, 61~Cyg~A and $\epsilon$~Ind~A are particularly 
noteworthy. Both are K5~V stars located within approximately 4~pc of the Sun, 
allowing their compact habitable zones to subtend unusually large angles 
on the sky.

For an Earth-sized planet placed at the Earth-equivalent insolation distance,
the current estimates give
$a_{\rm EEID}\simeq0.38$~AU and
$\theta_{\rm EEID}\simeq108$~mas for 61~Cyg~A, and
$a_{\rm EEID}\simeq0.50$~AU and
$\theta_{\rm EEID}\simeq137$~mas for $\epsilon$~Ind~A.
These separations correspond to approximately $1.1$ and $1.4$ times,
respectively, the nominal $3\lambda/D\simeq97$~mas IWA at 550~nm.
Under the adopted Earth-like assumptions, the corresponding reflected-light
contrasts are approximately $1.2\times10^{-9}$ for 61~Cyg~A and
$6.9\times10^{-10}$ for $\epsilon$~Ind~A.

These two systems illustrate the narrowness of the accessible parameter
space rather than representing straightforward detections. 61~Cyg~A offers
the more favorable planet-to-star contrast, but its Earth-equivalent orbit
lies only slightly outside the nominal IWA, where stellar leakage and
coronagraphic throughput are particularly critical. $\epsilon$~Ind~A
provides a larger angular separation, but its predicted Earth-like contrast
falls somewhat below the nominal $10^{-9}$ development goal. Their practical
detectability therefore depends on the delivered contrast as a function of
separation, near-IWA throughput, wavefront stability, stellar leakage,
planetary phase, and orbital geometry.

The search was also extended from late-K stars toward early-M stars to test
whether still cooler hosts provide more favorable targets. Although the
intrinsic reflected-light contrast continues to improve toward cooler stars,
their habitable zones move progressively inward and become more difficult to
resolve. Under the adopted 3.5mST IWA, this extension does not identify
additional nearby systems that clearly improve upon the best late-K cases.
The nearest late-K stars therefore define a narrow but scientifically
compelling regime in which the competing requirements of Earth-like
planet contrast and habitable-zone angular separation simultaneously
approach the capabilities of the 3.5mST.

The quoted angular separations represent favorable projected configurations
and should not be interpreted as single-visit detection guarantees.
For planets on inclined orbits, the projected separation varies throughout
the orbit and can fall inside the effective high-contrast working region.
The implications for discovery completeness, observing cadence, and
multi-epoch orbit determination are considered in
Section~\ref{sec:earthlike_multi_epoch}.

\subsection{Multi-Epoch Detection and Atmospheric Characterization}
\label{sec:earthlike_multi_epoch}

The favorable angular separations discussed in Section~\ref{sec:prime_earthlike_targets}. represent maximum or near-maximum projected configurations and do not imply that an
Earth-like planet will be observable during every visit. The projected
star--planet separation depends on both orbital phase and inclination. Even
when the physical orbit lies within the habitable zone, an inclined orbit
can carry the planet inside the effective high-contrast working region for a
substantial fraction of its orbital period. This effect is particularly
important for 61~Cyg~A and $\epsilon$~Ind~A, whose Earth-equivalent
insolation distances correspond to only approximately 1.1 and 1.4 times the
nominal IWA, respectively. A non-detection in a single visit therefore
provides only limited completeness for terrestrial planets in these systems
\cite{brown05}.

This geometry makes repeated observations an essential part of the search
strategy \cite{brown05,brown10}. For a circular orbit, the characteristic 
orbital period at the Earth-equivalent insolation distance can be estimated from

\begin{equation}
P_{\rm EEID}
\simeq
\left(
\frac{a_{\rm EEID}^{3}}{M_\star}
\right)^{1/2}
{\rm yr},
\label{eq:eeid_period}
\end{equation}

where $a_{\rm EEID}$ is expressed in AU and $M_\star$ in solar masses.
Rather than repeating observations at arbitrary intervals, successive visits
should be separated by a significant fraction of the expected orbital
period so that substantially different orbital phases and projected
separations are sampled. 
A cadence of order $P/6$--$P/4$ provides a useful starting point
for the present study, although the optimum cadence must ultimately
be evaluated through completeness simulations that include orbital
inclination, eccentricity, phase-dependent brightness, and the
separation-dependent coronagraphic response.

Once a candidate is detected, the observing strategy changes from discovery
to orbit determination. Multiple astrometric measurements distributed over the
orbit can constrain the semimajor axis, inclination, and orbital phase,
allowing subsequent observations to be scheduled near favorable projected
separations. This is especially valuable for planets whose orbits repeatedly
cross the effective inner working region. Multi-epoch imaging therefore
serves two purposes: it increases the probability of initial detection and,
after discovery, provides the orbital information required to determine
whether the planet is consistent with long-term residence in the habitable
zone and to predict the best epochs for spectroscopic follow-up.

The brightness of a detected terrestrial planet provides the next practical
constraint. At quadrature, its apparent magnitude can be estimated directly
from the host-star magnitude and the reflected-light contrast,

\begin{equation}
m_p
=
m_\star
-
2.5\log_{10}
\left(
\frac{F_p}{F_\star}
\right).
\label{eq:planet_magnitude}
\end{equation}

Using the Earth-like contrasts adopted in Section~\ref{sec:prime_earthlike_targets} 
gives apparent visual magnitudes of approximately $V\simeq27.5$ for an 
Earth-sized planet at the EEID of both 61~Cyg~A and $\epsilon$~Ind~A. 
These values illustrate an important distinction between intrinsic planet 
brightness and detectability: although the planets have comparable apparent 
magnitudes, their detection difficulty can differ substantially because 
their angular separations and residual stellar backgrounds are different.

The observing sequence should therefore be considered in three stages:
broadband discovery imaging, multi-epoch orbit determination, and
reflected-light spectroscopy. A broadband detection requires sufficient
signal-to-noise not only against photon noise but also against residual
stellar leakage, local and exozodiacal backgrounds, detector noise, and
the loss of coronagraphic throughput near the IWA. A reliable imaging
exposure time for 61~Cyg~A and $\epsilon$~Ind~A must consequently be derived
from an end-to-end, separation-dependent coronagraph model rather than from
planet magnitude alone \cite{nemati23,krist23}.

For atmospheric characterization, the $R=5000$ photon-budget calculation
developed in Section~\ref{sec:horizon} provides a useful first estimate. Under the adopted
Earth-like assumptions, the current model gives 5$\sigma$ molecular-band
template-detection times of approximately 8.7~hr for 61~Cyg~A and 12~hr for
$\epsilon$~Ind~A. These values should be regarded strictly as optimistic
photon-noise lower bounds. The calculation does not fully include correlated
residual speckles, detector noise, template mismatch, or the
separation-dependent loss of coronagraphic throughput and increased stellar
leakage close to the IWA. The actual spectroscopic integrations could
therefore be substantially longer, particularly for 61~Cyg~A at its small
projected separation.

A target-specific observing assessment should ultimately combine these
quantities in a single framework: the expected orbital period and revisit
cadence, the fraction of the orbit accessible outside the effective
high-contrast working region, the apparent planet brightness, the broadband
discovery exposure, and the spectroscopic characterization time. Such an
analysis will determine whether the narrow terrestrial-planet window
identified in Section~\ref{sec:earthlike_contrast} and 
~\ref{sec:earthlike_contrast} can be exploited in practice. The
program should therefore be regarded not as a general Earth--Sun twin survey,
but as a focused, multi-epoch experiment on a small number of exceptionally
nearby systems where the angular separation and reflected-light contrast of
an Earth-like planet simultaneously approach the performance limits of the
3.5mST.

%%%%%%%%%%%%%%%%%%%%%%%%%%%%%%%%%%%
\section{The Distance Horizon of Coronagraphic Imaging}
\label{sec:horizon}

\begin{figure}[tp]
\centering
\includegraphics[width=\textwidth]{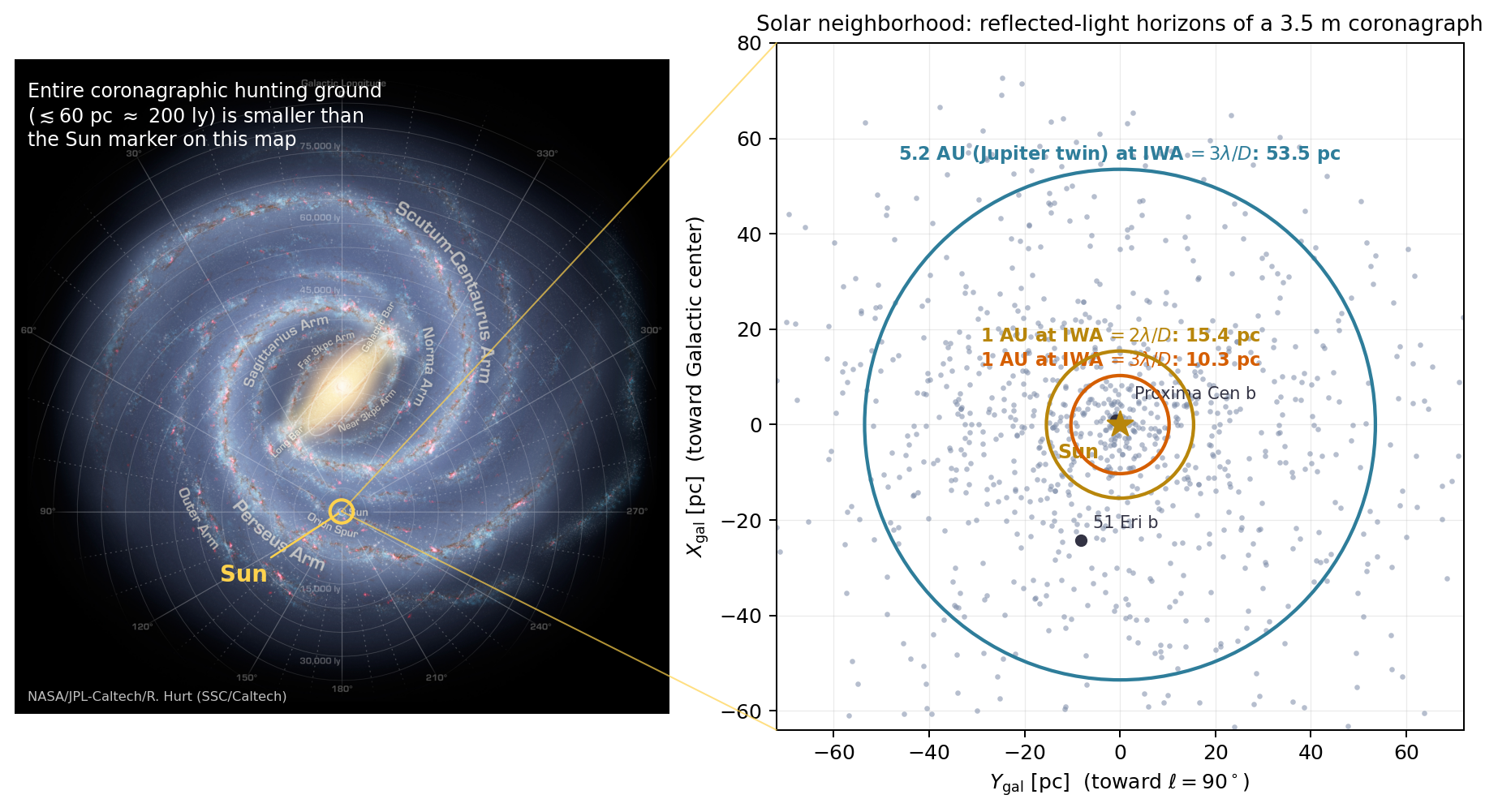}
\caption{The coronagraphic distance horizon in its Galactic context.
\emph{Left:} annotated face-on map of the Milky Way
(NASA/JPL-Caltech/R.~Hurt, SSC/Caltech) with the Sun marked. The
reflected-light hunting ground of any coronagraph, of order tens of
parsecs, is smaller than the Sun marker itself.
\emph{Right:} the solar neighborhood in face-on Galactic coordinates.
Gray points are the known planet-host stars within 90\,pc from the NASA
Exoplanet Archive (accessed 2026 July 11). The circles mark the
distances inside which a planet at 1\,AU (Earth twin) or at 5.2\,AU
(Jupiter twin) stays outside the inner working angle of a 3.5\,m
telescope at 550\,nm. The archive lists 64 known planet-host stars
inside the 10.3\,pc circle, 123 inside 15.4\,pc, and 652 inside
53.5\,pc. Proxima Centauri b and the directly imaged young giant
51~Eridani b \cite{macintosh15} are labeled.}
\label{fig:mwhorizon}
\end{figure}

A coronagraph does not survey the Galaxy. It surveys a small sphere
around the Sun, and the radius of that sphere follows directly from the
instrument parameters. Figure~\ref{fig:mwhorizon} places this horizon
on a face-on map of the Milky Way. The entire coronagraphic hunting
ground is smaller than the Sun marker on the map. Four factors define
the horizon.

\subsection{Angular geometry: the inner working angle}

The diffraction scale of a 3.5\,m aperture at 550\,nm is
\begin{equation}
\frac{\lambda}{D} = 206{,}265''\times\frac{550\,\mathrm{nm}}{3.5\,\mathrm{m}}
 = 0.0324''.
\end{equation}
An IWA of $3\lambda/D$ is $0.097''$, and an IWA of $2\lambda/D$ is
$0.065''$. A planet on an orbit of semi-major axis $a$ remains
resolvable outside the IWA only for stars closer than
\begin{equation}
d_{\max} = \frac{a/\mathrm{AU}}{\theta_{\rm IWA}/\mathrm{arcsec}}\;
\mathrm{pc}.
\label{eq:dmax}
\end{equation}
Equation~(\ref{eq:dmax}) gives the horizons drawn in
Figure~\ref{fig:mwhorizon} and listed in Table~\ref{tab:horizon}. This
distance is the maximum-elongation limit. A planet on a circular orbit
clears the inner working angle only over the fraction of its orbit where
the projected separation exceeds it, which the completeness analysis of
Section~\ref{sec:yield} treats in full. An
Earth twin at 1\,AU is geometrically accessible only within
10.3\,pc at $3\lambda/D$ and within 15.4\,pc at $2\lambda/D$. A Jupiter
twin at 5.2\,AU is accessible within 53.5\,pc at $3\lambda/D$. The
horizon scales linearly with wavelength. At 1.6\,$\mu$m the same
$3\lambda/D$ IWA grows to $0.28''$, and the Earth-twin horizon shrinks
to 3.5\,pc. Only the $\alpha$~Centauri system at 1.34\,pc
\cite{vanleeuwen07} remains inside it. At that distance a 1\,AU orbit
subtends $0.75''$, which is 23 diffraction widths at 550\,nm. The
nearest stars are therefore the easiest targets by a wide margin.

\begin{table}[tp]
\centering
\small
\begin{tabular}{lccc}
\toprule
Planet analog & Semi-major axis & $d_{\max}$ at $2\lambda/D$ &
$d_{\max}$ at $3\lambda/D$ \\
\midrule
Earth twin   & 1.0\,AU  & 15.4\,pc & 10.3\,pc \\
Jupiter twin & 5.2\,AU  & 80.3\,pc & 53.5\,pc \\
\bottomrule
\end{tabular}
\caption{Reflected-light distance horizons of a 3.5\,m coronagraph at
550\,nm from Equation~(\ref{eq:dmax}). The IWA values $0.065''$ and
$0.097''$ follow from the $0.0324''$ diffraction scale of the
aperture.}
\label{tab:horizon}
\end{table}

\subsection{Contrast floor: which planets are visible at all}

The IWA decides where the coronagraph can look. The contrast floor
decides what it can see there. A Jupiter twin in reflected light
requires roughly $10^{-9}$, and an Earth twin roughly $10^{-10}$
\cite{traub10}. A system that delivers $10^{-8}$ detects young
self-luminous giants and bright debris disks. A system at $10^{-9}$
adds mature giant planets in reflected light throughout the horizon of
Table~\ref{tab:horizon}. Earth analogs open up only beyond $10^{-10}$,
which is the regime targeted by the Habitable Worlds Observatory
concept \cite{hwo26}. The contrast floor therefore acts as a
selection on planet type, while the IWA acts as a selection on
distance.

\subsection{Photon budget}

Even a perfectly suppressed star leaves a very faint planet. The Sun
has an absolute $V$ magnitude of 4.81 \cite{willmer18}. An Earth twin
at $10^{-10}$ contrast is 25 magnitudes fainter, which places it near
$V\simeq29.8$ at 10\,pc. The planetary flux falls as the square of the
distance. Long integrations on a large, stable space aperture are
therefore mandatory, and the practical horizon for spectroscopy is
tighter than the geometric horizon of Equation~(\ref{eq:dmax}). The
diffuse zodiacal and exozodiacal backgrounds that raise this floor are
treated in Section~\ref{sec:backgrounds}.

\subsection{Target supply and the effect of binarity}

\begin{figure}[tp]
\centering
\includegraphics[width=\textwidth]{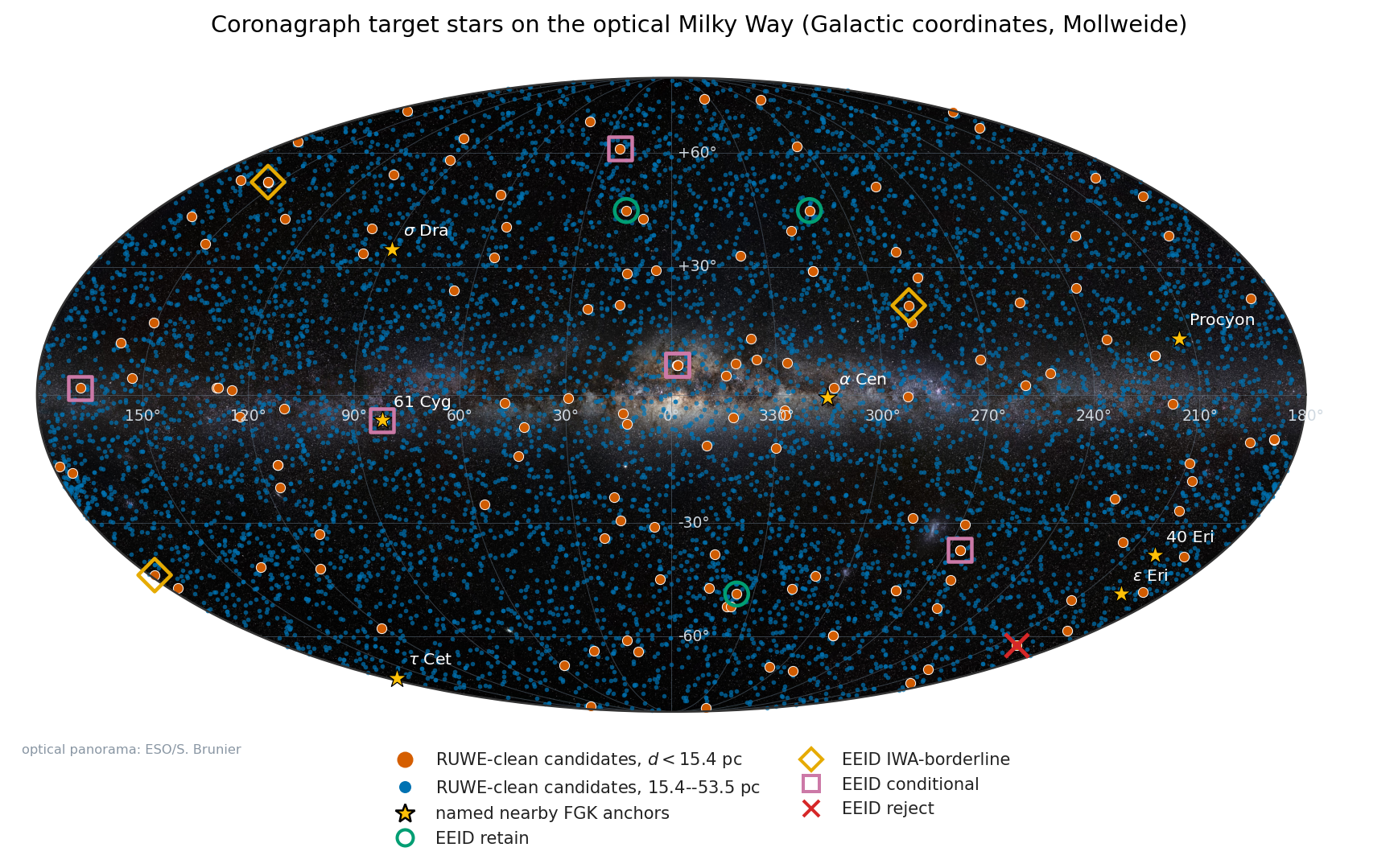}
\caption{The coronagraph target stars on the optical Milky Way, in
Galactic coordinates (Mollweide projection, optical panorama by
ESO/S.~Brunier). Points are RUWE-clean photometric candidates from
Gaia DR3 \cite{gaiadr3}. Orange marks candidates within the 15.4\,pc
geometric horizon and blue marks candidates from 15.4 to 53.5\,pc. Named nearby
anchors are starred. The targets spread almost uniformly across the sky.
The bright Milky Way band marks the high stellar density where
background-star confusion (Section~\ref{sec:backgrounds}) is worst. The
targets can therefore be drawn preferentially from the sparse
high-latitude sky. The open symbols show the assessment of the 13
preliminary EEID candidates in Table~\ref{tab:hztargets}. Green circles
mark retained targets. Gold diamonds mark IWA-borderline targets. Magenta
squares mark conditional targets and the red cross marks the rejected
spectroscopic binary.}
\label{fig:mollweide}
\end{figure}

\begin{figure}[tp]
\centering
\includegraphics[width=0.82\textwidth]{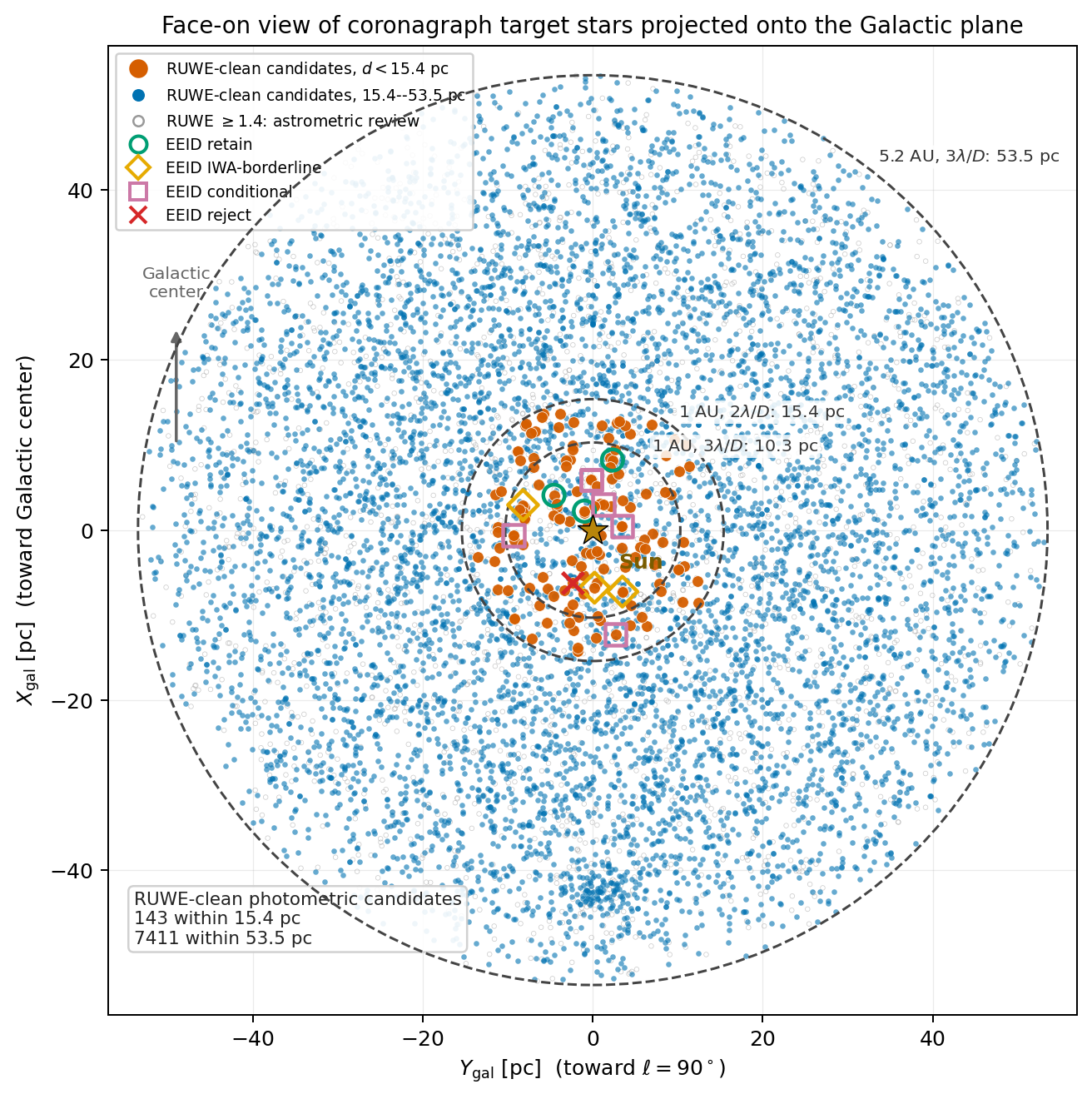}
\caption{The same target stars projected onto the Galactic plane, the
face-on view a virtual observer sees around the Sun. Axes are Galactic
$X$ (toward the Galactic center) and $Y$ (toward $\ell=90^\circ$).
Dashed circles are the reflected-light horizons of
Table~\ref{tab:horizon}. Colors follow Figure~\ref{fig:mollweide}. The
RUWE-clean photometric candidates number 143 within 15.4\,pc and 7{,}411 within
53.5\,pc. The overdensity near $X\simeq-45$\,pc is the Hyades open
cluster. Open symbols use the assessment scheme of
Figure~\ref{fig:mollweide}.}
\label{fig:faceon}
\end{figure}

The number of available stars grows with the cube of the horizon
radius. Only a fraction of them are usable coronagraph targets.
Table~\ref{tab:targets} follows the list through three successive
filters, and Figures~\ref{fig:mollweide} and~\ref{fig:faceon} map the
survivors across the sky and in the Galactic plane.

The first filter is the stellar census. Gaia DR3 \cite{gaiadr3}
contains 338 stars within 10.3\,pc and 65{,}586 within 53.5\,pc down to
the M-dwarf regime. These counts are lower limits at the smallest
radii, because the nearest and brightest stars saturate the Gaia
detectors. The volume-complete census of the 10\,pc sphere contains
541 objects in 339 systems \cite{reyle21}.

The second filter is spectral type. Reflected-light detection of an
Earth analog is a program for FGK host stars, because their habitable
zone lies near 1\,AU where the geometry of Section~\ref{sec:horizon}
places it outside the inner working angle. M dwarfs make up roughly
four-fifths of the census, and their habitable zones lie near 0.1\,AU.
Holding a 0.1\,AU orbit outside a $3\lambda/D$ inner working angle at
550\,nm requires a distance below 1\,pc from
Equation~(\ref{eq:dmax}), a condition that no M dwarf meets. The tens
of thousands of nearby M dwarfs are therefore closed to habitable-zone
reflected-light imaging with a 3.5\,m aperture, although their wider
giant planets remain accessible.

The third filter is multiplicity. A close stellar companion floods the
coronagraphic dark hole and removes the star from the target list.
About 46\% of solar-type stars host a companion \cite{raghavan10}, and
about 27\% of M dwarfs do \cite{winters19}. The intermediate-separation
pairs that fall within the search region should be dropped. A Gaia
renormalized unit weight error below 1.4 serves only as a data-driven flag
for a well-fit five-parameter astrometric solution \cite{lindegren21}.
The RUWE cut does not prove that a star is single because unresolved
companions can remain within the nominal range \cite{stassun21}. The cut leaves 39 photometrically
selected, RUWE-clean FGK candidates within 10.3\,pc, 143 within
15.4\,pc, and 7{,}411 within 53.5\,pc (Table~\ref{tab:targets}). The
15.4\,pc sample includes 11 objects typed as early M stars by SIMBAD and
must therefore be spectroscopically reclassified before a final FGK count
is quoted. Wide, well-resolved
binaries remain usable through observation of a single component and
are not counted here as losses.

\begin{table}[tp]
\centering
\small
\begin{tabular}{llrrrr}
\toprule
Horizon & Enables & All stars$^{a}$ & FGK & RUWE-clean FGK$^{b}$ & M dwarfs \\
\midrule
3.5\,pc  & Earth twin, 1.6\,$\mu$m          & 14       & 3     & 2     & 11 \\
10.3\,pc & Earth twin, $3\lambda/D$, 550\,nm & 338      & 61    & 39    & 268 \\
15.4\,pc & Earth twin, $2\lambda/D$, 550\,nm & 1{,}176  & 211   & 143   & 929 \\
53.5\,pc & Jupiter twin, $3\lambda/D$, 550\,nm & 65{,}586 & 9{,}560 & 7{,}411 & 53{,}889 \\
\bottomrule
\end{tabular}
\caption{The coronagraph target-star funnel within each reflected-light
horizon of Table~\ref{tab:horizon}. Counts are from a Gaia DR3 query
(accessed 2026 July 11) for main-sequence stars with parallax
signal-to-noise above 5. FGK dwarfs are selected by absolute magnitude
$2.0 < M_G < 8.0$ and colour $0.3 < (G_{\rm BP}-G_{\rm RP}) < 2.0$;
M dwarfs by $M_G > 8.5$. $^{a}$Total is a lower limit at the smallest
radii, where bright nearby stars saturate Gaia. The volume-complete
10\,pc census has 541 objects in 339 systems \cite{reyle21}.
$^{b}$Photometric candidates with Gaia RUWE $<1.4$. The threshold is an astrometric-quality cut and does not prove that a source is single.}
\label{tab:targets}
\end{table}

The result sets the scale of the two programs. The Earth-analog
reflected-light survey addresses a reviewed subset of nearby FGK stars
within the 10--15\,pc horizon. The giant-planet survey at $10^{-9}$
contrast opens a much larger photometric candidate pool within 53.5\,pc, where
the photon budget rather than the target supply becomes the limiting
factor. Because the count grows as the cube of the horizon, every gain
in inner working angle or contrast multiplies the accessible sample.
This leverage is the central argument for pushing the diffraction
performance of the segmented 3.5\,m aperture to its limit.

\subsection{Diffuse backgrounds and false positives}
\label{sec:backgrounds}

The horizon and the target list assume that a resolved planet can be
detected against the local sky. Two of the relevant backgrounds are set
by dust. The local zodiacal light, sunlight scattered by interplanetary
dust, is the dominant diffuse foreground of any space-based visible
observation. Its $V$-band surface brightness is near
23.3\,mag\,arcsec$^{-2}$ at the ecliptic poles and stays within about
one magnitude of that value over most of the sky \cite{leinert98}. The
brightness rises toward the ecliptic plane. Therefore the ecliptic
latitude of a target enters the exposure-time and scheduling
calculation. The exozodiacal light, starlight scattered by dust in the
target system's own habitable zone, is frequently the larger term for
Earth-analog imaging. The LBTI HOSTS survey measured a median
habitable-zone dust level near 3 zodis for Sun-like stars, with a 95\%
upper limit of 27 zodis, where one zodi is the solar-system
habitable-zone dust surface density \cite{ertel20}. Exozodiacal dust
adds photon noise, and its clumps can mimic a planet. Its level is one
of the parameters that sets the exposure time and the planet yield of a
direct-imaging mission \cite{stark14}.

\begin{figure}[tp]
\centering
\includegraphics[width=0.66\textwidth]{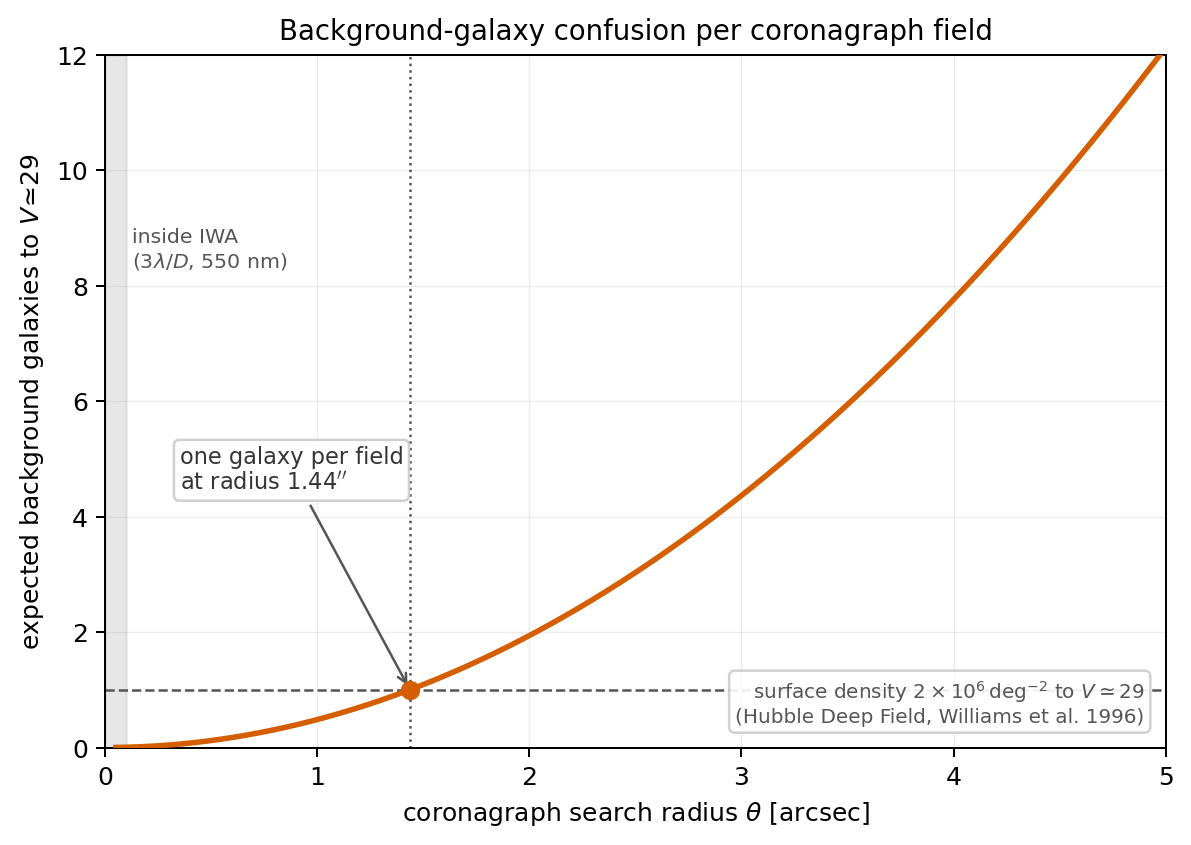}
\caption{Background-galaxy confusion. The curve is the expected number of
background galaxies inside a coronagraphic search region of radius
$\theta$, down to the $V\simeq29$ depth needed to detect an Earth analog,
from the Hubble Deep Field surface density of $2\times10^{6}$ galaxies
per square degree \cite{williams96}. One galaxy falls in the field, on
average, at a radius of $1.4''$. The shaded strip marks the region inside
the $3\lambda/D$ inner working angle at 550\,nm. The brighter
giant-planet program works at much shallower depth and encounters far
fewer background galaxies.}
\label{fig:bggal}
\end{figure}

A third background is discrete. A field star or a background galaxy that
projects close to the host star at a planet-like separation appears as a
candidate. The contamination rate scales with the surface density of
background sources. It therefore rises sharply for targets at low
Galactic latitude, as Figure~\ref{fig:mollweide} makes visible through
the bright stellar band. Faint background galaxies set the deepest layer
of this contamination. Their surface density reaches $2\times10^{6}$ per
square degree down to $V\simeq29$, the depth needed to detect an Earth
analog \cite{williams96}. Figure~\ref{fig:bggal} converts this density
into an expected count per coronagraph field. A search region of about
$1.4''$ radius already contains one such galaxy on average. Background
galaxies are therefore a real confusion source at the deepest contrast,
and they thin out for the brighter giant-planet program at shallower
depth. A nearby target star shows a large proper motion that a bound
planet shares and a background source does not. A second epoch then
separates the two by common proper motion \cite{bowler16}. The
reflected-light spectrum of a planet also differs from a stellar or
galactic spectrum. Multi-epoch astrometry and low-resolution
spectroscopy together remove essentially all such false positives.

\begin{figure}[tp]
\centering
\includegraphics[width=\textwidth]{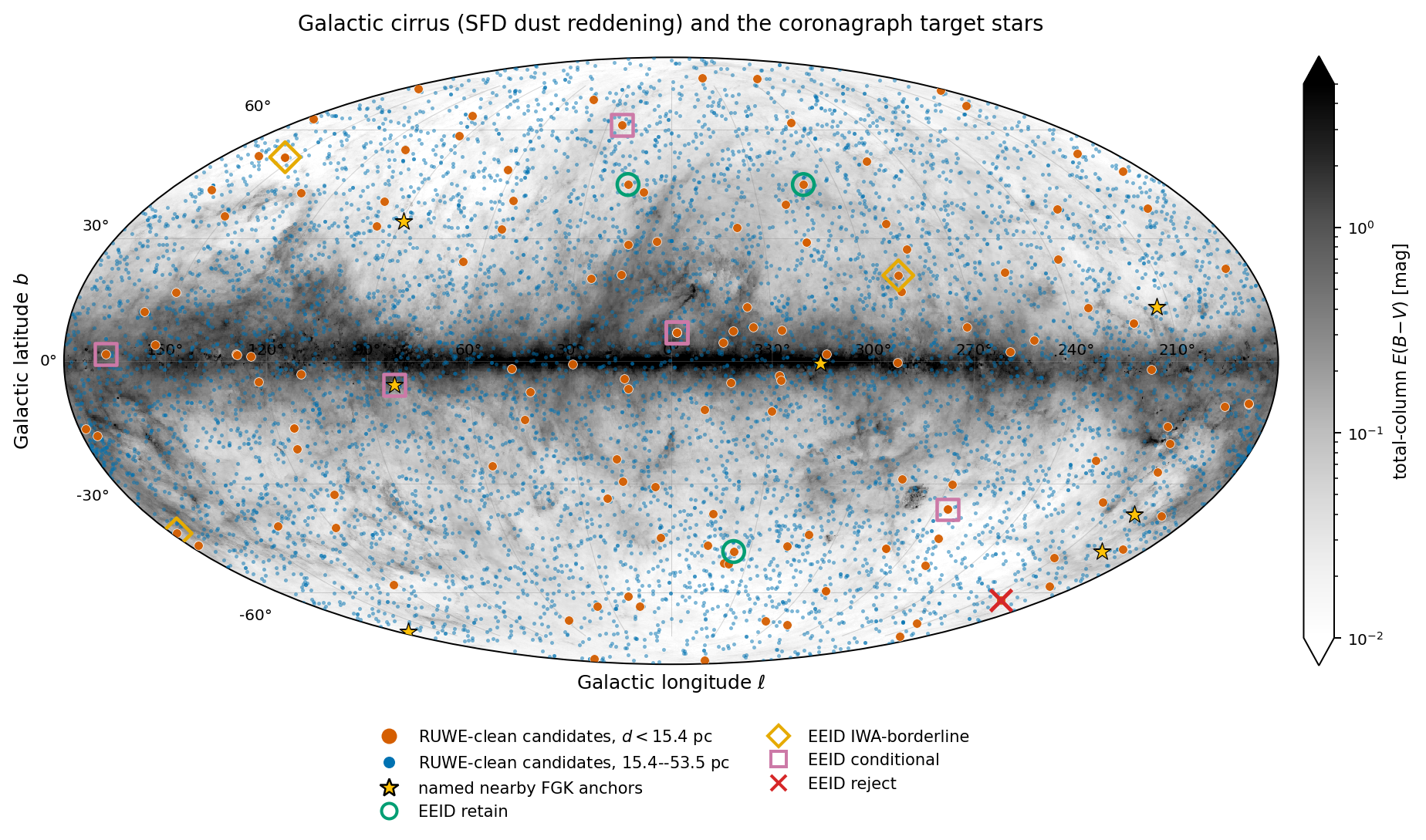}
\caption{The Galactic cirrus in Galactic coordinates (Mollweide
projection), from the Schlegel-Finkbeiner-Davis reddening map
\cite{sfd98}, with the coronagraph target stars of
Figure~\ref{fig:mollweide} overplotted. The gray scale is the
total-column $E(B\!-\!V)$ to infinity on a logarithmic stretch. The dust
forms a narrow band along the Galactic plane and thin high-latitude
filaments while the photometric candidates fill the clean high-latitude sky. The
map shows the full Galactic column. A target within 10--50\,pc lies
inside the nearly dust-free Local Bubble \cite{lallement19} and suffers
almost no extinction. Named nearby anchors are starred, including
$\alpha$~Centauri near the plane at $b=-0.7^\circ$. Open symbols show the
assessment of the 13 preliminary EEID candidates in
Table~\ref{tab:hztargets}.}
\label{fig:cirrus}
\end{figure}

Galactic cirrus, the diffuse emission and scattered light of
interstellar dust, is the background most prominent on an all-sky map,
yet it has little effect on the point-source coronagraph.
Figure~\ref{fig:cirrus} shows the standard all-sky reddening map
\cite{sfd98} with the target stars overplotted. The dust concentrates
in the Galactic plane, and the targets can be scheduled toward the clean
high-latitude sky. Three facts make cirrus negligible here. Its
scattered optical light is faint and smooth on the sub-arcsecond scale
of the coronagraphic field, far below the zodiacal and speckle floors.
Its thermal emission peaks in the far infrared near 100\,$\mu$m, away
from the reflected-light bands. Its extinction is negligible for the
sample, because the targets lie inside the Local Bubble, a cavity
largely cleared of dust within roughly 70\,pc where the reddening toward
nearby stars stays near zero \cite{lallement19}. The map records the
total dust column to infinity, and a target at 10--50\,pc intercepts
only the small near-side fraction of it.

\begin{table}[tp]
\centering
\small
\begin{tabular}{p{0.20\textwidth}p{0.44\textwidth}p{0.28\textwidth}}
\toprule
Background & Effect on the coronagraph & Mitigation \\
\midrule
Local zodiacal light & Dominant diffuse foreground, $V\simeq23.3$\,mag\,arcsec$^{-2}$ at the ecliptic poles & Favor high ecliptic latitude, schedule accordingly \\
Exozodiacal light & Often the largest noise term for Earth analogs (median 3 zodis), and its clumps can mimic planets & Pre-measure the exozodi level, multi-roll subtraction \\
Background stars and galaxies & Confusion and false positives, stellar confusion worst at low Galactic latitude, of order one background galaxy per field to $V\simeq29$ & Common proper motion, planet spectrum \\
Galactic cirrus (ISM) & Negligible: faint and smooth at sub-arcsecond scale, thermal in the far infrared, no extinction inside the Local Bubble & None needed within $\sim$70\,pc \\
\bottomrule
\end{tabular}
\caption{Sky backgrounds and false-positive sources for coronagraphic
imaging with the 3.5\,m telescope, and how each is handled. Diffuse
dust backgrounds set the photon-noise floor of
Section~\ref{sec:horizon}, while discrete sources are removed by
astrometry and spectroscopy.}
\label{tab:backgrounds}
\end{table}

\section{Direct Imaging of Exoplanets around Nearby Stars}
\label{sec:imaging}

Direct imaging is the core coronagraphic program. Ground-based
adaptive-optics systems have imaged young self-luminous giants, most
famously the four planets of HR~8799, three found in 2008 and a fourth in 2010
\cite{marois08,marois10}. The 51~Eridani
system added a cooler, more Jupiter-like young planet at 29\,pc
\cite{macintosh15}. In 2025 the James Webb Space Telescope detected a
sub-Jovian planet in the young TWA~7 disk with a coronagraph, at a mass
comparable to Saturn. This is among the lowest-mass planets yet
detected by direct imaging \cite{lagrange25}. The technique is
steadily advancing toward lower masses and smaller separations.

\begin{figure}[tp]
\centering
\includegraphics[width=0.86\textwidth]{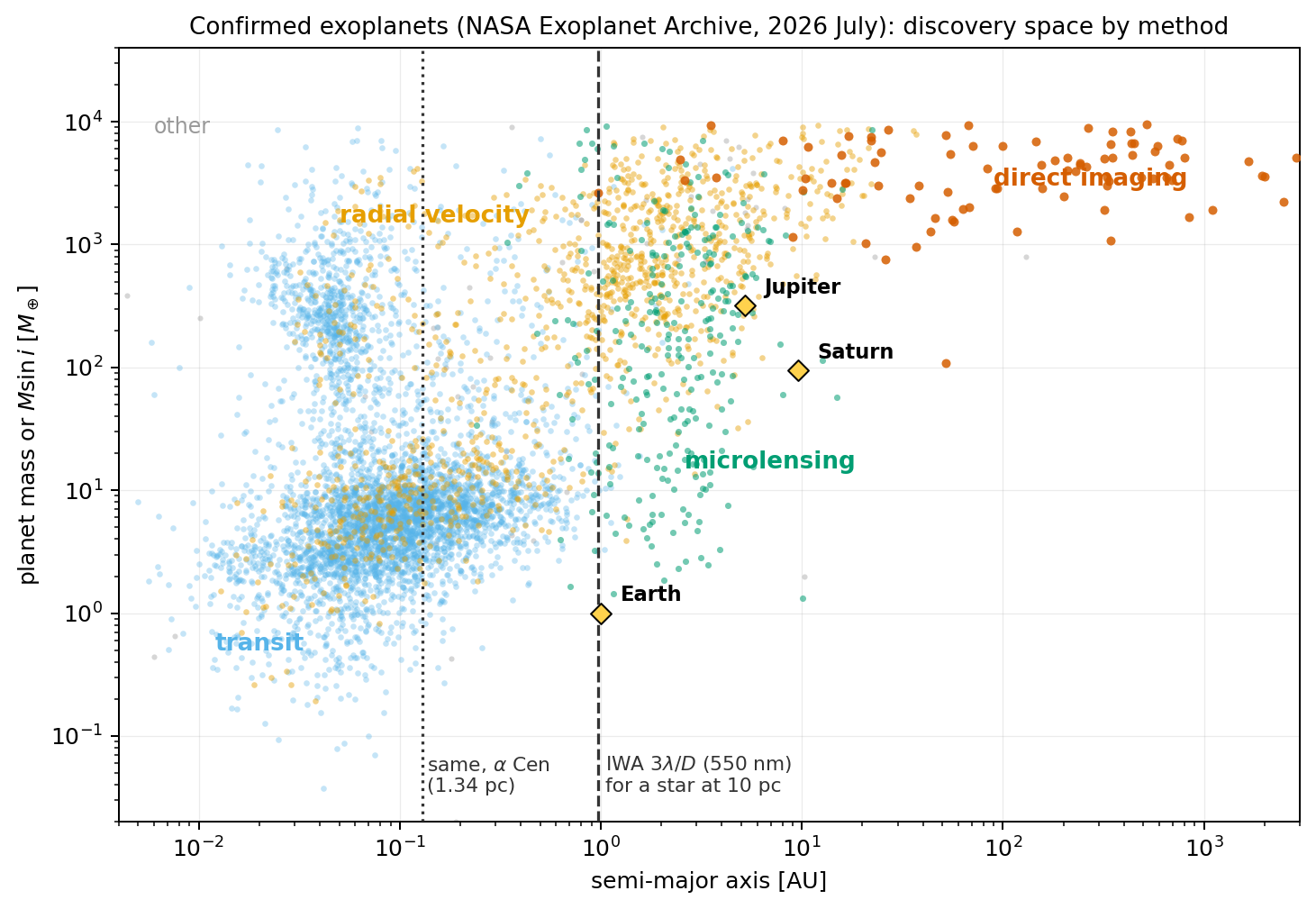}
\caption{Confirmed exoplanets from the NASA Exoplanet Archive (6{,}319
planets, accessed 2026 July 11) in the mass versus semi-major-axis
plane, colored by discovery method. Earth, Jupiter, and Saturn are
marked. The dashed line shows the projected separation of the
$3\lambda/D$ inner working angle of a 3.5\,m telescope at 550\,nm for a
star at 10\,pc. The dotted line shows the same angle at the 1.34\,pc
distance of $\alpha$~Centauri \cite{vanleeuwen07}. A space coronagraph
on the \facility\ opens the reflected-light region between the
close-in transit and radial-velocity populations and the wide-orbit
self-luminous population reached by current direct imaging.}
\label{fig:discovery}
\end{figure}

Figure~\ref{fig:discovery} shows why a space coronagraph on a 3.5\,m
aperture occupies a distinct region of discovery space. The confirmed
planet population from the NASA Exoplanet Archive divides cleanly by
method. Transit and radial-velocity discoveries crowd the region inside
1\,AU. Directly imaged planets sit at tens to hundreds of AU. The
reflected-light region between them, where the solar-system giants
actually reside, is nearly empty. For a star at 10\,pc the
$3\lambda/D$ IWA of the 3.5\,m telescope at 550\,nm corresponds to
1\,AU. The whole outer solar-system analog region around nearby stars
is therefore inside reach. For $\alpha$~Centauri the same IWA
corresponds to 0.13\,AU, and orbits comparable to that of Earth are
resolved with ample margin.

The space platform brings two structural advantages. It removes
atmospheric turbulence, which sets the speckle floor of ground-based
high-contrast imaging. It also permits integrations of essentially
unlimited length under a stable thermal environment. The specific
programs are the following.

\begin{itemize}
\item \textbf{Reflected-light detection of mature giants.} Jupiter and
Saturn analogs around stars within roughly 50\,pc, at contrasts near
$10^{-9}$ \cite{traub10,galicher23}. These planets are invisible to
transit surveys at such separations and only partially constrained by
long-baseline radial-velocity monitoring.
\item \textbf{Small separations around the nearest stars.} For systems
inside a few parsecs the IWA reaches well inside 1\,AU. The program
can search the habitable zones of $\alpha$~Centauri~A and B and image
any giant planets there directly.
\item \textbf{Orbit determination.} Repeated imaging epochs across the
mission lifetime measure orbital motion directly. Combined with radial
velocities this yields dynamical masses, which calibrate evolutionary
models of giant planets \cite{gaudi21}.
\end{itemize}
%%%%%%%%%%%%%%%%%%%%%%% B.5.1

\subsection{Discovery space and illustrative population scaling}
\label{sec:yield}

\begin{figure}[tp]
\centering
\includegraphics[width=0.78\textwidth]{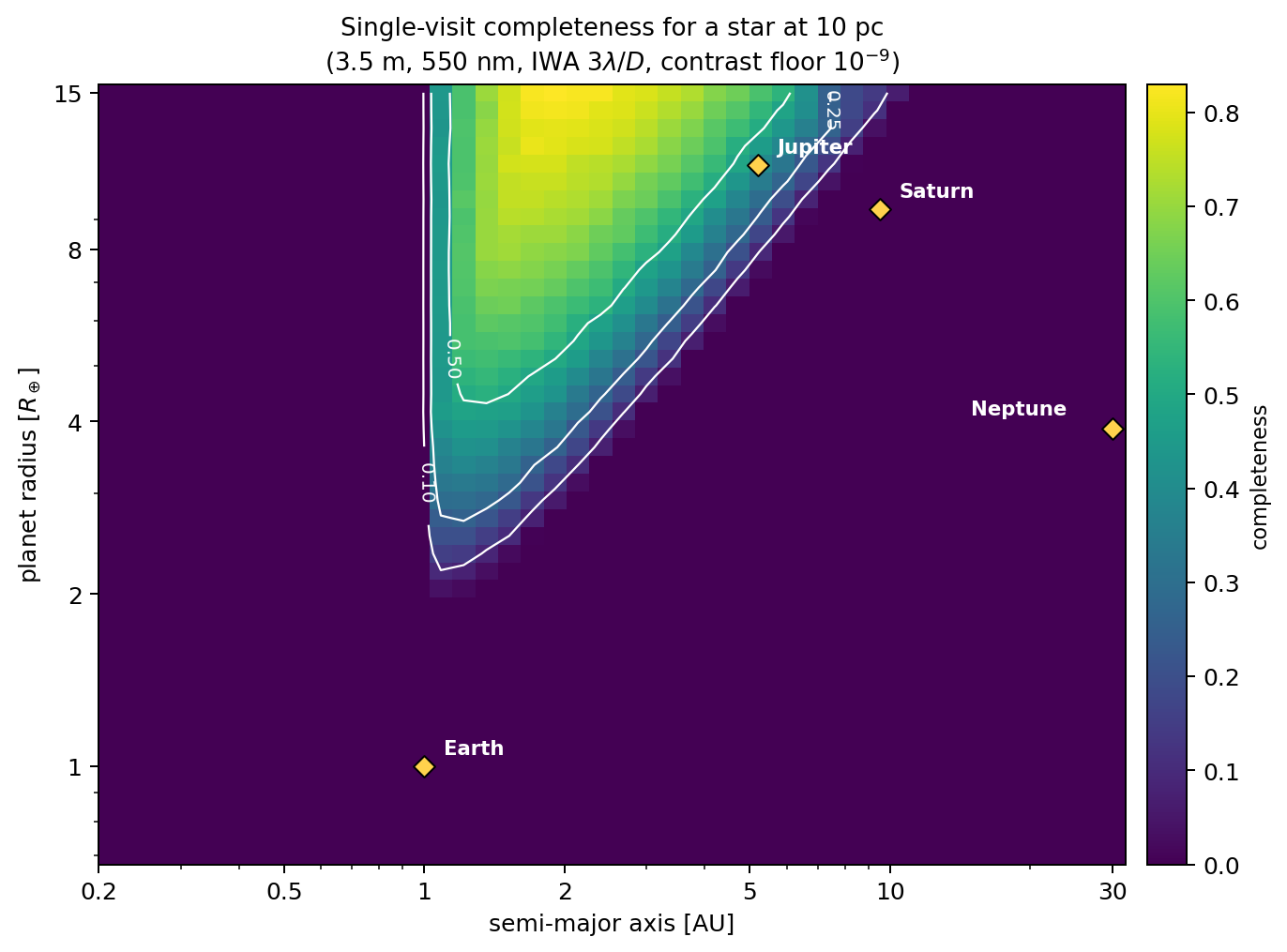}
\caption{Geometric single-visit completeness for a star at 10\,pc in the
planet-radius versus semi-major-axis plane, from a Monte-Carlo integration
over inclination and orbital phase with the $3\lambda/D$ inner working
angle and the $10^{-9}$ contrast floor at 550\,nm. Solar-system planets
are marked. The calculation omits separation-dependent throughput,
finite stellar diameter, exozodiacal emission, detector noise, and the
time required for each visit. The result is a discovery-space diagnostic
rather than a mission-yield prediction.}
\label{fig:completeness}
\end{figure}

The two instrument parameters of Section~\ref{sec:perf} map onto the
planets the mission can find. A planet is detectable when its orbit puts
the projected separation outside the inner working angle and its
instantaneous reflected contrast above the floor. Averaging over random
inclination and orbital phase gives the single-visit completeness across
the plane of planet radius and semi-major axis.
Figure~\ref{fig:completeness} shows it for a star at 10\,pc. The accessible
region is the mature giants and large sub-Neptunes between roughly 1 and
10\,AU. The inner edge is the working angle and the lower edge is the
contrast floor. Jupiter is therefore reached at moderate completeness,
while Earth, small and close, is not reached at all.

\begin{figure}[tp]
\centering
\includegraphics[width=0.74\textwidth]{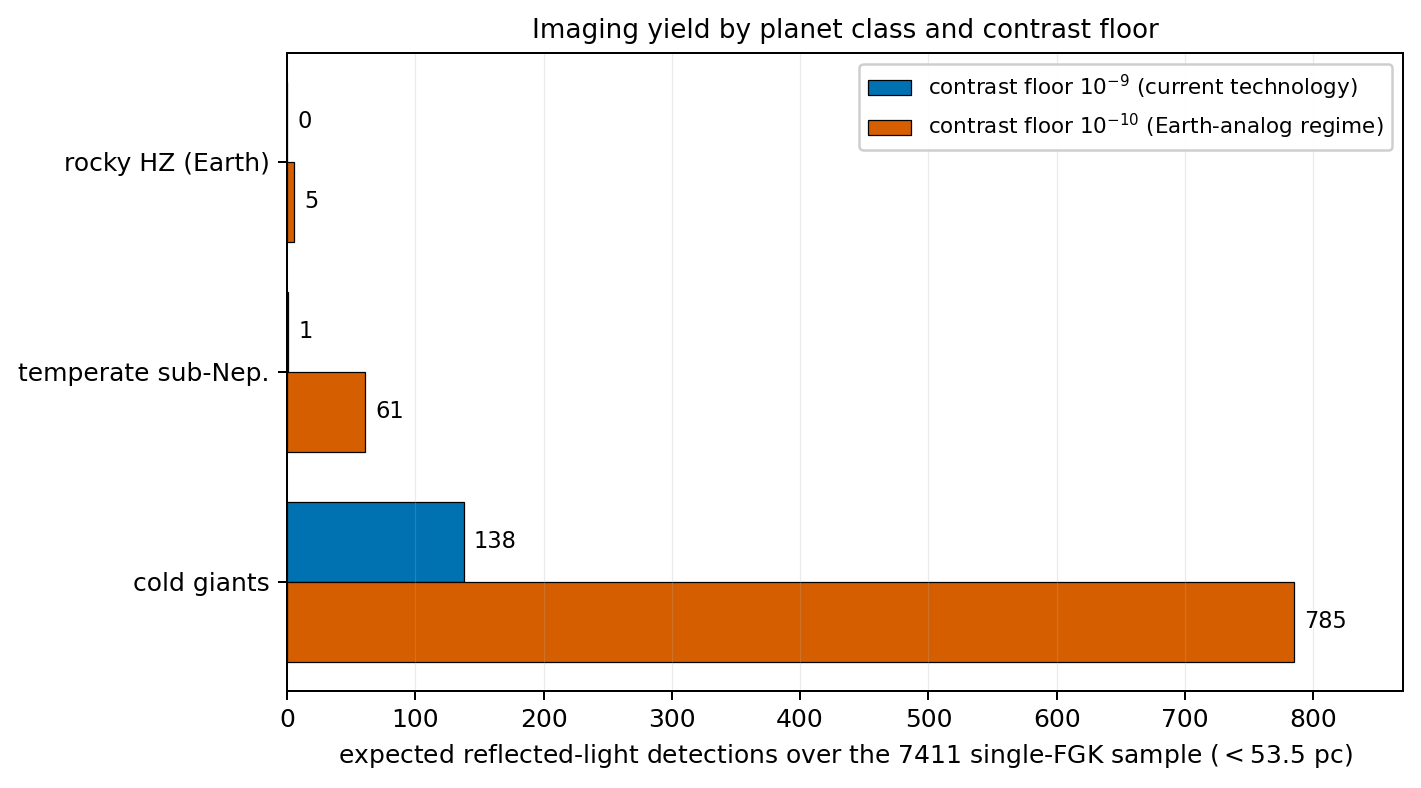}
\caption{Illustrative accessible populations over the 7411 RUWE-clean
photometric candidates within 53.5\,pc. The adopted occurrence rates are
0.15 for cold giants \cite{fernandes19,fulton21}. The calculation adopts
fiducial values of 0.30 for temperate sub-Neptunes and rocky
habitable-zone planets because published estimates depend on the class
boundaries and completeness model \cite{fulton17,bryson21}. The bars apply geometric completeness and a hard
contrast threshold. The bars do not include exposure time or target
optimization and must not be interpreted as mission detections.}
\label{fig:yieldbar}
\end{figure}

Figure~\ref{fig:yieldbar} multiplies geometric completeness by adopted
occurrence rates across the 7411 RUWE-clean photometric candidates.
Cold giants occur around roughly 15\% of FGK stars
\cite{fernandes19,fulton21}. Small-planet occurrence depends on the
adopted radius, period, and insolation intervals
\cite{fulton17,bryson21}. The calculation uses 0.30 for each
small-planet class as a fiducial assumption rather than a direct
measurement. The resulting counts show how strongly an
idealized accessible population changes with contrast. The counts are
not expected detections. A defensible mission yield requires
separation-dependent coronagraph throughput, stellar leakage,
astrophysical backgrounds, target-specific exposure times, revisits,
reliability, and optimization under a fixed mission-time budget.

\section{Reflected-Light Spectroscopy and Atmospheric Characterization}
\label{sec:spectra}

Detection is only the entry point. The coronagraph feeds the mission's
$R=5000$ optical and near-infrared spectrographs and the combination
turns direct imaging into atmospheric physics. At $R=5000$ one
resolution element spans 60\,km\,s$^{-1}$. The absorption bands of
water vapor and methane that dominate giant-planet reflection spectra
are sampled across many resolution elements rather than recorded as
single broad depressions. Reflected-light models show the bands, cloud
and haze signatures, and phase-dependent brightness that encodes
scattering properties \cite{cahoy10}. A
sequence of observations across orbital phase samples the atmosphere at
changing illumination and constrains cloud structure and vertical
mixing \cite{robinson18}.

The coronagraphic reflected-light spectroscopy follows the same
$0.3$--$3.0\,\mu$m optical and near-infrared spectroscopic baseline
defined in Volume~I.

\begin{figure}[tp]
\centering
\includegraphics[width=\textwidth]{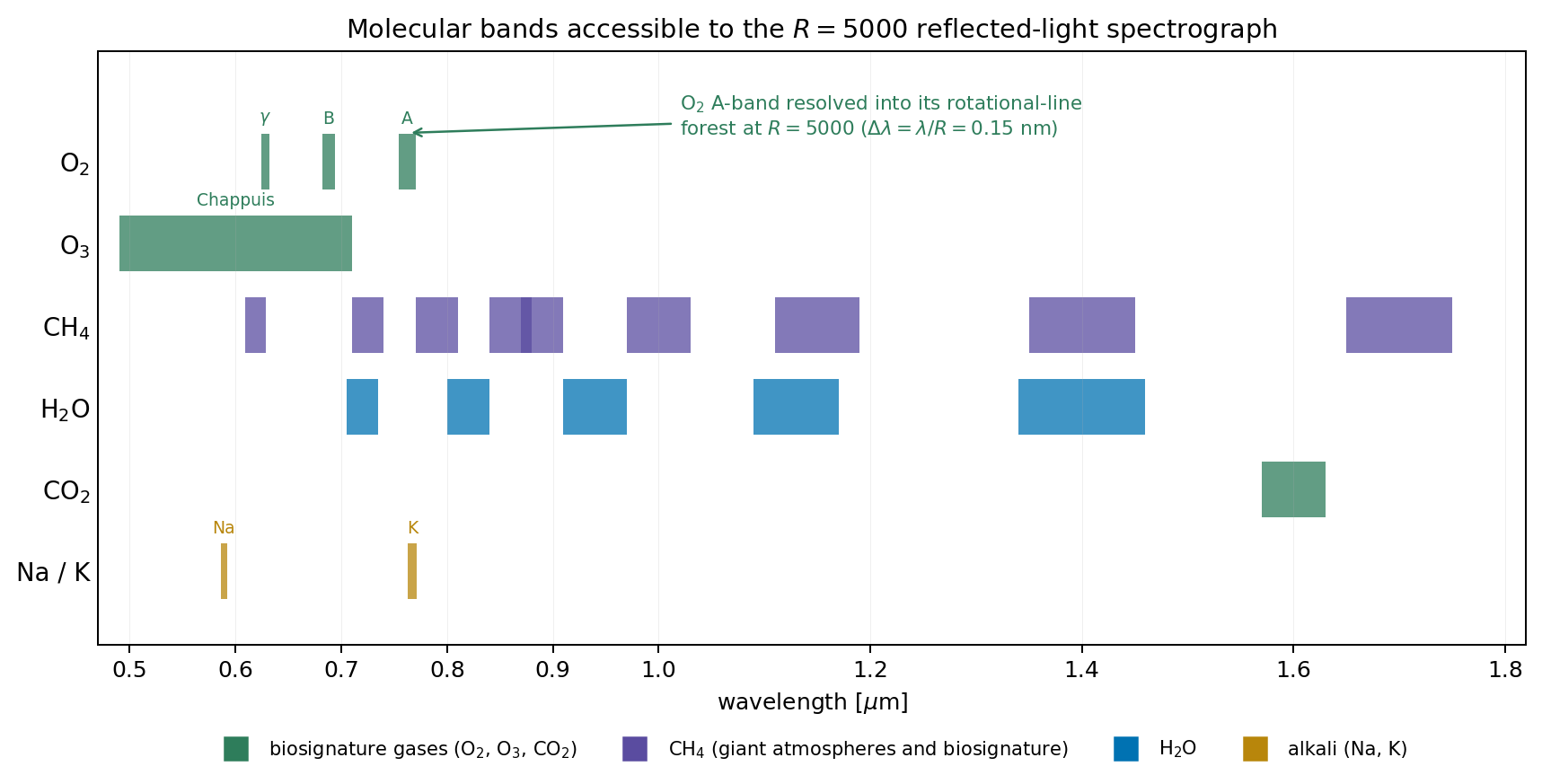}
\caption{Molecular absorption bands accessible to the $R=5000$
reflected-light spectrograph across $0.5$--$1.8\,\mu$m, at their true
wavelengths, grouped by what they diagnose. Biosignature gases (O$_2$,
O$_3$, CO$_2$) and the atmospheric tracers CH$_4$ and H$_2$O all fall in
the coronagraph band. Band positions from the biosignature and
giant-planet literature \cite{schwieterman18,karkoschka94}.}
\label{fig:bands}
\end{figure}

Figure~\ref{fig:bands} shows which molecules the $R=5000$ band reaches
and where. Methane and water shape giant-planet reflection spectra
across the optical and near-infrared range
\cite{karkoschka94,cahoy10},
and the biosignature gases of a terrestrial atmosphere, oxygen, ozone,
methane, and carbon dioxide, fall in the same window
\cite{schwieterman18}. The resolving power matters most for the narrow
features. The oxygen A-band at 0.76\,$\mu$m spans many $R=5000$
resolution elements. The spectrum samples its structured envelope but
does not fully resolve the individual rotational lines. The
cross-correlation calculation of Section~\ref{sec:specetc} therefore
combines resolved band structure rather than a fully separated line
forest.

For the nearest targets the same machinery extends toward smaller
planets. Reflected-light spectroscopy of super-Earths and
sub-Neptunes within a few parsecs can distinguish cloudy hydrogen
envelopes from thin secondary atmospheres \cite{damiano23}. The
$R=5000$ resolving power also supports template matching across a
molecular band. One resolution element spans about
60\,km\,s$^{-1}$. Typical orbital velocities do not cleanly separate
the planet spectrum from the stellar spectrum at this resolution. A
Doppler-separation gain would require a dedicated high-resolution mode
near $R=100{,}000$ \cite{snellen15} and is not included here.
Reflected-light measurements complement published transmission spectra,
which probe the upper atmosphere along the terminator, because reflected light
probes the dayside at depth \cite{robinson18}. The two geometries
together give a far more complete picture than either alone.

\subsection{Exposure time to read the atmosphere}
\label{sec:specetc}

\begin{figure}[tp]
\centering
\includegraphics[width=0.78\textwidth]{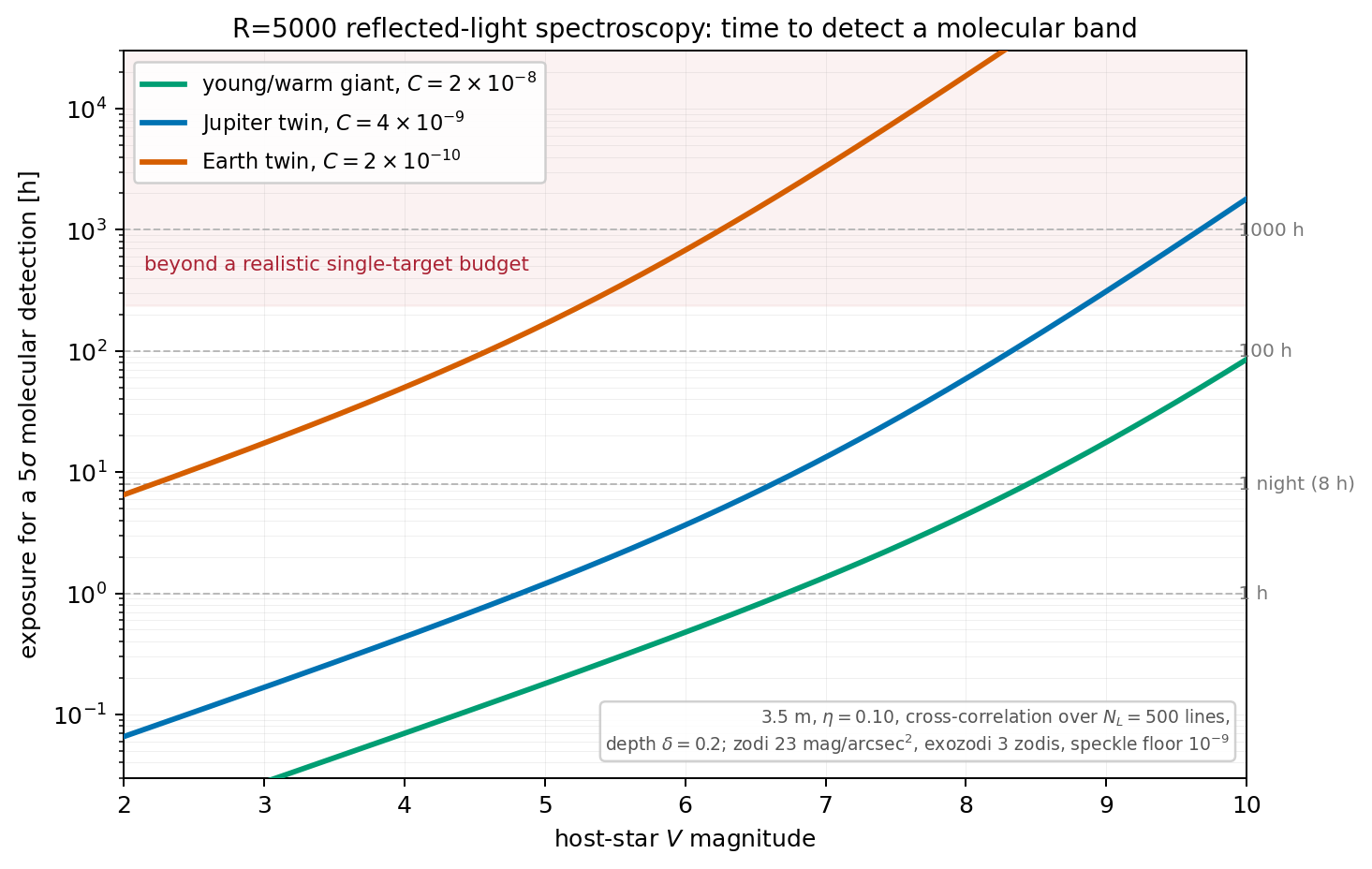}
\caption{Optimistic photon-noise time for a $5\sigma$ template detection of a
molecular band with the $R=5000$ spectrograph, as a function of
host-star $V$ magnitude, for three planet-to-star contrasts. The photon
budget uses a 3.5\,m aperture, an end-to-end throughput of 0.10, a
post-processed speckle floor of $10^{-9}$, the local zodiacal light at
$V=23$\,mag\,arcsec$^{-2}$ \cite{leinert98}, and a 3-zodi exozodi
\cite{ertel20}, with an idealized band of 500 lines of mean depth 0.2.
The calculation omits correlated residual speckles, detector noise,
template mismatch, and separation-dependent coronagraph throughput.}
\label{fig:specetc}
\end{figure}

The resolving power sets the exposure time as much as the detection
does. At $R=5000$ an individual molecular line of a reflected-light
planet holds far less than one photon. A band can be measured by
combining many spectral elements with a molecular template.
Figure~\ref{fig:specetc} gives the resulting time for a $5\sigma$
detection of a molecular band as a function of host brightness and
planet-to-star contrast, from a photon budget that includes the planet,
the residual starlight speckle floor, the local zodiacal light
\cite{leinert98} and the exozodiacal light \cite{ertel20}. The times are
optimistic photon-noise lower bounds. The model assigns 500 lines a
common mean depth and assumes that the speckle field subtracts to photon
noise. The calculation omits wavelength-dependent planet spectra,
template mismatch, detector noise, correlated speckles, and the loss of
coronagraph throughput near the IWA. The curves are sensitivity scalings
rather than observing-time predictions. Bright reflected-light giants
remain the appropriate spectroscopic targets. Earth-analog spectroscopy
belongs to the stretch-goal regime and to larger observatories such as
HWO \cite{hwo26}.

\begin{figure}[tp]
\centering
\includegraphics[width=0.78\textwidth]{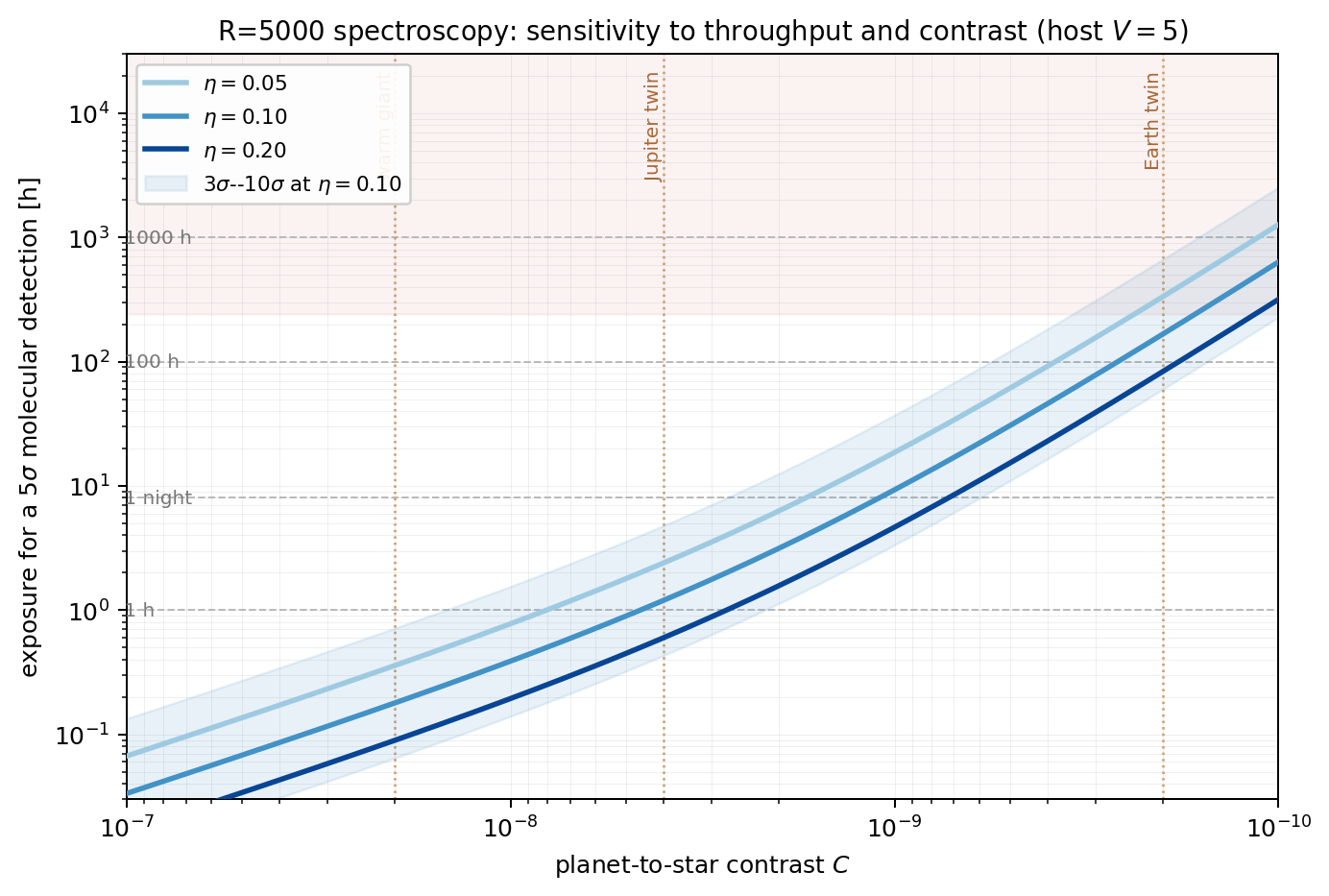}
\caption{Sensitivity of the $R=5000$ molecular-detection time to the
design parameters, for a fixed $V=5$ host. Curves are three end-to-end
throughputs, and the shaded band spans a $3\sigma$ to $10\sigma$
detection at $\eta=0.10$. Dotted lines mark the reflected-light contrasts
of a warm giant, a Jupiter twin, and an Earth twin. The time grows as the
inverse square of the contrast, which is why the Earth-twin regime lies
hundreds of times deeper than the giant-planet regime.}
\label{fig:specetcsens}
\end{figure}

The exposure time scales with the two levers of the design. It falls
inversely with the end-to-end throughput and inversely with the square of
the delivered contrast. Figure~\ref{fig:specetcsens} shows both
dependences at a fixed $V=5$ host. A factor-of-two gain in throughput, or
a factor-of-ten gain in contrast, moves a target from the overnight
regime to the one-hour regime. This is the quantitative form of the
argument in Section~\ref{sec:context}, in which contrast selects the
planet class and the photon budget selects how long its atmosphere takes
to read.

Table~\ref{tab:hztargets} applies the same lower-bound model to the 13
preliminary EEID candidates of Section~\ref{sec:catalog}. The numerical
times isolate photon scaling and must not be used as exposure-time
requests. The calculation gives especially short times for nearby K
dwarfs because compact EEIDs raise the reflected-light contrast. The
same targets lie close to the IWA where the omitted throughput and
stellar-leakage terms can dominate. Figures~\ref{fig:mollweide}
and~\ref{fig:cirrus} show the target assessments.

\begin{table}[tp]
\centering
\small
\setlength{\tabcolsep}{5pt}
\begin{tabular}{r l l r r r r r r l}
\toprule
\textbf{\#} & \textbf{Name} & \textbf{SpT} & \textbf{d\,(pc)} & \textbf{V} & $\boldsymbol{a_{\rm EEID}}$ & $\boldsymbol{\theta_{\rm EEID}}$ & $\boldsymbol{C_{\oplus}}$ & $\boldsymbol{t_{\rm char}}$ & \textbf{Assessment} \\
 & & & & & (AU) & (mas) & ($10^{-10}$) & (h) & \\
\midrule
1 & 61 Cyg A & K5V & 3.50 & 5.21 & 0.38 & 108 & 12.1 & 8.7 & conditional$^{a}$ \\
2 & eps Ind & K5V & 3.64 & 4.69 & 0.50 & 137 & 6.9 & 12 & retain$^{b}$ \\
3 & 36 Oph B & K1V & 5.95 & 5.03 & 0.59 & 99 & 5.0 & 31 & conditional$^{c}$ \\
4 & 36 Oph A & K2V & 5.95 & 5.08 & 0.59 & 99 & 5.0 & 34 & conditional$^{c}$ \\
5 & ksi Boo A & G7Ve & 6.75 & 4.68 & 0.75 & 111 & 3.1 & 49 & conditional$^{d}$ \\
6 & 61 Vir & G6.5V & 8.53 & 4.74 & 0.91 & 107 & 2.1 & 112 & retain$^{e}$ \\
7 & HD 102365 & G2V & 9.32 & 4.88 & 0.91 & 98 & 2.1 & 132 & borderline$^{f}$ \\
8 & kap01 Cet & G5V & 9.28 & 4.85 & 0.94 & 101 & 2.0 & 141 & borderline$^{g}$ \\
9 & zet Dor & F9VFe-0. & 11.69 & 4.71 & 1.22 & 104 & 1.2 & 322 & conditional$^{h}$ \\
10 & 36 UMa & F8V & 12.95 & 4.67 & 1.29 & 100 & 1.0 & 382 & borderline$^{i}$ \\
11 & lam Ser & G0-V & 11.92 & 4.42 & 1.43 & 120 & 0.9 & 417 & retain$^{j}$ \\
12 & lam Aur & G1.5IV-V & 12.56 & 4.71 & 1.32 & 106 & 1.0 & 448 & conditional$^{k}$ \\
13 & tau01 Eri & F7V & 14.28 & 4.46 & 1.65 & 115 & 0.6 & 768 & reject$^{l}$ \\
\bottomrule
\end{tabular}
\caption{The 13 preliminary targets whose Earth-equivalent-insolation distance (EEID) lies outside the nominal $3\lambda/D=97$\,mas inner working angle at 550\,nm, ordered by the modeled spectroscopic exposure time. The maximum-separation cut does not establish observability across the full habitable zone or at every orbital phase. ``Retain'' marks a sound candidate. ``Borderline'' marks a target within a few mas of the nominal IWA. ``Conditional'' identifies a target that requires further review of binarity, disks, or luminosity class. ``Reject'' identifies a target that is inconsistent with the nominal single-star sample. $C_\oplus$ assumes an Earth analog at $a_{\rm EEID}$ with a geometric albedo of 0.3 at quadrature. $t_{\rm char}$ is the modeled time for a $5\sigma$ cross-correlation detection of a molecular band at $R=5000$. The estimate is not an achieved instrumental sensitivity.}
\label{tab:hztargets}
\vspace{2pt}
\begin{minipage}{0.98\textwidth}
\footnotesize
$^{a}$The 61~Cyg~AB binary has a separation of about $28''$. Retention requires a companion-light simulation.
$^{b}$The known planet host is a strong nearby target. The field review must include the distant substellar companions.
$^{c}$36~Oph~A and B form a nearly equal-brightness pair separated by about $5.1''$. The pair is one problematic binary system rather than two independent single stars. Both EEIDs are also IWA-borderline.
$^{d}$Xi~Boo~A has a companion at about $5.2''$.
$^{e}$Known three-planet host and a sound target.
$^{f}$The EEID lies only $0.6$\,mas outside the nominal IWA. Performance uncertainty can reverse the flag.
$^{g}$Only $3.7$\,mas outside the nominal IWA and magnetically active.
$^{h}$The target is a very wide binary. Use requires the companion and circumstellar environment to satisfy the stray-light and background budgets.
$^{i}$Only $2.3$\,mas outside the nominal IWA and a wide-binary component.
$^{j}$Known planet host and a sound target, subject to ordinary activity/RV review.
$^{k}$The wide-binary component has a SIMBAD luminosity class of G1.5IV--V rather than an unambiguous dwarf.
$^{l}$The target is a known spectroscopic binary with a period of about 958 days. The target should be removed from a nominal single-star list unless a dedicated binary-coronagraph analysis demonstrates usability.
\end{minipage}
\end{table}

\section{Planet Formation and Young Systems}
\label{sec:formation}

Young planetary systems are the natural early targets of the module,
because newly formed giant planets remain warm and self-luminous for
tens of millions of years. At those ages the planet-to-star contrast is
far more favorable than in reflected light, and the science does not
require the deepest contrast floor. Observations of nearby young moving
groups can therefore begin as soon as the module is commissioned.

The program addresses planet formation directly. Imaging of young
giants at different ages traces their cooling curves and
discriminates between hot-start and cold-start formation histories
\cite{zurlo24}. Coronagraphic imaging of circumstellar and debris
disks maps the belts, gaps, and asymmetries that planets carve into
their birth environment. The TWA~7 detection shows this synergy in a
single system, where the planet was found inside the structure of the
very disk it shapes \cite{lagrange25}. A stable space platform with a
3.5\,m aperture extends such disk imaging to fainter and more compact
structures than current facilities reach.

The favorable contrast of young systems sets the entry point of the
program. A young giant of a few Jupiter masses at an age of 10\,million
years is self-luminous at a planet-to-star flux ratio near $10^{-6}$ in
the near infrared, four orders of magnitude brighter than the same planet
in mature reflected light, and comfortably above the contrast floor of
Section~\ref{sec:perf} \cite{zurlo24}. Scattered-light debris disks around
the nearest stars reach surface brightnesses within reach of the same
observations, and the exozodiacal dust that limits the terrestrial program
(Section~\ref{sec:backgrounds}) becomes, for these targets, a science
signal in its own right. The nearby young moving groups therefore give the
module its first science while the wavefront control matures toward the
reflected-light floor.

\section{Observing Plan and First-Light Targets}
\label{sec:ops}

The catalog of Section~\ref{sec:catalog} already contains known planets,
and the archive gives their orbits. Table~\ref{tab:firstlight} ranks the
known planets of the catalog stars by their maximum angular separation and
lists the reflected-light contrast that follows from the measured or
assumed radius. Six of them fall between the inner and outer working
angles at 550\,nm and clear the adopted $10^{-9}$ threshold. These six are
first-light candidates whose existence and approximate orbits are already
known. Final selection requires predicted phase, radius or albedo,
separation-dependent throughput, and stellar leakage. A successful
recovery would test characterization methods rather than blind discovery.
The five widest known planets, including the directly
imaged cold giant $\epsilon$~Indi~A~b, project beyond the $20\lambda/D$
outer working angle and reflect too little light to clear the floor. They
remain out of reach in reflected light.

\begin{table}[tp]\centering\small\setlength{\tabcolsep}{5pt}
\begin{tabular}{l r r r r r c}
\toprule
\textbf{Planet} & $\boldsymbol{a}$ & $\boldsymbol{\theta_{\max}}$ & $\boldsymbol{M}$ & $\boldsymbol{R}$ & $\boldsymbol{C}$ & \textbf{img?} \\
 & (AU) & (mas) & ($M_\oplus$) & ($R_\oplus$) & ($10^{-9}$) & \\
\midrule
eps Ind A b & 15.76 & 4330 & 2066 & 12.7 & 0.1 & n \\
HD 115404 A c & 11.36 & 1034 & 3280 & 12.5 & 0.2 & n \\
HD 222237 b & 10.80 & 943 & 1650 & 12.8 & 0.2 & n \\
47 UMa d & 11.60 & 835 & 521 & 13.5 & 0.2 & n \\
HD 140901 c & 11.80 & 774 & 572 & 13.4 & 0.2 & n \\
HD 219134 h & 3.11 & 476 & 108 & 12.7 & 2.9 & Y \\
55 Cnc d & 5.60 & 445 & 1232 & 13.0 & 0.9 & n \\
47 UMa c & 3.60 & 259 & 172 & 14.2 & 2.7 & Y \\
47 UMa b & 2.10 & 151 & 804 & 13.2 & 6.8 & Y \\
HD 192310 c & 1.18 & 134 & 24 & 5.2 & 3.4 & Y \\
GJ 414 A c & 1.40 & 118 & 54 & 8.4 & 6.2 & Y \\
HD 147513 b & 1.32 & 102 & 385 & 13.7 & 18.7 & Y \\
55 Cnc f & 0.80 & 64 & 45 & 7.6 & 15.5 & n \\
HD 219134 g & 0.38 & 57 & 11 & 3.3 & 13.2 & n \\
61 Vir d & 0.48 & 56 & 23 & 5.1 & 20.0 & n \\
HD 69830 d & 0.63 & 50 & 18 & 4.5 & 8.6 & n \\
\bottomrule
\end{tabular}
\caption{Known planets orbiting the catalog stars, ranked by maximum angular separation $\theta_{\max}=a/d$ (the 16 widest of 48 known planets around the 22 hosts). $C$ is the reflected-light contrast at quadrature for the listed or assumed radius ($^{g}$: no measured radius, a gas giant is assumed to be one Jupiter radius). The last column flags the 6 planets that a reflected-light coronagraph can actually image: their separation falls between the $3\lambda/D$ inner ($97$\,mas) and $20\lambda/D$ outer ($648$\,mas) working angles at 550\,nm and their contrast clears the $10^{-9}$ floor. The widest planets project beyond the outer working angle and reflect too little light to clear the floor. They are therefore not reflected-light targets. Data: NASA Exoplanet Archive.}\label{tab:firstlight}
\end{table}

\begin{figure}[tp]
\centering
\includegraphics[width=0.76\textwidth]{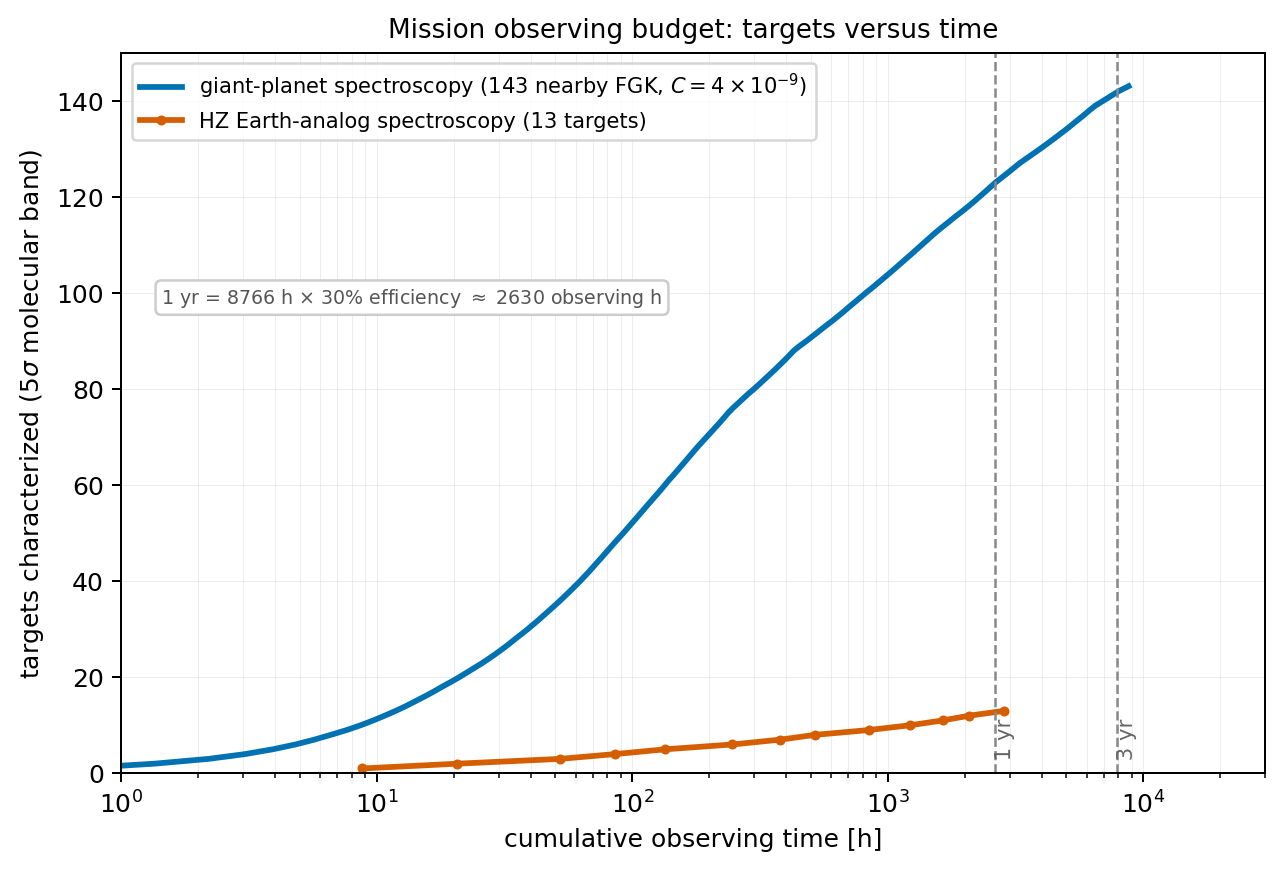}
\caption{Illustrative cumulative ETC scaling for hypothetical planets,
ordered by photon-noise time. The giant curve assigns a
$C=4\times10^{-9}$ Jupiter analog to each of the 143 input stars. The
EEID curve assigns an Earth analog to each preliminary EEID candidate.
Neither curve applies planet occurrence, coronagraph throughput near the
IWA, correlated residuals, confirmation visits, or operational overhead.
The curves do not represent a mission yield or schedule.}
\label{fig:ops-exo}
\end{figure}

Figure~\ref{fig:ops-exo} shows an ETC scaling rather than a mission
timeline. The calculation places a Jupiter analog around every one of
the 143 input stars and orders the hypothetical spectra by photon-noise
time. Planet occurrence is not applied. A real program first searches
the candidate stars and confirms detected companions. Spectroscopy then
follows only for planets with adequate separation and brightness. The
EEID curve provides a stretch-goal comparison and does not define a
scheduled Earth-analog campaign.

\section{Demographics and Classification}
\label{sec:demographics}

\begin{figure}[tp]
\centering
\includegraphics[width=0.82\textwidth]{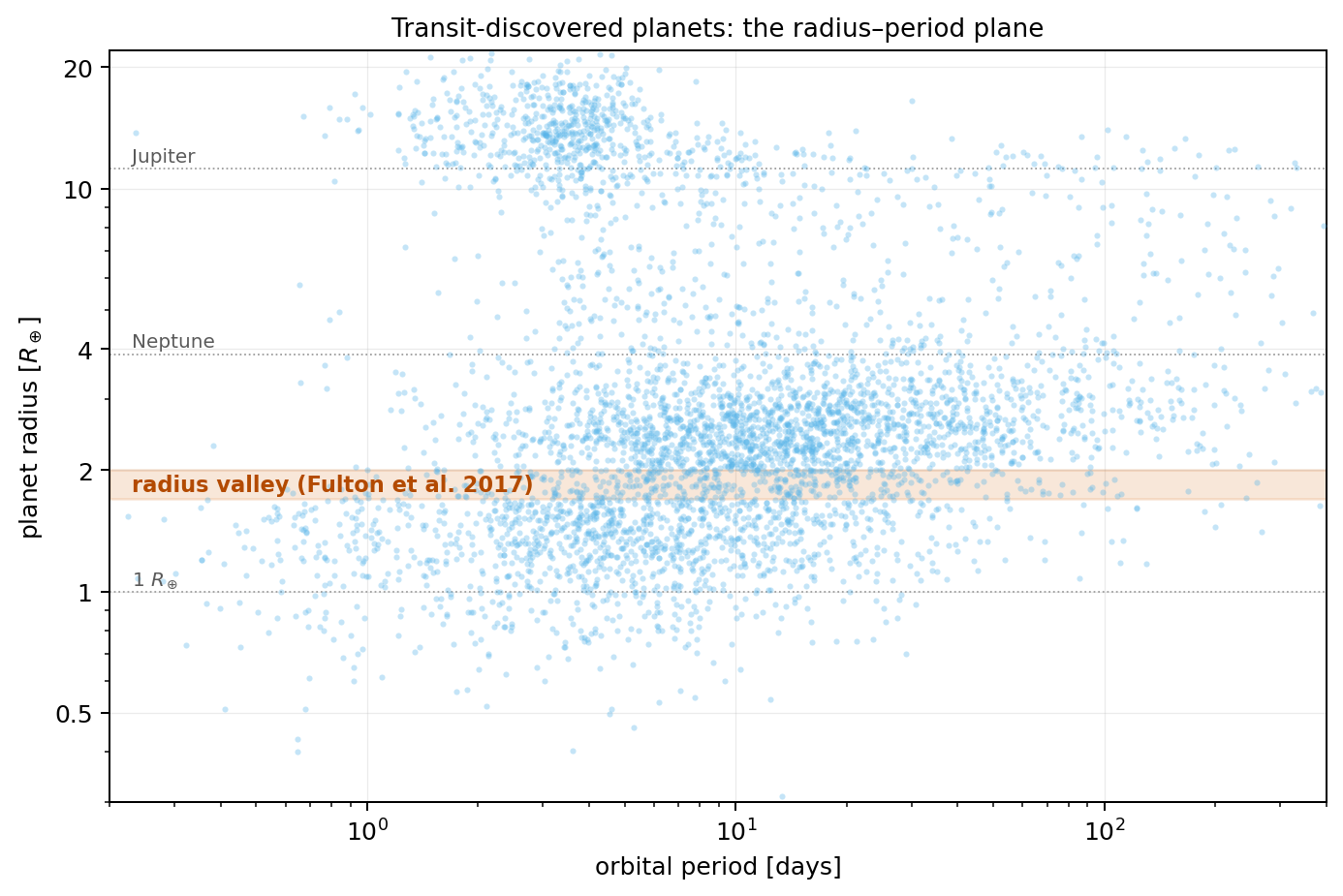}
\caption{Transit-discovered planets from the NASA Exoplanet Archive in
the radius versus orbital-period plane. The shaded band marks the
radius valley at 1.7--2.0\,$R_\oplus$ identified by the
California-Kepler Survey \cite{fulton17}. Occurrence-rate work built
on such samples underlies current estimates of $\eta_\oplus$
\cite{bryson21}. The published population provides a comparison sample for
the coronagraphic survey and does not represent a forecast of 3.5ST transit
detections.}
\label{fig:radiusvalley}
\end{figure}

Exoplanet science has moved from discovery to statistics. More than
6{,}300 confirmed planets are now cataloged \cite{nea} and classified by
mass, radius, composition, and system architecture
\cite{kopparapu18,howe25}. The Kepler mission monitored roughly
150{,}000 stars for four years and established that small planets
inside 1\,AU are extremely common \cite{borucki10}. Bias-corrected
occurrence rates from that sample now constrain the frequency of rocky
habitable-zone planets around Sun-like stars, the quantity
$\eta_\oplus$ \cite{bryson21}. Figure~\ref{fig:radiusvalley} shows the
sharpest demographic feature of the transit population, the valley near
1.7--2.0\,$R_\oplus$ that separates super-Earths from sub-Neptunes
\cite{fulton17}.

The \facility\ measures the population of giant planets at separations
accessible to its coronagraph \cite{gaudi21,zurlo24}. Published Kepler
results and catalogued TESS discoveries provide comparison samples for
the inner planetary systems \cite{borucki10,bryson21,nea}. The transit
population in Figure~\ref{fig:radiusvalley} supplies observational context
for the direct-imaging program. The observing plan in this volume assigns
no independent transit search or transmission-spectroscopy campaign.

The comparison must include the selection function of each survey.
Transit samples depend on orbital alignment, observing duration, and
photometric completeness. The coronagraphic sample depends on angular
separation, planet-to-star contrast, and orbital phase. Occurrence rates
can be compared only after these detection probabilities and the host-star
distributions have been included in the analysis. Published transit and
radial-velocity measurements of the same systems can then test whether
inner planets preferentially coexist with outer giants
\cite{bryson21,howe25}.

\section{Habitability and Biosignatures}
\label{sec:habitability}

\begin{figure}[tp]
\centering
\includegraphics[width=0.82\textwidth]{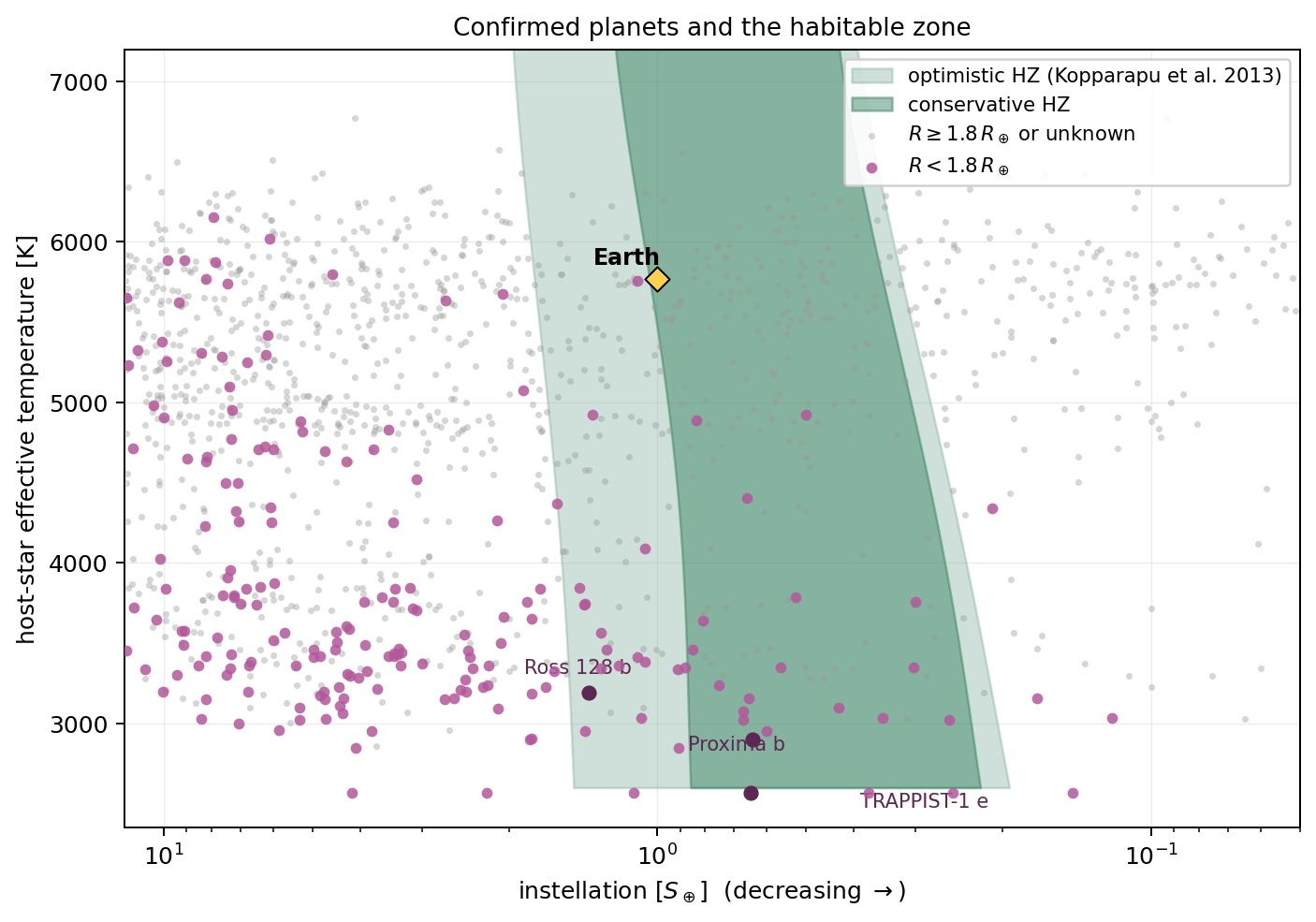}
\caption{Confirmed planets from the NASA Exoplanet Archive in the
instellation versus host-star temperature plane. The green bands are
the optimistic and conservative habitable zones of Kopparapu et al.
\cite{kopparapu13}. Planets smaller than $1.8\,R_\oplus$ are
highlighted, and Earth, Proxima Centauri b, Ross 128 b, and
TRAPPIST-1 e are labeled. The best-characterized temperate small
planets orbit M dwarfs, where stellar activity complicates
habitability (Section~\ref{sec:activity}).}
\label{fig:hz}
\end{figure}

The habitability program asks whether a planet possesses the physical
and chemical conditions required to support liquid water and,
ultimately, life. The first-order tool is the habitable zone, the
range of orbital distances in which stellar irradiation permits liquid
surface water \cite{seager13,kopparapu13}. Figure~\ref{fig:hz} places
the confirmed planet population on the instellation versus stellar
temperature plane with the standard habitable-zone boundaries.
Residence in the zone is necessary but far from sufficient.
Atmospheric composition, greenhouse balance, geology, and magnetic
protection all enter a full assessment \cite{robinson18,hong23}.

Observationally the program rests on atmospheric spectroscopy. Gases
such as O$_2$, O$_3$, CH$_4$, CO$_2$, and H$_2$O are the canonical
biosignature candidates, and the simultaneous presence of oxygen and
methane is especially significant because the pair is hard to maintain
abiotically \cite{schwieterman18,claudi19}. Published transmission spectra
provide comparative constraints on
atmospheric absorption along the terminator. The 3.5ST program measures
reflected-light spectra through the coronagraph. These spectra sample the
illuminated hemisphere and add
sensitivity to surface pressure, clouds, and even ultraviolet-band
biosignatures \cite{damiano23}. Interpretation should confront
false positives and false negatives, because abiotic photochemistry can
mimic biosignatures and living planets can hide them
\cite{schwieterman18,robinson18}. Multiple indicators observed in
context are therefore required.

The \facility\ contributes at three levels. First, its coronagraphic
spectroscopy of nearby giants and sub-Neptunes builds the reference
atlas of reflected-light spectra against which future terrestrial
detections will be interpreted. Second, its long-term monitoring
capability tracks temporal variability, including clouds, seasons, and
weather, on directly imaged planets \cite{robinson18}. Third, it
serves as a working precursor for the Habitable Worlds Observatory
class of missions, which require the $10^{-10}$ contrast regime to
reach Earth analogs \cite{hwo26}. Experience with
segmented-aperture wavefront control on this mission feeds directly into
that technology path.

\section{Stellar Activity as a Calibration Requirement}
\label{sec:activity}

Stellar activity is retained as a supporting calibration measurement rather
than as an independent time-domain science program. Flares, coronal mass
ejections, and high-energy radiation can modify an atmosphere and can also
introduce variability into reflected-light spectra
\cite{airapetian17escape,chen21}. The activity state of the host star must
therefore enter the interpretation of atmospheric features and the assessment
of biosignature false positives.

The 3.5ST measures the required activity indicators during observations of
the coronagraphic target sample. Long-baseline imaging provides a reference
level for rotational modulation and flare occurrence. Selected $R=5000$
spectra measure chromospheric line variability when the activity state could
affect a planet-spectrum interpretation. These data constrain the stellar
contamination model and establish whether a putative molecular feature is
stable across epochs. They do not define a separate flare-monitoring survey,
and no standalone time-domain yield is assigned in the five-year allocation.

This restriction keeps the scientific scope focused on direct imaging and
reflected-light spectroscopy. It also preserves the physical role of stellar
activity in the habitability analysis without presenting the 3.5ST as a
general transient or multi-messenger observatory.

\section{Relation to the International Landscape}
\label{sec:landscape}

The exoplanet field has turned from discovery to characterization, and
the two largest programs have converged on the same destination. The
United States Astro2020 decadal survey \cite{astro2020} and the
European Voyage~2050 plan \cite{voyage2050} both assign high priority
to the characterization of temperate terrestrial planets. The roads
to that destination differ by method, and each
method suits a different planet population.

The United States has committed its flagship to a large reflected-light
coronagraph. The Habitable Worlds Observatory, the roughly 6\,m descendant
of the LUVOIR and HabEx studies, is designed to image and characterize
Earth analogs around Sun-like stars at the $10^{-10}$ contrast level
\cite{astro2020,hwo26}. The Roman Space Telescope Coronagraph Instrument is
a 2.4\,m technology pathfinder that demonstrates the $10^{-8}$ contrast
regime and seeks a post-processed level of a few $\times10^{-9}$
\cite{bailey23,llopsayson25}. Ground-based Extremely Large Telescopes add
high-resolution spectroscopy for the nearest systems, but atmospheric
turbulence and telluric absorption remain part of their measurement model.

The comparison below places the 3.5ST alongside the facilities that set the
relevant direct-imaging and high-contrast benchmarks.

\begin{table}[tp]
\centering
\caption{Scope comparison for the high-contrast exoplanet program. The
comparison records the primary measurement and the complementary role of
each facility.}
\label{tab:exoplanet_comparison}
\begin{tabularx}{\textwidth}{@{}p{34mm}p{31mm}X@{}}
\hline
\textbf{Facility} & \textbf{Primary regime} & \textbf{Relationship to the 3.5ST}\\
\hline
Roman Coronagraph & Technology demonstration near $10^{-8}$ contrast &
Provides the closest flight reference for high-contrast control. The 3.5ST
adds a larger aperture, a systematic nearby-star sample, and repeated
optical/NIR characterization.\\
Habitable Worlds Observatory & Approximately 6\,m flagship at the
$10^{-10}$ contrast scale & Targets a larger Earth-analog population. The
3.5ST provides a nearer-term reflected-light bridge and a wavefront-control
precursor for the flagship program.\\
Ground-based ELTs & Adaptive-optics imaging and high-resolution spectroscopy &
Provide very high spectral resolution for the nearest systems. The 3.5ST
provides an atmosphere-free PSF, stable time series, and uninterrupted
moderate-resolution spectroscopy.\\
3.5ST & 3.5\,m space coronagraph with repeated nearby-star observations &
Targets mature giants, warm sub-Neptunes, and the few nearest FGK habitable
zones while measuring stellar activity and reflected-light spectra.\\
\hline
\end{tabularx}
\end{table}

A 3.5\,m reflected-light coronagraph sits between the Roman pathfinder and
the Habitable Worlds Observatory. At this aperture a full survey of
Sun-like-star Earth twins is out of reach, which is the reason the
flagship is sized at 6\,m. The aperture is instead matched to three
roles. It characterizes the reflected-light atmospheres of nearby mature
giants and warm sub-Neptunes, the population that Section~\ref{sec:specetc}
shows to be reachable in hours. It reaches the habitable zones of the few
nearest and brightest FGK stars. It matures segmented-aperture wavefront
control and reflected-light spectroscopy on real targets as a risk
reduction for the flagship. The nearby systems observed by this mission
provide a reflected-light reference sample for future flagship concepts.
The present program does not include a separate thermal-infrared observing
mode.
\section{Program Integration and Roadmap}
\label{sec:roadmap}

The coronagraphic programs share targets, instruments, and calibration
observations. Published radial velocities and Gaia astrometry provide
orbital constraints for target selection and companion confirmation.
External transit measurements provide radii for transiting planets in
the same systems. A bulk density requires both the mass and radius of
the same planet. Reflected-light brightness alone leaves a degeneracy
between radius and albedo. The imaging visits and external monitoring
data supply the stellar-activity reference required by
Section~\ref{sec:activity}. The program uses the published transit
measurements as complementary data.

The roadmap separates demonstrated capability from development goals.
The baseline program uses $10^{-8}$-class performance for young
self-luminous giants, debris disks, and selected known companions. The
development program targets calibrated $10^{-9}$ performance for mature
giant discovery and follow-up of the brightest detections. Earth-analog
work at $10^{-10}$ is a stretch goal rather than a mission promise. The
stretch program positions the mission as a wavefront-control and
target-selection precursor for dedicated flagship observatories
\cite{quanz22,hwo26}.

\section{Five-Year Observing Schedule}
\label{sec:schedule}

The exposure model of Section~\ref{sec:specetc} and the target census of
Section~\ref{sec:horizon} together define the inputs to a five-year observing
plan. Two facts shape the allocation. The first is that the Sun governs when a
coronagraphic target is reachable. The current observatory concept permits
solar elongations from $90^\circ$ to $180^\circ$. A Roman-type external baffle
and telescope tube provide Sun rejection rather than a JWST-type deployed
sunshield. Reflected-light imaging at $10^{-9}$ contrast also favors low
zodiacal foreground. A given star therefore receives target-specific windows
rather than continuous access.

The second fact is that search, confirmation, and characterization have
different cadences. A search visit identifies candidates. Confirmation
visits reject background sources and constrain orbital motion. Spectroscopy
is reserved for confirmed planets that meet the brightness and separation
limits. EEID observations remain an unscheduled stretch activity because
the present ETC does not establish their feasibility.

\begin{figure}[tp]
\centering
\includegraphics[width=\textwidth]{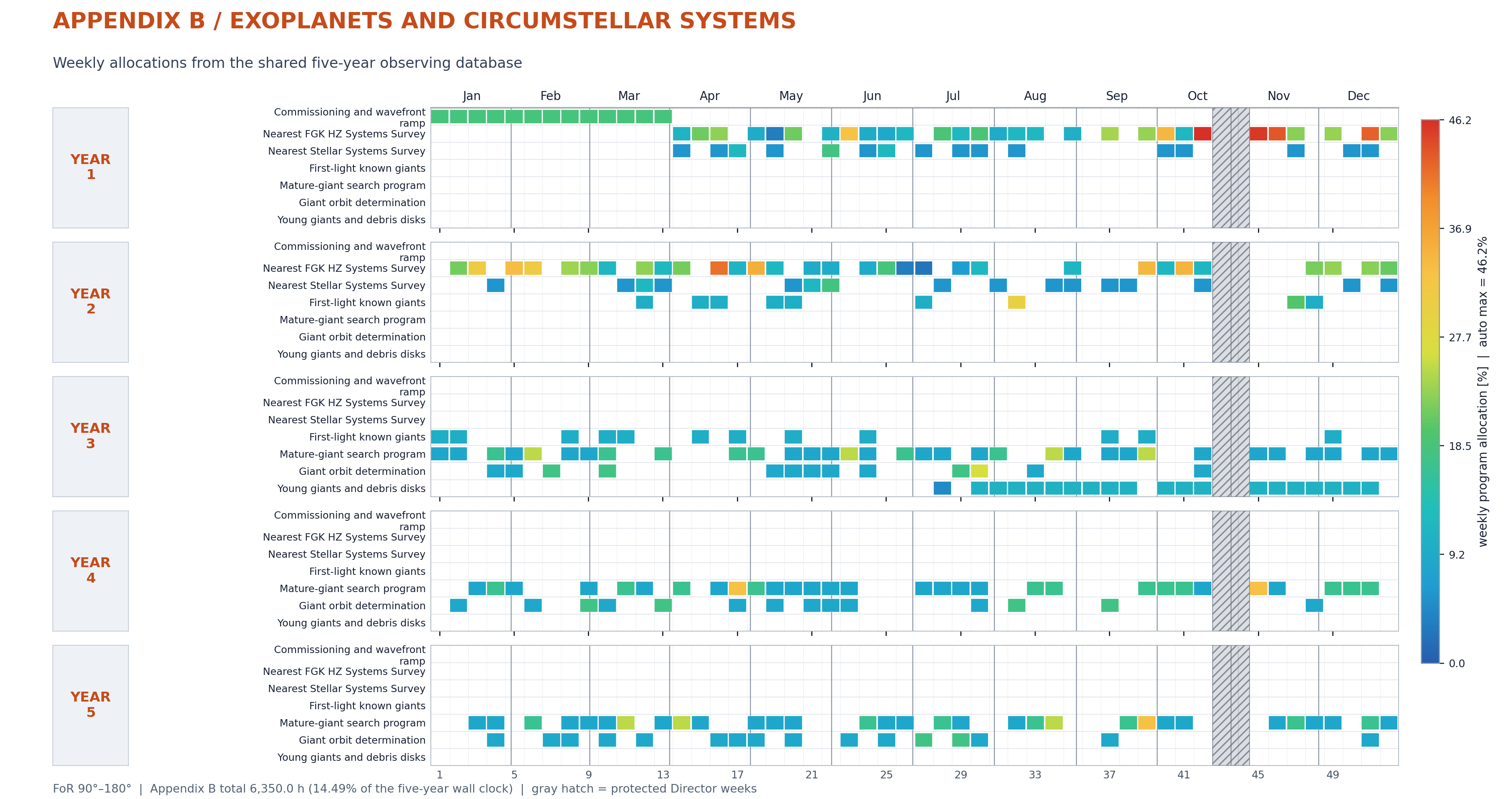}
\caption{Weekly Volume III allocation generated from the shared observing
database. Color gives the fraction of one 168-hr week assigned to each
program. The color scale uses the largest Volume III allocation and gray
hatching marks Director weeks 43 and 44 in every mission year. Named targets
are scheduled only within their $90^\circ$--$180^\circ$ solar-elongation
windows and under the mission-wide weekly capacity. The calculation does not
yet impose planet-phase posteriors or measured coronagraph overheads.}
\label{fig:exotimeline}
\end{figure}

Table~\ref{tab:exotimeline} gives the current database-driven allocation and
Figure~\ref{fig:exotimeline} shows the corresponding weekly assignments. The
6350~hr total equals 14.49 per cent of the five-year wall clock. The allocation
demonstrates seasonal and shared-capacity feasibility but does not establish
planet completeness. A validated target sequence will require target-specific
integration times, reference-star visits, dark-hole maintenance, roll
diversity, acquisition, calibration, confirmation epochs, and orbital-phase
constraints.

\begin{table}[tp]
\centering
\sffamily\small
\begin{tabularx}{\textwidth}{@{}
  >{\raggedright\arraybackslash}p{3.5cm}
  >{\raggedright\arraybackslash}p{2.15cm}
  >{\raggedright\arraybackslash}p{2.4cm}
  >{\raggedright\arraybackslash}X@{}}
\toprule
\textbf{Mission phase} & \textbf{Nominal period} & \textbf{Science time} & \textbf{Main deliverable}\\
 & mission year & hr & \\
\midrule
Commissioning and wavefront ramp & Year 1 & 400 hr & Demonstrate a calibrated dark hole near $10^{-8}$ and measure the achieved contrast, throughput, bandwidth, and stability.\\
Nearest Stellar Systems Survey & Years 1--2 & 400 hr & Obtain a reference atlas of the 20 nearest stellar components and search for giant planets, circumstellar disks, and reflected-light companions.\\
Nearest FGK HZ Systems Survey & Years 1--2 & 2000 hr & Survey the 20 nearest FGK main-sequence stars with priority given to fully accessible conservative habitable zones.\\
First-light known giants & Years 2--3 & 400 hr & Attempt reflected-light recovery of the candidates in Table~\ref{tab:firstlight} after phase and throughput review.\\
Young giants and debris disks & Year 3 & 350 hr & Observe nearby moving-group giants and scattered-light disks after commissioning establishes wavefront-control performance.\\
Mature-giant search program & Years 3--5 & 2000 hr & Search a reviewed stellar sample after $10^{-9}$ performance is demonstrated. Confirmed detections receive spectroscopic follow-up.\\
Giant orbit determination & Years 3--5 & 800 hr & Repeat epochs across the mission to measure orbital arcs and dynamical masses.\\
\midrule
\textbf{Volume III total} & \textbf{Years 1--5} & \textbf{6350 hr} & \textbf{14.49 per cent of the five-year wall clock.}\\
\bottomrule
\end{tabularx}
\caption{Database-driven five-year reference allocation for Volume III. The
mature-giant search begins only after on-orbit performance demonstrates the
required contrast. Earth-analog EEID observations remain an unallocated
feasibility study. Stellar-activity measurements use the wide-field program
and do not consume the coronagraph allocation.}
\label{tab:exotimeline}
\end{table}

\section{Target Catalog: RUWE-clean Photometric Candidates within 15.4\,pc}
\label{sec:catalog}

Table~\ref{tab:catalog} lists the 143 Gaia photometric candidates with
RUWE below 1.4 inside the 15.4\,pc geometric horizon of
Section~\ref{sec:horizon}. The catalog is an input sample rather than a
set of 143 confirmed single FGK dwarfs. SIMBAD classifies 11 as early-M stars. Several have
IV--V luminosity classes, and known wide or spectroscopic binaries survive
the RUWE cut.
The candidates form the input sample mapped in
Figures~\ref{fig:mollweide} and~\ref{fig:faceon}, ordered by distance.
The catalog combines Gaia DR3 astrometry and photometry with SIMBAD
identifiers and spectral types and with the confirmed-planet history from
the NASA Exoplanet Archive. To keep it compact the table is laid out in
two blocks, with star $k$ on the left and star $k+72$ on the right. For
each star it lists the distance, brightness, and spectral type, together
with quantities used to plan a reflected-light observation. The
Earth-equivalent-insolation distance
$a_{\rm EEID}=\sqrt{L/L_\odot}$\,AU is a representative orbit, not the
full habitable zone. Its maximum angular separation
$\theta_{\rm EEID}=a_{\rm EEID}/d$ supplies a preliminary geometric flag.
The final column of each block marks the 13 stars for which this separation
nominally clears the $3\lambda/D$ inner working angle at 550\,nm. The flag does
not include orbital projection, phase completeness, IWA uncertainty, or
companion-light contamination. Table~\ref{tab:hztargets} records the
resulting target-by-target assessment. One entry is rejected and several
entries are conditional or IWA-borderline. The remainder of the input sample may
still offer accessible outer habitable-zone or giant-planet orbits.
Twenty-two
of the stars already host known planets, most from radial-velocity
surveys, and these systems provide the first characterization targets for
the coronagraph. Coordinates, luminosity, mass, the reflected-light
contrast, and the $R=5000$ exposure time of every star are in the
machine-readable catalog. A few wide, resolved pairs such as
36~Ophiuchi~A and~B appear as separate entries. Each is observed one
component at a time, in line with the multiplicity policy of
Section~\ref{sec:horizon}.

\begin{landscape}
\footnotesize
\setlength{\tabcolsep}{3.6pt}
\begin{longtable}{r l r r l r r c l @{\;\vrule\;} r l r r l r r c l}
\caption{Gaia photometric coronagraph candidates within 15.4\,pc, ordered by distance (143 objects). The two-block layout places object $k$ on the left and object $k+72$ on the right. $d$ is the distance in parsecs. $a_{\rm HZ}=\sqrt{L/L_\odot}$\,AU is more precisely the Earth-equivalent-insolation distance (EEID) and $\theta_{\rm HZ}=a_{\rm HZ}/d$ is the maximum angular separation in mas. HZ? is a preliminary geometric flag set to Y when $\theta_{\rm HZ}$ exceeds the nominal $3\lambda/D$ inner working angle at 550\,nm ($97$\,mas). The flag does not establish full-orbit or full-HZ accessibility. The Gaia DR3 photometric cut and $\mathrm{RUWE}<1.4$ select an astrometrically clean input sample rather than confirmed single FGK dwarfs. SIMBAD types include 11 early-M stars, several IV--V objects, and known binaries. Spectral type and $V$ are from SIMBAD. Planets are from the NASA Exoplanet Archive. RV denotes radial velocity and tr denotes transit.}\label{tab:catalog}\\
\toprule
\textbf{\#} & \textbf{Name} & \textbf{d} & \textbf{V} & \textbf{SpT} & $\boldsymbol{a_{\rm HZ}}$ & $\boldsymbol{\theta_{\rm HZ}}$ & \textbf{HZ?} & \textbf{Pl.} & \textbf{\#} & \textbf{Name} & \textbf{d} & \textbf{V} & \textbf{SpT} & $\boldsymbol{a_{\rm HZ}}$ & $\boldsymbol{\theta_{\rm HZ}}$ & \textbf{HZ?} & \textbf{Pl.} \\
 & & (pc) & & & (AU) & (mas) & &  &  & & (pc) & & & (AU) & (mas) & &  \\
\midrule
\endfirsthead
\multicolumn{18}{l}{\footnotesize\itshape Table~\ref{tab:catalog} continued}\\
\toprule
\textbf{\#} & \textbf{Name} & \textbf{d} & \textbf{V} & \textbf{SpT} & $\boldsymbol{a_{\rm HZ}}$ & $\boldsymbol{\theta_{\rm HZ}}$ & \textbf{HZ?} & \textbf{Pl.} & \textbf{\#} & \textbf{Name} & \textbf{d} & \textbf{V} & \textbf{SpT} & $\boldsymbol{a_{\rm HZ}}$ & $\boldsymbol{\theta_{\rm HZ}}$ & \textbf{HZ?} & \textbf{Pl.} \\
 & & (pc) & & & (AU) & (mas) & &  &  & & (pc) & & & (AU) & (mas) & &  \\
\midrule
\endhead
\bottomrule
\endfoot
1 & 61 Cyg B & 3.50 & 6.03 & K7V & 0.30 & 85 & n &  & 73 & HD 82106 & 12.79 & 7.20 & K3V & 0.50 & 39 & n &  \\
2 & 61 Cyg A & 3.50 & 5.21 & K5V & 0.38 & 108 & Y &  & 74 & HD 158633 & 12.79 & 6.22 & K0 & 0.74 & 58 & n &  \\
3 & eps Ind & 3.64 & 4.69 & K5V & 0.50 & 137 & Y & 1 (RV) & 75 & HD 14412 & 12.83 & 6.34 & G8V & 0.66 & 52 & n &  \\
4 & AX Mic & 3.97 & 6.68 & M1V & 0.28 & 71 & n &  & 76 & HD 147513 & 12.89 & 5.38 & G5V & 1.02 & 79 & n & 1 (RV) \\
5 & HD 88230 & 4.87 & 6.61 & K6VeFe-1 & 0.30 & 61 & n &  & 77 & 36 UMa B & 12.92 & 8.08 & K7Ve & 0.30 & 23 & n &  \\
6 & HD 131977 & 5.89 & 5.72 & K4V & 0.48 & 81 & n &  & 78 & HD 36003 & 12.93 & 7.64 & K5V & 0.46 & 35 & n &  \\
7 & eta Cas B & 5.93 & 7.51 & K7Ve & 0.26 & 43 & n &  & 79 & HD 40307 & 12.93 & 7.15 & K2.5V & 0.51 & 39 & n & 5 (RV) \\
8 & 36 Oph B & 5.95 & 5.03 & K1V & 0.59 & 99 & Y &  & 80 & 36 UMa & 12.95 & 4.67 & F8V & 1.29 & 100 & Y &  \\
9 & 36 Oph A & 5.95 & 5.08 & K2V & 0.59 & 99 & Y &  & 81 & HD 172051 & 13.03 & 5.86 & G6V & 0.83 & 64 & n &  \\
10 & V2215 Oph & 5.95 & 6.34 & K5V & 0.38 & 64 & n &  & 82 & HD 27274 & 13.04 & 7.63 & K4.5Vk: & 0.45 & 34 & n &  \\
11 & HD 191408 & 6.01 & 5.32 & K2.5V & 0.55 & 91 & n &  & 83 & HD 211970 & 13.13 & 8.99 & K7Vk: & 0.29 & 22 & n & 1 (RV) \\
12 & HD 79210 & 6.33 & 6.98 & K7V & 0.30 & 47 & n &  & 84 & HD 170657 & 13.18 & 6.81 & K2V & 0.59 & 44 & n &  \\
13 & HD 219134 & 6.54 & 5.57 & K3V & 0.54 & 83 & n & 6 (RV) & 85 & HD 166348 & 13.21 & 8.37 & M0V & 0.43 & 33 & n &  \\
14 & ksi Boo A & 6.75 & 4.68 & G7Ve & 0.75 & 111 & Y &  & 86 & HD 29697 & 13.21 & 7.98 & K4V & 0.38 & 29 & n &  \\
15 & HD 4628 & 7.44 & 5.74 & K2.5V & 0.56 & 75 & n &  & 87 & HD 128165 & 13.23 & 6.92 & K3 & 0.52 & 39 & n &  \\
16 & TW PsA & 7.60 & 6.48 & K4Ve & 0.43 & 57 & n &  & 88 & 58 Eri & 13.24 & 5.50 & G2.5IV-V & 0.97 & 74 & n &  \\
17 & 107 Psc & 7.64 & 5.24 & K1V & 0.68 & 90 & n &  & 89 & HD 214749 & 13.28 & 7.84 & K4.5Vk & 0.42 & 31 & n &  \\
18 & HD 157881 & 7.71 & 7.56 & K7V & 0.32 & 42 & n &  & 90 & BD+62 780 & 13.49 & 8.38 & K7V & 0.30 & 22 & n &  \\
19 & p Eri A & 8.19 & 5.68 & K2V & 0.61 & 75 & n &  & 91 & HD 10436 & 13.52 & 8.41 & K5Vbe & 0.34 & 25 & n &  \\
20 & p Eri B & 8.20 & 5.80 & K2V & 0.58 & 71 & n &  & 92 & HD 145417 & 13.60 & 7.54 & K3VFe-1. & 0.45 & 33 & n &  \\
21 & HD 217357 & 8.23 & 7.87 & K7+Vk & 0.30 & 36 & n &  & 93 & HD 22496 & 13.60 & 8.57 & K5V & 0.31 & 23 & n & 1 (RV) \\
22 & 61 Vir & 8.53 & 4.74 & G6.5V & 0.91 & 107 & Y & 3 (RV) & 94 & HD 154577 & 13.73 & 7.41 & K2.5Vk: & 0.47 & 35 & n &  \\
23 & HD 50281 & 8.74 & 6.59 & K3.5V & 0.46 & 53 & n &  & 95 & HD 166 & 13.77 & 6.07 & G8V & 0.81 & 59 & n &  \\
24 & 41 Ara & 8.79 & 5.48 & G9V & 0.68 & 78 & n &  & 96 & 47 UMa & 13.89 & 4.87 & G1-VFe-0 & 1.05 & 76 & n & 3 (RV) \\
25 & HD 192310 & 8.81 & 5.72 & K2+V & 0.65 & 73 & n & 2 (RV) & 97 & HD 23356 & 13.94 & 7.08 & K2V & 0.55 & 39 & n &  \\
26 & HD 32147 & 8.84 & 6.21 & K3+V & 0.55 & 62 & n &  & 98 & HD 5133 & 13.94 & 7.17 & K2.5Vk: & 0.54 & 39 & n &  \\
27 & AK Lep & 8.89 & 6.15 & K2.5V(k) & 0.55 & 62 & n &  & 99 & HD 211415 & 14.07 & 5.39 & G0V & 1.08 & 77 & n &  \\
28 & HD 103095 & 9.17 & 6.45 & K1V\_Fe-1 & 0.47 & 51 & n &  & 100 & HD 120467 & 14.07 & 8.17 & K6Va & 0.42 & 30 & n &  \\
29 & kap01 Cet & 9.28 & 4.85 & G5V & 0.94 & 101 & Y &  & 101 & HD 61606 & 14.08 & 7.17 & K3V & 0.56 & 40 & n &  \\
30 & HD 102365 & 9.32 & 4.88 & G2V & 0.91 & 98 & Y &  & 102 & BD-03 2002 & 14.09 & 8.93 & K7V & 0.32 & 22 & n &  \\
31 & 20 Crt & 9.56 & 5.98 & K0V & 0.59 & 62 & n &  & 103 & HD 110315 & 14.11 & 7.93 & K4.5V & 0.43 & 30 & n &  \\
32 & 61 UMa & 9.58 & 5.34 & G8V & 0.79 & 82 & n &  & 104 & 18 Sco & 14.14 & 5.50 & G2Va & 1.04 & 73 & n &  \\
33 & HD 151288 & 9.85 & 8.11 & K7.5Ve & 0.33 & 33 & n &  & 105 & HD 188088 & 14.14 & 6.21 & K3VaCN1 & 0.87 & 62 & n &  \\
34 & 12 Oph & 9.89 & 5.77 & K1V & 0.69 & 70 & n &  & 106 & HD 150689 & 14.22 & 7.51 & K3V & 0.48 & 34 & n &  \\
35 & HD 232979 & 9.91 & 8.65 & M0.5V & 0.28 & 29 & n &  & 107 & HD 234078 & 14.26 & 8.25 & K7V & 0.30 & 21 & n &  \\
36 & HD 10780 & 10.04 & 5.63 & K0V & 0.73 & 73 & n &  & 108 & HD 173818 & 14.26 & 8.77 & K5V & 0.30 & 21 & n &  \\
37 & HD 122064 & 10.07 & 6.17 & K3V & 0.52 & 52 & n &  & 109 & tau01 Eri & 14.28 & 4.46 & F7V & 1.65 & 115 & Y &  \\
38 & HD 103932 & 10.17 & 6.96 & K4+V & 0.47 & 46 & n &  & 110 & HD 144579 & 14.36 & 6.67 & G8V & 0.64 & 44 & n &  \\
39 & alf Men & 10.21 & 5.07 & G7V & 0.94 & 92 & n &  & 111 & pi.01 UMa & 14.44 & 5.64 & G0.5V & 0.99 & 68 & n &  \\
40 & HD 17925 & 10.36 & 6.05 & K1V & 0.64 & 62 & n &  & 112 & HD 184489 & 14.44 & 9.33 & M0V & 0.30 & 21 & n &  \\
41 & HD 154363 & 10.46 & 7.71 & K4/5V & 0.34 & 32 & n &  & 113 & HD 97584 & 14.46 & 7.65 & K4V & 0.48 & 33 & n &  \\
42 & HD 111631 & 10.65 & 8.47 & M0V & 0.33 & 31 & n &  & 114 & BD-08 2582 & 14.50 & 9.48 & K7V & 0.25 & 17 & n &  \\
43 & HD 147379 & 10.77 & 7.90 & M1-Ve & 0.23 & 21 & n & 1 (RV) & 115 & BD+30 2512 & 14.52 & 8.02 & K6V & 0.34 & 24 & n &  \\
44 & HD 115404 & 10.99 & 6.66 & K2V & 0.52 & 48 & n & 2 (RV) & 116 & 111 Tau B & 14.54 & 7.92 & K4V & 0.43 & 30 & n &  \\
45 & HD 11507 & 11.06 & 8.88 & M0V & 0.28 & 26 & n &  & 117 & HD 221503 & 14.55 & 8.61 & K6V & 0.35 & 24 & n &  \\
46 & HD 166620 & 11.10 & 6.40 & K2V & 0.60 & 54 & n &  & 118 & HD 120036A & 14.58 & 8.85 & K6V & 0.32 & 22 & n &  \\
47 & 54 Psc & 11.11 & 5.88 & K0.5V & 0.73 & 66 & n & 1 (RV) & 119 & 111 Tau & 14.58 & 4.85 & F8V & 1.29 & 88 & n &  \\
48 & HD 74576 & 11.19 & 6.56 & K3V & 0.59 & 52 & n &  & 120 & HD 120036B & 14.58 & 9.25 & K7V & 0.28 & 19 & n &  \\
49 & HD 28343 & 11.23 & 8.30 & M0.5V & 0.38 & 34 & n &  & 121 & w Her & 14.59 & 5.39 & G0V & 1.12 & 77 & n &  \\
50 & 11 LMi & 11.23 & 5.34 & G8Va & 0.92 & 82 & n &  & 122 & HD 196761 & 14.67 & 6.37 & G7.5IV-V & 0.75 & 51 & n &  \\
51 & HD 85512 & 11.28 & 7.65 & K6Vk: & 0.42 & 38 & n &  & 123 & BD+40 45 & 14.69 & 8.37 & K7V & 0.30 & 20 & n &  \\
52 & V2689 Ori & 11.42 & 8.90 & K6V & 0.24 & 21 & n &  & 124 & HD 19305 & 14.72 & 9.06 & M0V & 0.35 & 24 & n &  \\
53 & HD 222237 & 11.45 & 7.09 & K3+V & 0.47 & 41 & n & 1 (RV) & 125 & nu.02 Lup & 14.74 & 5.65 & G2-V & 1.01 & 68 & n & 3 (RV) \\
54 & HD 38B & 11.52 & 8.99 & M0.5V & 0.28 & 24 & n &  & 126 & HD 142709 & 14.74 & 8.06 & K4V & 0.41 & 28 & n &  \\
55 & HD 38A & 11.52 & 8.83 & K6V & 0.25 & 22 & n &  & 127 & psi Ser & 14.79 & 5.70 & G2.5V & 1.00 & 68 & n &  \\
56 & HD 99279B & 11.68 & 8.60 & K7-V(k) & 0.30 & 26 & n &  & 128 & HD 205390 & 14.81 & 7.16 & K1V & 0.55 & 37 & n &  \\
57 & CD-57 1079 & 11.69 & 8.93 & K7Vk & 0.26 & 22 & n &  & 129 & HD 144628 & 14.83 & 7.12 & K2V & 0.57 & 38 & n &  \\
58 & zet Dor & 11.69 & 4.71 & F9VFe-0. & 1.22 & 104 & Y &  & 130 & HD 218511 & 14.84 & 8.39 & K6V & 0.40 & 27 & n &  \\
59 & HD 125072 & 11.82 & 6.66 & K3IV & 0.59 & 50 & n &  & 131 & b Aql & 14.92 & 4.97 & G7IVHdel & 0.87 & 58 & n &  \\
60 & HD 97101 & 11.88 & 7.74 & K7V & 0.30 & 25 & n & 2 (RV) & 132 & 20 LMi & 14.93 & 5.37 & G3VaHdel & 1.17 & 78 & n & 1 (RV) \\
61 & lam Ser & 11.92 & 4.42 & G0-V & 1.43 & 120 & Y & 1 (RV) & 133 & HD 84117 & 14.95 & 4.94 & F9V & 1.40 & 94 & n &  \\
62 & zet02 Ret & 12.04 & 5.23 & G1V & 1.00 & 83 & n &  & 134 & HD 200779 & 15.05 & 8.27 & K6V & 0.43 & 28 & n &  \\
63 & zet01 Ret & 12.04 & 5.54 & G2.5VHde & 0.87 & 72 & n &  & 135 & HD 4391 & 15.05 & 5.79 & G3V & 0.97 & 64 & n &  \\
64 & HD 72673 & 12.16 & 6.38 & K1V & 0.64 & 53 & n &  & 136 & BD+63 137 & 15.07 & 8.42 & K7V & 0.30 & 20 & n &  \\
65 & HD 37394 & 12.27 & 6.23 & K0V & 0.68 & 56 & n &  & 137 & YY Gem & 15.08 & 9.27 & M0.5VeFe & 0.33 & 22 & n &  \\
66 & HD 196877 & 12.34 & 8.82 & K7V & 0.29 & 24 & n &  & 138 & HD 38858 & 15.21 & 5.97 & G2V & 0.90 & 59 & n &  \\
67 & lam Aur & 12.56 & 4.71 & G1.5IV-V & 1.32 & 106 & Y &  & 139 & HD 140901 & 15.25 & 6.01 & G7IV & 0.91 & 60 & n & 2 (RV) \\
68 & HD 69830 & 12.58 & 5.95 & G8:V & 0.78 & 62 & n & 3 (RV) & 140 & nu. Phe & 15.26 & 4.96 & F9VFe+0. & 1.41 & 93 & n &  \\
69 & rho01 Cnc & 12.59 & 5.73 & K0IV-V & 0.74 & 59 & n & 5 (RV) & 141 & HD 24916 & 15.28 & 8.01 & K4V & 0.43 & 28 & n &  \\
70 & HD 190007 & 12.72 & 7.08 & K5V & 0.39 & 31 & n & 1 (RV) & 142 & HD 21197 & 15.32 & 7.84 & K4V & 0.47 & 31 & n &  \\
71 & HD 104304 & 12.77 & 5.55 & G8IV & 0.95 & 74 & n &  & 143 & HD 4967 & 15.36 & 8.95 & K5V & 0.30 & 19 & n &  \\
72 & HD 101581 & 12.78 & 7.76 & K4.5Vk: & 0.42 & 32 & n & 2 (tr) &  &  &  &  &  &  &  &  &  \\
\end{longtable}
\end{landscape}

\vspace{2mm}
\noindent{\small\sffamily\color{muted} Planet and host-star data in Figures~\ref{fig:mwhorizon}--\ref{fig:hz} are drawn from the NASA Exoplanet Archive Planetary Systems Composite Parameters table (6{,}319 confirmed planets, accessed 2026 July 11). The face-on Galactic map is by NASA/JPL-Caltech/R.~Hurt (SSC/Caltech). Habitable-zone boundaries in Figure~\ref{fig:hz} use the polynomial coefficients of Kopparapu et al.\ (2013).}

\clearpage
\definecolor{Blue1}{HTML}{1FABD5}
\definecolor{Blue2}{HTML}{1D8DB0}
\definecolor{Blue3}{HTML}{116E8A}
\backmatter
\MakeBackCover

\begin{thebibliography}{99}\small

%%%%%%%%%%%%%%%%%% Nearset
\bibitem{Henry2018} Henry, T.~J., Jao, W.-C., Winters, J.~G., Dieterich, S.~B., Finch, C.~T., Ianna, P.~A., Riedel, A.~R., Silverstein, M.~L., Subasavage, J.~P., \& Vrijmoet, E.~H. 2018, The Solar Neighborhood XLIV: RECONS Discoveries within 10 Parsecs, AJ, 156, 102.
\bibitem{Gaia2022}
Gaia Collaboration, Vallenari, A., Brown, A.~G.~A., et al. 2022,
Gaia Data Release 3: Summary of the contents and survey properties,
Astronomy \& Astrophysics, 674, A1.
\bibitem{Lubin2016}
Lubin, P. 2016,
A Roadmap to Interstellar Flight,
Journal of the British Interplanetary Society,
69, 40--72.
\bibitem{Traub2010} Traub, W.~A., \& Oppenheimer, B.~R. 2010, Direct Imaging of Exoplanets, in \textit{Exoplanets}, ed. S. Seager, University of Arizona Press, 111--156.
\bibitem{LUVOIR2019}
The LUVOIR Team. 2019, The LUVOIR Mission Concept Study Final Report, NASA, arXiv:1912.06219.
\bibitem{Macintosh2015} Macintosh, B., Graham, J.~R., Barman, T., et al. 2015, Discovery and spectroscopy of the young Jovian planet 51 Eridani b with the Gemini Planet Imager, Science, 350, 64--67.
\bibitem{Nielsen2019} Nielsen, E.~L., De Rosa, R.~J., Macintosh, B., et al. 2019, The Gemini Planet Imager Exoplanet Survey: Giant Planet and Brown Dwarf Demographics from 10--100 au, Astronomical Journal, 158, 13.
\bibitem{Bowler2016} Bowler, B.~P. 2016, Imaging Extrasolar Giant Planets, Annual Review of Astronomy and Astrophysics, 54, 271--305.
\bibitem{Currie2022} Currie, T., Biller, B., Lagrange, A.-M., Marois, C., Guyon, O., Nielsen, E.~L., Bonnefoy, M., \& De Rosa, R.~J. 2022, Direct Imaging and Spectroscopy of Extrasolar Planets, arXiv:2205.05696.
\bibitem{HabEx2020} Gaudi, B.~S., Seager, S., Mennesson, B., et al. 2020, The Habitable Exoplanet Observatory (HabEx) Mission Concept Study Final Report, NASA, arXiv:2001.06683.
\bibitem{reyle21} Reyl\'{e} C., Jardine K., Fouqu\'{e} P., et al., The 10 parsec sample in the Gaia era, A\&A 650:A201 (2021).
\bibitem{gaiadr3} Gaia Collaboration (Vallenari A., et al.), Gaia Data Release 3: summary of the content and survey properties, A\&A 674:A1 (2023).
\bibitem{Golovin2023} Golovin, A., Reffert, S., Just, A., Jordan, S., Vani, A., \& Jahrei{\ss}, H. 2023, The Fifth Catalogue of Nearby Stars (CNS5), A\&A, 670, A19.
\bibitem{Akeson2013} Akeson, R.~L., Chen, X., Ciardi, D., et al. 2013, The NASA Exoplanet Archive: Data and Tools for Exoplanet Research, PASP, 125, 989.
\bibitem{Hartkopf2001} Hartkopf, W.~I., Mason, B.~D., \& Worley, C.~E. 2001, The 2001 US Naval Observatory Double Star CD-ROM. II. The Fifth Catalog of Orbits of Visual Binary Stars, AJ, 122, 3472.
\bibitem{Guyon2006} Guyon, O., Pluzhnik, E. A., Kuchner, M. J., Collier Cameron, A., Ridgway, S. T., \& Woodruff, R. A. 2006, Theoretical Limits on Extrasolar Terrestrial Planet Detection with Coronagraphs, ApJS, 167, 81.
\bibitem{Beichman2020} Beichman, C., Ygouf, M., Llop-Sayson, J., Mawet, D., Yung, Y., Choquet, E., Kervella, P., Boccaletti, A.,
Belikov, R., Lissauer, J.~J., Quarles, B., Lagage, P.-O., Dicken, D., Hu, R., Mennesson, B., Ressler, M., Serabyn, E., Krist, J., Bendek, E., Leisenring, J., \& Pueyo, L. 2020, Searching for Planets Orbiting Alpha Centauri A with the James Webb Space Telescope,
PASP, 132, 018002.
\bibitem{Shields2016} Shields, A. L., Ballard, S., \& Johnson, J. A. 2016, The Habitability of Planets Orbiting M-Dwarf Stars,
Phys. Rep., 663, 1.
\bibitem{Sirbu2017} Sirbu, D., Thomas, S.~J., Belikov, R., \& Bendek, E. 2017, Techniques for High-Contrast Imaging in Multi-Star Systems II: Multi-Star Wavefront Control, ApJ, 849, 142.
\bibitem{Thomas2015} Thomas, S.~J., Belikov, R., \& Bendek, E.~A. 2015, Techniques for High-Contrast Imaging in Multi-Star Systems I: Super-Nyquist Wavefront Control, ApJ, 810, 81.
\bibitem{Holman1999} Holman, M.~J., \& Wiegert, P.~A. 1999, Long-Term Stability of Planets in Binary Systems, AJ, 117, 621.
\bibitem{kopparapu13} Kopparapu, R.~K., Ramirez, R., Kasting, J.~F., Eymet, V., Robinson, T.~D., Mahadevan, S., Terrien, R.~C., Domagal-Goldman, S., Meadows, V., \& Deshpande, R. 2013, Habitable Zones Around Main-Sequence Stars: New Estimates, ApJ, 765, 131.
\bibitem{Stark2014}
Stark, C.~C., Roberge, A., Mandell, A., \& Robinson, T.~D. 2014,
Maximizing the ExoEarth Candidate Yield from a Future Direct Imaging Mission,
ApJ, 795, 122.
\bibitem{Schwieterman2018}
Schwieterman, E.~W., Kiang, N.~Y., Parenteau, M.~N., et al. 2018,
Exoplanet Biosignatures: A Review of Remotely Detectable Signs of Life,
Astrobiology, 18, 663.
\bibitem{Tuchow2024}
Tuchow, N.~W., Stark, C.~C., \& Mamajek, E.~E. 2024,
HPIC: The Habitable Worlds Observatory Preliminary Input Catalog,
AJ, 167, 139.
\bibitem{Stark2019}
Stark, C.~C., Belikov, R., Bolcar, M.~R., et al. 2019,
The ExoEarth Yield Landscape for Future Direct Imaging Space Telescopes,
JATIS, 5, 024009.
\bibitem{kopparapu14} Kopparapu, R.~K., Ramirez, R.~M., SchottelKotte, J., Kasting, J.~F., Domagal-Goldman, S., \& Eymet, V. 2014, Habitable Zones Around Main-Sequence Stars: Dependence on Planetary Mass, ApJL, 787, L29.
\bibitem{pecaut13} Pecaut, M.~J., \& Mamajek, E.~E.
2013, Intrinsic Colors, Temperatures, and Bolometric Corrections of
Pre-Main-Sequence Stars, ApJS, 208, 9.
\bibitem{DanchiLopez2013}
Danchi, W.~C., \& Lopez, B. 2013,
Effect of Metallicity on the Evolution of the Habitable Zone from the
Pre-Main Sequence to the Asymptotic Giant Branch and the Search for Life,
ApJ, 769, 27
\bibitem{GuimondCowan2019}
Guimond, C.~M., \& Cowan, N.~B. 2019,
Three Direct Imaging Epochs Could Constrain the Orbit of Earth 2.0
inside the Habitable Zone,
AJ, 157, 197.
\bibitem{Feng2018}
Feng, Y.~K., Robinson, T.~D., Fortney, J.~J., et al. 2018,
Characterizing Earth Analogs in Reflected Light:
Atmospheric Retrieval Studies for Future Space Telescopes,
AJ, 155, 200.
\bibitem{KawashimaRugheimer2019}
Kawashima, Y., \& Rugheimer, S. 2019,
Theoretical Reflectance Spectra of Earth-Like Planets through Their
Evolutions: Impact of Clouds on the Detectability of Oxygen, Water,
and Methane with Future Direct Imaging Missions,
AJ, 157, 213.
\bibitem{Catling2018}
Catling, D.~C., Krissansen-Totton, J., Kiang, N.~Y., et al. 2018,
Exoplanet Biosignatures: A Framework for Their Assessment,
Astrobiology, 18, 709.
\bibitem{Meadows2018}
Meadows, V.~S., Reinhard, C.~T., Arney, G.~N., et al. 2018,
Exoplanet Biosignatures: Understanding Oxygen as a Biosignature
in the Context of Its Environment,
Astrobiology, 18, 630.
\bibitem{KrissansenTotton2018}
Krissansen-Totton, J., Olson, S., \& Catling, D.~C. 2018,
Disequilibrium Biosignatures over Earth History and Implications
for Detecting Exoplanet Life,
Sci. Adv., 4, eaao5747.

%%%%%%%%%%%%%%%%%%%% B.3
\bibitem{galicher23} Galicher R., Mazoyer J., Imaging exoplanets with coronagraphic instruments, C.~R.~Physique 24:69 (2023).
\bibitem{zurlo24} Zurlo A., Direct imaging of exoplanets, arXiv:2404.05797 (2024).
\bibitem{bailey23} Bailey V.~P., Bendek E., Monacelli B., et al., Nancy Grace Roman Space Telescope Coronagraph Instrument overview and status, Proc.~SPIE 12680:126800T (2023), doi:10.1117/12.2679036.
\bibitem{llopsayson25} Llop-Sayson J., Bailey V.~P., Hom J., et al., Roman coronagraph simulations of exozodi observations in the presence of wavefront errors, JATIS 11:045009 (2025), doi:10.1117/1.JATIS.11.4.045009.
\bibitem{traub10} Traub W.~A., Oppenheimer B.~R., Direct imaging of exoplanets, in Seager S. (ed.), \emph{Exoplanets}, University of Arizona Press, Tucson, pp.~111--156 (2010).
\bibitem{deshler24} Deshler N., Haffert S., Ashok A., Quantum limits of exoplanet detection and localization, Phys.~Rev.~A 113:042406 (2026), doi:10.1103/g6kr-mqdj.
\bibitem{macintosh15} Macintosh B., Graham J.~R., Barman T., et al., Discovery and spectroscopy of the young jovian planet 51~Eri~b with the Gemini Planet Imager, Science 350:64 (2015).
\bibitem{vanleeuwen07} van Leeuwen F., Validation of the new Hipparcos reduction, A\&A 474:653 (2007).
\bibitem{hwo26} NASA, Habitable Worlds Observatory, NASA Science, \url{https://science.nasa.gov/astrophysics/programs/habitable-worlds-observatory/} (accessed 2026 July 28).
\bibitem{willmer18} Willmer C.~N.~A., The absolute magnitude of the Sun in several filters, ApJS 236:47 (2018).
\bibitem{raghavan10} Raghavan D., McAlister H.~A., Henry T.~J., et al., A survey of stellar families: multiplicity of solar-type stars, ApJS 190:1 (2010).
\bibitem{winters19} Winters J.~G., Henry T.~J., Jao W.-C., et al., The Solar Neighborhood. XLV. The stellar multiplicity rate of M dwarfs within 25\,pc, AJ 157:216 (2019).
\bibitem{lindegren21} Lindegren L., Klioner S.~A., Hern\'{a}ndez J., et al., Gaia Early Data Release 3: the astrometric solution, A\&A 649:A2 (2021).
\bibitem{stassun21} Stassun K.~G., Torres G., Parallax systematics and photocenter motions of benchmark eclipsing binaries in Gaia EDR3, ApJL 907:L33 (2021), doi:10.3847/2041-8213/abdaad.
\bibitem{leinert98} Leinert C., Bowyer S., Haikala L.~K., et al., The 1997 reference of diffuse night sky brightness, A\&AS 127:1 (1998).
\bibitem{ertel20} Ertel S., Defr\`{e}re D., Hinz P., et al., The HOSTS survey for exozodiacal dust: observational results from the complete survey, AJ 159:177 (2020).
\bibitem{stark14} Stark C.~C., Roberge A., Mandell A., Robinson T.~D., Maximizing the exoEarth candidate yield from a future direct imaging mission, ApJ 795:122 (2014).
\bibitem{williams96} Williams R.~E., Blacker B., Dickinson M., et al., The Hubble Deep Field: observations, data reduction, and galaxy photometry, AJ 112:1335 (1996).
\bibitem{bowler16} Bowler B.~P., Imaging extrasolar giant planets, PASP 128:102001 (2016).
\bibitem{sfd98} Schlegel D.~J., Finkbeiner D.~P., Davis M., Maps of dust infrared emission for use in estimation of reddening and cosmic microwave background radiation foregrounds, ApJ 500:525 (1998).
\bibitem{lallement19} Lallement R., Babusiaux C., Vergely J.~L., et al., Gaia-2MASS 3D maps of Galactic interstellar dust within 3\,kpc, A\&A 625:A135 (2019).
\bibitem{marois08} Marois C., Macintosh B., Barman T., et al., Direct imaging of multiple planets orbiting the star HR~8799, Science 322:1348 (2008).
\bibitem{marois10} Marois C., Zuckerman B., Konopacky Q.~M., Macintosh B., Barman T., Images of a fourth planet orbiting HR~8799, Nature 468:1080 (2010).
\bibitem{lagrange25} Lagrange A.-M., Wilkinson C., M\^{a}lin M., et al., Evidence for a sub-Jovian planet in the young TWA~7 disk, Nature 642:905 (2025), doi:10.1038/s41586-025-09150-4.
\bibitem{gaudi21} Gaudi B.~S., Meyer M.~R., Christiansen J.~L. (eds.), \emph{ExoFrontiers: Big Questions in Exoplanetary Science}, IOP Publishing, Bristol (2021).

%%%%%%%%%B.4
\bibitem{trauger07}
Trauger J.~T., Traub W.~A.,
A laboratory demonstration of the capability to image an Earth-like extrasolar planet,
Nature 446:771--773 (2007).
\bibitem{lawson13}
Lawson P.~R., Belikov R., Cash W., Clampin M., Glassman T.,
Guyon O., Kasdin N.~J., Kern B.~D., Lyon R., Mawet D.,
Moody D., Samuele R., Serabyn E., Sirbu D., Trauger J.,
Survey of experimental results in high-contrast imaging for future exoplanet missions,
Proc. SPIE 8864:88641F (2013).
\bibitem{nemati23}
Nemati B., Krist J.~E., Poberezhskiy I., Kern B.~D.,
The Analytical Performance Model and Error Budget for the Roman Coronagraph Instrument,
J. Astron. Telesc. Instrum. Syst. 9:034007 (2023).
\bibitem{brown05}
Brown R.~A.,
Single-Visit Photometric and Obscurational Completeness,
ApJ 624:1010--1024 (2005).
\bibitem{brown10}
Brown R.~A., Soummer R.,
New Completeness Methods for Estimating Exoplanet Discoveries by Direct Detection,
ApJ 715:122--131 (2010).
\bibitem{krist23}
Krist J.~E., Steeves J.~B., Dube B.~D., Riggs A.~J.~E.,
Kern B.~D., Marx D.~S., Cady E.~J., Zhou H.,
Poberezhskiy I.~Y., Baker C.~W., et al.,
End-to-End Numerical Modeling of the Roman Space Telescope Coronagraph,
J. Astron. Telesc. Instrum. Syst. 9:045002 (2023).


%%%%%%%%%%%%B.5.1
\bibitem{fernandes19} Fernandes R.~B., Mulders G.~D., Pascucci I., et al., Hints for a turnover at the snow line in the giant planet occurrence rate, ApJ 874:81 (2019).
\bibitem{fulton21} Fulton B.~J., Rosenthal L.~J., Hirsch L.~A., et al., California Legacy Survey. II. Occurrence of giant planets beyond the ice line, ApJS 255:14 (2021).
\bibitem{fulton17} Fulton B.~J., Petigura E.~A., Howard A.~W., et al., The California-Kepler Survey. III. A gap in the radius distribution of small planets, AJ 154:109 (2017).
\bibitem{bryson21} Bryson S., Kunimoto M., Kopparapu R.~K., et al., The occurrence of rocky habitable-zone planets around solar-like stars from Kepler data, AJ 161:36 (2021).
\bibitem{cahoy10} Cahoy K.~L., Marley M.~S., Fortney J.~J., Exoplanet albedo spectra and colors as a function of planet phase, separation, and metallicity, ApJ 724:189 (2010), doi:10.1088/0004-637X/724/1/189.
\bibitem{robinson18} Robinson T.~D., Characterizing exoplanet habitability, in \emph{Handbook of Exoplanets}, Springer, Cham, pp.~1--34 (2018).
\bibitem{karkoschka94} Karkoschka E., Spectrophotometry of the Jovian planets and Titan at 300- to 1000-nm wavelength: the methane spectrum, Icarus 111:174 (1994).
\bibitem{schwieterman18} Schwieterman E.~W., Kiang N.~Y., Parenteau M.~N., et al., Exoplanet biosignatures: a review of remotely detectable signs of life, Astrobiology 18:663 (2018).
\bibitem{damiano23} Damiano M., Hu R., Mennesson B., Reflected spectroscopy of small exoplanets. III. Probing the UV band to measure biosignature gases, AJ 166:157 (2023).
\bibitem{snellen15} Snellen I., de Kok R.~J., Birkby J.~L., et al., Combining high-dispersion spectroscopy with high contrast imaging. Probing rocky planets around our nearest neighbors, A\&A 576:A59 (2015), doi:10.1051/0004-6361/201425018.
\bibitem{nea} NASA Exoplanet Archive, Planetary Systems Composite Parameters table, NASA Exoplanet Science Institute, IPAC/Caltech, \url{https://exoplanetarchive.ipac.caltech.edu} (accessed 2026 July 11).
\bibitem{kopparapu18} Kopparapu R.~K., H\'{e}brard E., Belikov R., et al., Exoplanet classification and yield estimates for direct imaging missions, ApJ 856:122 (2018).
\bibitem{howe25} Howe A.~R., Becker J.~C., Stark C.~C., Adams F.~C., Architecture classification for extrasolar planetary systems, AJ 169:149 (2025).
\bibitem{borucki10} Borucki W.~J., Koch D., Basri G., et al., Kepler planet-detection mission: introduction and first results, Science 327:977 (2010).
\bibitem{seager13} Seager S., Exoplanet habitability, Science 340:577 (2013).
\bibitem{hong23} Hong S.~E., Kwon R.~Y., Kim Y., Kang H., Kim M., Exoplanets and habitability, PKAS 38:13 (2023).
\bibitem{claudi19} Claudi R., Alei E., Biosignatures search in habitable planets, Galaxies 7:82 (2019).
\bibitem{airapetian17escape} Airapetian V.~S., Glocer A., Khazanov G.~V., et al., How hospitable are space weather affected habitable zones? The role of ion escape, ApJL 836:L3 (2017), doi:10.3847/2041-8213/836/1/L3.
\bibitem{ridgway23} Ridgway R.~J., Zamyatina M., Mayne N.~J., et al., 3D modelling of the impact of stellar activity on tidally locked terrestrial exoplanets: atmospheric composition and habitability, MNRAS 518:2472 (2023).
\bibitem{chen21} Chen H., Zhan Z., Youngblood A., et al., Persistence of flare-driven atmospheric chemistry on rocky habitable zone worlds, Nature Astronomy 5:298 (2021), doi:10.1038/s41550-020-01264-1.
\bibitem{airapetian16} Airapetian V.~S., Glocer A., Gronoff G., H\'{e}brard E., Danchi W., Prebiotic chemistry and atmospheric warming of early Earth by an active young Sun, Nature Geoscience 9:452 (2016).
\bibitem{kobayashi23} Kobayashi K., Ise J., Aoki R., et al., Formation of amino acids and carboxylic acids in weakly reducing planetary atmospheres by solar energetic particles from the young Sun, Life 13:1103 (2023).
\bibitem{namekata26} Namekata K., France K., Chae J., et al., Discovery of multi-temperature coronal mass ejection signatures from a young solar analogue, Nature Astronomy 10:64 (2026), doi:10.1038/s41550-025-02691-8.
\bibitem{gallet17} Gallet F., Bouvier J., Improved angular momentum evolution model for solar-like stars. II. Exploring the mass dependence, A\&A 577:A98 (2015), doi:10.1051/0004-6361/201525660.
\bibitem{astro2020} National Academies of Sciences, Engineering, and Medicine, \emph{Pathways to Discovery in Astronomy and Astrophysics for the 2020s}, The National Academies Press, Washington, DC (2021).
\bibitem{voyage2050} European Space Agency, Voyage 2050 sets sail. ESA chooses future science mission themes, \url{https://www.esa.int/Science_Exploration/Space_Science/Voyage_2050_sets_sail_ESA_chooses_future_science_mission_themes} (2021).
\bibitem{tinetti18} Tinetti G., Drossart P., Eccleston P., et al., A chemical survey of exoplanets with Ariel, Experimental Astronomy 46:135 (2018).
\bibitem{rauer14} Rauer H., Catala C., Aerts C., et al., The PLATO 2.0 mission, Experimental Astronomy 38:249 (2014).
\bibitem{quanz22} Quanz S.~P., Ottiger M., Fontanet E., et al., Large interferometric space missions for exoplanet direct imaging and biosignature detection (LIFE~I), A\&A 664:A21 (2022).
\bibitem{romanfor} NASA Goddard Space Flight Center, Roman observatory technical information, \url{https://roman.gsfc.nasa.gov/science/observatory_technical.html} (accessed 2026 July 28).
\bibitem{roy2026III} Roy A., et al., The Lazuli Space Observatory: Architecture \& Capabilities, arXiv:2601.02556 (2026).
\bibitem{wevers2026III} Wevers T., et al., The Lazuli Space Observatory: Opportunities for time-domain and multi-messenger astronomy, arXiv:2606.17136 (2026).
\end{thebibliography}
\end{document}